\documentclass[aps,prl,twocolumn,superscriptaddress,footinbib,notitlepage,longbibliography,amsmath,amssymb,floatfix]{revtex4-2}
\usepackage{times}
\usepackage{bm,graphicx,booktabs,mathtools,amsthm,multirow}
\usepackage{xcolor}
\usepackage[colorlinks=true,linkcolor=blue!45!black,citecolor=blue!45!black,urlcolor=blue!45!black]{hyperref}
\usepackage[expansion=false]{microtype}
\graphicspath{{figures/}}

\newcommand{\F}{\mathbb F_2}
\newcommand{\wt}{\operatorname{wt}}
\newcommand{\rk}{\operatorname{rank}}

\newcommand{\dist}{\operatorname{dist}}
\newcommand{\row}{\operatorname{row}}
\newcommand{\cod}[3]{[\![#1,#2,#3]\!]}
\newcommand{\XX}{\overline X}
\newcommand{\ZZ}{\overline Z}
\newcommand{\YY}{\overline Y}
\newcommand{\PP}{\overline P}
\newcommand{\ket}[1]{\lvert #1\rangle}
\newcommand{\IB}{\mathcal I_B}
\newcommand{\IH}{\mathcal I_H}
\newcommand{\IF}{\mathcal I_F}
\newcommand{\Sg}{\mathcal S}
\newcommand{\wmax}{w_{\max}}
\newcommand{\Dmax}{\Delta_{\max}}
\newcommand{\Rq}{\mathcal R_q}
\newcommand{\one}{\mathbf 1}
\newcommand{\src}{\mathrm{src}}

\makeatletter
\newcounter{headingbookmark}
\newcommand{\@headingmark}[2]{\stepcounter{headingbookmark}\pdfbookmark[#1]{#2}{heading.\theheadingbookmark}}
\newcommand{\majorheading}[1]{\par\addvspace{2.2ex}\@headingmark{1}{#1}\noindent{\large\bfseries #1\par}\nobreak\vspace{0.6ex}\@afterheading}
\newcommand{\mainheading}[1]{\par\addvspace{1.6ex}\@headingmark{2}{#1}\noindent{\normalsize\bfseries #1\par}\nobreak\vspace{0.3ex}\@afterheading}
\makeatother

\theoremstyle{plain}
\newtheorem{theorem}{Theorem}
\newtheorem{lemma}{Lemma}
\newtheorem{proposition}{Proposition}

\newenvironment{datatable}{\begin{table*}[p]}{\end{table*}}

\begin{document}
\renewcommand{\figurename}{Fig.}
\renewcommand{\tablename}{Tab.}
\renewcommand{\thetable}{\arabic{table}}

\title{A modular quantum computer based on Cornucopia codes}

\author{Zhide Lu}
\affiliation{Shanghai Qi Zhi Institute, Shanghai 200232, China}

\author{Weikang Li}
\affiliation{Center for Quantum Information, IIIS, Tsinghua University, Beijing 100084, China}
\affiliation{Hefei National Laboratory, Hefei 230088, China}

\author{Yinghao Ma}
\affiliation{Center for Quantum Information, IIIS, Tsinghua University, Beijing 100084, China}

\author{Hanyu Yan}
\affiliation{Center for Quantum Information, IIIS, Tsinghua University, Beijing 100084, China}

\author{Bingrun Wang}
\affiliation{Center for Quantum Information, IIIS, Tsinghua University, Beijing 100084, China}

\author{Yixuan Ma}
\affiliation{Center for Quantum Information, IIIS, Tsinghua University, Beijing 100084, China}

\author{Zheng-Zhi Sun}
\affiliation{Center for Quantum Information, IIIS, Tsinghua University, Beijing 100084, China}
\affiliation{Hefei National Laboratory, Hefei 230088, China}

\author{Dong-Ling Deng}
\email{dldeng@tsinghua.edu.cn}
\affiliation{Center for Quantum Information, IIIS, Tsinghua University, Beijing 100084, China}
\affiliation{Shanghai Qi Zhi Institute, Shanghai 200232, China}
\affiliation{Hefei National Laboratory, Hefei 230088, China}

\begin{abstract}\noindent
\bfseries\boldmath
Ultra-high-rate quantum error correction codes store many logical qubits in a single code block with few physical qubits.
A fault-tolerant quantum computer, however, also requires a universal set of logical operations on individually addressable logical qubits, and the ancilla overhead of these operations may erode the advantage of the high encoding rate.
Here, we propose the Cornucopia architecture, in which a universal set of logical operations is implemented at low overhead for the Cornucopia codes. In particular, we focus on the $\cod{2844}{1426}{18}$ Cornucopia code as an example and introduce a modular quantum computer built from such code blocks connected by bridges. Each block stores 1{,}426 logical qubits and serves as a universal logical processor. Exploiting the code's translation symmetry, we construct a family of reusable gadgets that individually address 1{,}422 logical qubits within each block. These gadgets support three universal logical instruction sets: basic, high-throughput and fixed-extractor. Excluding magic-state factories, a module with sufficient workspace to execute any instruction in the basic set requires an average of 4.1 physical qubits per logical qubit, compared with 3.0 for memory alone.
The architecture thus preserves the resource advantage of ultra-high-rate quantum memory while enabling universal logical computation, providing a route towards low-overhead fault-tolerant quantum computers.
\end{abstract}

\maketitle

\noindent 
\noindent Quantum computers promise speedups over the best known classical algorithms for problems ranging from factorization~\cite{Shor1997PolynomialTime} to the simulation of molecules for chemistry~\cite{AspuruGuzik2005Simulated}. Instances of practical interest, however, involve thousands of logical qubits and billions of logical operations~\cite{Gidney2021FactorRSA,Beverland2022Practical}, and therefore require logical error rates many orders of magnitude below the physical error rates of current hardware, which are of order $10^{-3}$. Quantum error correction can close this gap by spreading logical information over many physical qubits and repeatedly measuring parity checks whose outcomes reveal errors without disturbing the encoded state~\cite{Shor1995Scheme,Terhal2015Quantum}.
Error-corrected logical qubits have now been demonstrated experimentally~\cite{Dasu2026Computing,Tham2026Breakeven,Gupta2024Encoding,Lacroix2025Scaling,GoogleQuantumAIandCollaborators2025Quantum,He2025Experimental,Rosenfeld2025Magic,Wang2026Demonstration,Bluvstein2024Logical,SalesRodriguez2025Experimental,Bluvstein2026Faulttolerant}, but a reliable quantum memory is only the first step towards a useful fault-tolerant quantum computer.
A practical architecture must also specify how individual logical qubits are addressed, which logical operations are available, how many of them can be executed in parallel, how separate hardware modules exchange quantum information and how non-Clifford resources are supplied~\cite{Litinski2019GameOfSurfaceCodes,Yoder2025Tour,Bhardwaj2026Highrate}. The physical-qubit overhead of all these components, rather than that of the memory alone, determines the cost of fault-tolerant quantum computation.

The surface code has long been the leading candidate for such an architecture because of its high error threshold and geometrically local checks, with logical operations implemented by lattice surgery~\cite{Fowler2012Surface,Horsman2012Surface,Litinski2019GameOfSurfaceCodes}. Its low encoding rate, however, entails a physical-qubit overhead that grows rapidly with the code distance. Quantum low-density parity-check (qLDPC) codes reduce this overhead by encoding many logical qubits into a single code block with bounded-weight checks~\cite{Gottesman2014Faulttolerant,Breuckmann2021Quantum,Zhou2025Opportunities,Bravyi2024Highthreshold,Xu2024Constantoverhead,Liang2025Generalized,Liang2025Planar,Liu2026LargeLanguageModel,Choe2026Barbell,Nixon2026Vine,Gu2026Nearestneighbour,Khesin2026Mirror,Lee2026Logical,Zhao2026UltraHighRate}. Because logical qubits that share a code block cannot, in general, be addressed individually by transversal operations, logical operations are instead implemented by qLDPC surgery, in which an ancillary system is coupled to the code block to measure selected logical Pauli operators~\cite{Vuillot2019Code,Cohen2022Lowoverhead,Cross2024Improved,Swaroop2024Universal,Ide2025Faulttolerant,He2025Extractors,Hamada2026Boundeddepth,Williamson2026Lowoverhead,Cowtan2026Parallel,Gu2026QGPU,Zhou2026Parallel}. 
Recent work has developed universal logical instruction sets for specific qLDPC codes: the bicycle architecture equips bivariate bicycle modules, each encoding $12$ logical qubits, with logical processing units and inter-module adapters~\cite{Yoder2025Tour}; mitten codes and canonical lifted-product codes use group-covariant canonical bases to derive their instruction sets from a few reusable seed gadgets~\cite{Bhardwaj2026Highrate,Zheng2026Logical}.
For ultra-high-rate codes, the code symmetry has been used to measure a Pauli product together with all of its translates at low overhead~\cite{Maskara2026Efficient}.
 
Cornucopia codes~\cite{Lu2026Quantum} are a class of high-rate qLDPC codes with an encoding rate exceeding $1/2$ and weight-$12$ checks, constructed by co-designing the code structure with the capabilities of reconfigurable neutral-atom arrays~\cite{Bluvstein2026Faulttolerant}. Their checks are specified by coordinate permutations that decompose into collective column translations and a few row swaps, so that a complete syndrome extraction cycle consists of $12$ layers of transversal gates between blocks of atoms that move as rigid units. In particular, a single $\cod{2844}{1426}{18}$ code block encodes $1{,}426$ logical qubits using $4{,}266$ physical qubits and achieves an extrapolated logical error rate of $2.6\times10^{-16}$ per logical qubit per cycle under circuit-level noise at a physical error rate of $10^{-3}$~\cite{Lu2026Quantum}. 
However, turning such a block into a logical processor poses several challenges: individually addressing more than a thousand logical qubits requires gadgets that are reused across the block rather than constructed one by one; the representatives of a canonical logical basis are much heavier than the code distance, which raises the ancilla cost of every surgery operation; every merged code must provably preserve the code distance; and interconnects between blocks must remain inexpensive.

\begin{figure*}[t]
\centering\includegraphics[width=\textwidth]{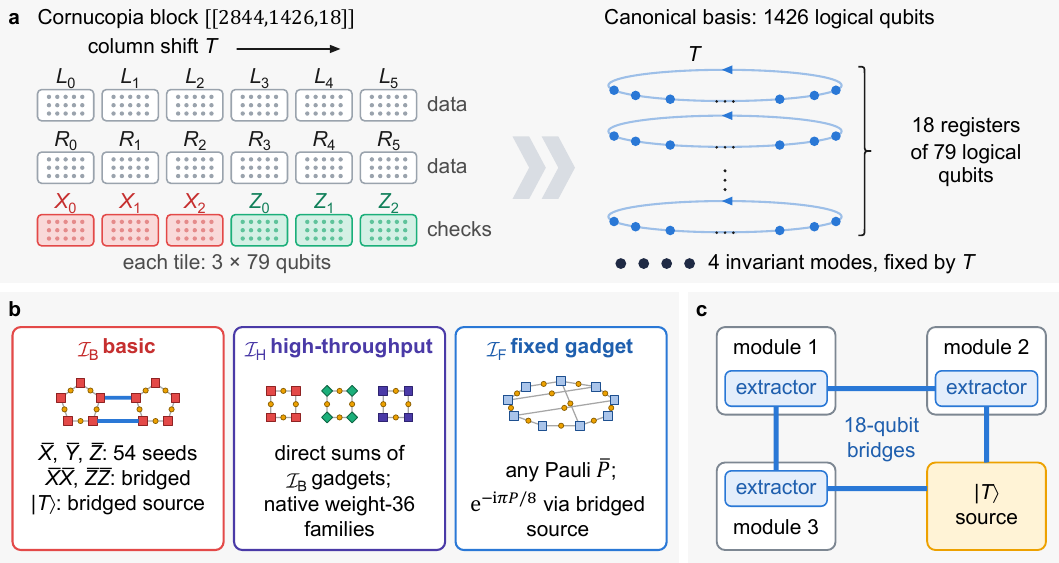}
\caption{
\textbf{A modular quantum computer based on Cornucopia codes.} \textbf{a},~One $\cod{2844}{1426}{18}$ Cornucopia block~\cite{Lu2026Quantum}: twelve data tiles $L_j,R_j$ and six check tiles, each a $3\times79$ array; every check acts on one qubit in each data tile, located by a row permutation and a column translation. The column translation $T$ of the data tiles is a code automorphism; in our canonical logical basis it cyclically shifts each of the $18$ registers of $79$ logical qubits and fixes the four invariant modes, which together make up the $1{,}426$ logical qubits.
\textbf{b},~Three instruction sets: $\IB$ is generated by $54$ reusable seed gadgets, their translates and matching bridges; $\IH$ executes compatible $\IB$ gadgets in parallel as direct sums and measures native weight-$36$ operator families; $\IF$ is a single fixed extractor that can measure any logical Pauli product.
\textbf{c},~Modules, each a Cornucopia block with its extractor, are joined by $18$-qubit bridges between their extractors to form a modular quantum computer; magic-state sources attach through bridges of the same kind.
}
\label{fig:overview}
\end{figure*}

Here, we address these challenges and develop the Cornucopia architecture (Fig.~\ref{fig:overview}), a modular quantum computer in which each $\cod{2844}{1426}{18}$ code block serves as a universal logical processor. The architecture has three properties.
First, the logical qubits are individually addressable: each single-qubit Pauli measurement and each two-qubit $\XX\XX$ and $\ZZ\ZZ$ measurement on the $1{,}422$ logical qubits of the $18$ registers introduced below has its own gadget.
Second, the processor is parallel and expressive: parallel surgery can execute many logical measurements and magic-state injections simultaneously in a single merged code, and a fixed extractor can measure any logical Pauli product.
Third, the architecture is modular: modules and magic-state sources are joined by matching bridges of $18$ ancilla data qubits and $17$ ancilla checks.
Throughout, every surgery operation is certified at the level of its merged code~\cite{Cross2024Improved,Williamson2026Lowoverhead,Bhardwaj2026Highrate}: 
the merged code is constructed explicitly, measures exactly the target logical operators and provably preserves the code distance, which we establish with exact certificates.

We achieve these properties with four techniques. First, we organize the logical qubits by the symmetry-adapted standard form~\cite{Gottesman1997Stabilizer,Maskara2026Efficient}, which applies Gottesman's standard form to whole orbits of the code's translation symmetry and yields $18$ cyclic registers of $79$ qubits and four invariant modes, so that a gadget built for one logical qubit serves its entire register. We then reduce the weight of the logical representatives by optimizing which orbits serve as pivots and by symmetry-preserving transformations that mix the registers. 
Second, two gadgets are joined by a matching bridge of only $d$ ancilla data qubits, which preserves the code distance~\cite{Cross2024Improved,Swaroop2024Universal}, so that two-logical-qubit measurements, links between modules and magic-state injection all reuse the same gadgets.
Third, we introduce a ``load-aware'' construction that chooses the cycle checks of every gadget such that each ancilla data qubit takes part in only a few of them.
Fourth, we construct a single fixed extractor and, exploiting the translation symmetry, prove with an exact certificate that it can measure any logical Pauli product without reducing the code distance.
On this foundation we develop basic, high-throughput and fixed-gadget logical instruction sets.

\begin{figure*}[t]
\centering\includegraphics[width=\textwidth]{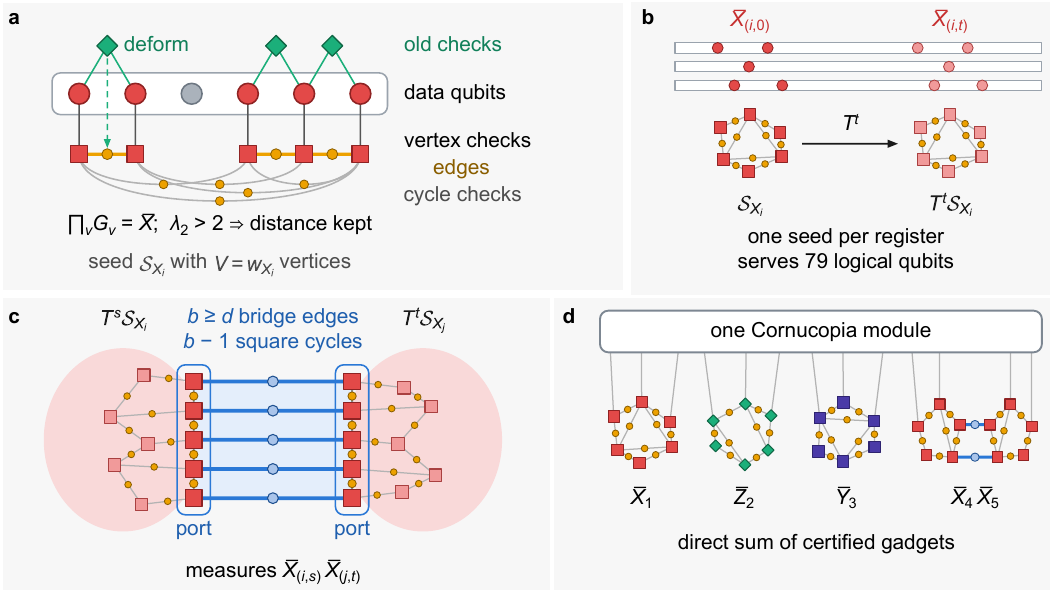}
\caption{
\textbf{Surgery toolkit.} 
\textbf{a},~Schematic of a seed gadget $\Sg_{X_i}$ measuring $\XX$: one vertex check (square) is attached to each data qubit in the support of the target (red), old checks (diamonds) that anticommute locally with the target are deformed onto skeleton edges (orange; every edge is an ancilla data qubit), and the product of all vertex checks equals $\XX$. Expansion edges raise the spectral gap of the gadget graph to $\lambda_2>2$, which keeps the code distance, and cycle checks on a complete cycle basis ensure that the gadget encodes no additional logical qubits.
\textbf{b},~Translation covariance: $T^t$ maps the seed gadget for $\XX_{(i,0)}$ to one for $\XX_{(i,t)}$, so that one seed per register serves its $79$ logical qubits.
\textbf{c},~A matching bridge of $b\ge d$ edges between fixed ports of two translated seeds, completed by $b-1$ square cycle checks, measures the product $\XX_{(i,s)}\XX_{(j,t)}$ of their targets.
\textbf{d},~A direct sum of multiple compatible certified gadgets on one module measures these independent targets in parallel in a single merged code.
}
\label{fig:toolkit}
\end{figure*}

\vspace{.2cm}
\noindent\textbf{\large{}Processor model and instruction sets} \\
\noindent 
A Cornucopia code~\cite{Lu2026Quantum} is defined on twelve data tiles $L_0,\dots,L_5,R_0,\dots,R_5$ and six check tiles $X_0,X_1,X_2,Z_0,Z_1,Z_2$, each a $3\times q$ array of qubits (Fig.~\ref{fig:overview}\textbf{a}). Every check couples tiles through coordinate permutations that combine a row map with a column shift (Methods); for $q=79$ these yield the $\cod{2844}{1426}{18}$ code with weight-$12$ checks. The column translation $T$, which shifts every data tile by one column, maps the support of each check onto that of the check one column further, because column shifts commute with the coordinate permutations; it is therefore an automorphism of the code. We call the $79$ qubits at a fixed tile and row a \emph{fibre}; the data qubits form $36$ fibres, each of which is a single orbit of $T$.

\begin{figure*}[t]
\centering\includegraphics[width=\textwidth]{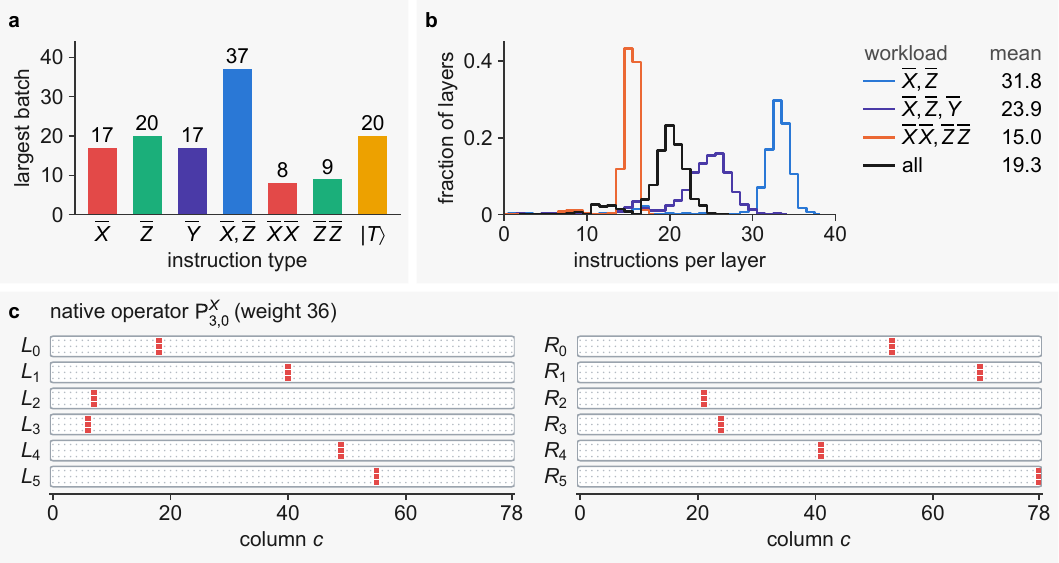}
\caption{
\textbf{High-throughput instructions.} 
\textbf{a},~Largest verified parallel batch for each instruction type, that is, the number of instructions executed in a single merged code, including mixed single-qubit $\XX$ and $\ZZ$ measurements on distinct logical qubits and simultaneous magic-state injections ($\ket T$).
\textbf{b},~Distribution of the number of instructions per layer when random workloads of $3{,}000$ instructions are scheduled into compatible parallel layers, for single-qubit $\XX$ and $\ZZ$ measurements, single-qubit measurements including $\YY$, bridged products $\XX\XX$ and $\ZZ\ZZ$, and all five basic types mixed (all); for each type, the histogram combines three independent workloads, and the legend gives their mean number of instructions per layer.
\textbf{c},~Support of the native logical operator $\mathsf P^X_{3,0}$ of weight $36$: one column of three qubits in each of the twelve data tiles.
}
\label{fig:parallel}
\end{figure*}

A module of the Cornucopia processor is one Cornucopia code block $\cod{2844}{1426}{18}$ together with its logical gadgets. Its logical qubits are labelled in a fixed canonical basis, constructed below, as $(i,t)$ with register $i\in\{0,\dots,17\}$ and position $t\in\mathbb Z_{79}$ for the $1{,}422$ moving logical qubits, together with four invariant modes. In this work, the instruction sets address the moving logical qubits, which we also call register qubits; the invariant modes complete the basis but carry no gadgets of their own and are reached only by the fixed extractor. 
Following refs.~\cite{Yoder2025Tour,Bhardwaj2026Highrate}, an instruction is a logical operation together with the physical gadget that realizes it. 
Apart from the shift automorphism and the transversal CNOT gate between identical modules, each logical instruction is a logical Pauli-product measurement (PPM) or is compiled from such measurements. 
A PPM of a logical operator $\PP$ is performed by code surgery~\cite{Horsman2012Surface,Vuillot2019Code,Cohen2022Lowoverhead,Cross2024Improved,Williamson2026Lowoverhead,Ide2025Faulttolerant}: ancilla qubits and new checks are attached to the code such that the resulting merged code contains $\PP$ as a product of new checks, which are measured for $O(d)$ syndrome rounds before the ancilla qubits are measured out. 
We call a gadget \emph{certified} if its merged code measures exactly $\PP$ and no other logical operator, has maximum check weight $\wmax\le16$ and maximum qubit degree $\Dmax\le16$, and has distance at least the code distance $d=18$.

We consider three types of logical instruction sets (Fig.~\ref{fig:overview}\textbf{b} and Extended Data Tab.~\ref{tab:isa}). The basic set $\IB$ consists of the single-qubit measurements $\XX_j$, $\YY_j$ and $\ZZ_j$, the two-qubit measurements $\XX_j\XX_k$ and $\ZZ_j\ZZ_k$ for all register qubits $j\neq k$, and the injection of a magic state $\ket T$ into any register qubit; each of these instructions has its own certified gadget, so that each register qubit is individually addressable. 
The high-throughput set $\IH$ executes compatible $\IB$ instructions simultaneously in a single merged code and measures native logical-operator families that return many independent outcomes. The fixed-gadget set $\IF$ measures an arbitrary logical Pauli product with a single ancilla system whose connections are selected in software, and implements the rotation $e^{-i\pi\PP/8}$ by consuming a magic state. These sets are complemented by the shift automorphism $\overline U^{s}$, implemented by the column translation $T^s$ of all the data tiles, and by joint measurements between modules. Each type of logical instruction set is universal (Methods), and the three sets are complementary~\cite{Bhardwaj2026Highrate}: $\IB$ minimizes the workspace per operation, $\IH$ maximizes the number of operations per logical cycle, and $\IF$ offers the most expressive operation on fixed hardware.

\vspace{.2cm}
\noindent\textbf{\large{}Canonical logical basis} \\
\noindent 
A logical basis is canonical if its $X$- and $Z$-type representatives $x_j$ and $z_j$ satisfy $x_j\cdot z_l=\delta_{jl}$ for all $j$ and $l$; this pairing certifies that all $2k$ representatives are independent and nontrivial. Unlike for lifted-product codes with square-invertible base matrices~\cite{Bhardwaj2026Highrate}, such a basis cannot be read off from codewords of factor codes. It follows instead from the symmetry-adapted standard form~\cite{Maskara2026Efficient}, which applies Gottesman's standard form~\cite{Gottesman1997Stabilizer} to whole symmetry orbits: the $36$ fibres are partitioned into two pivot sets $P_X$ and $P_Z$ of nine fibres each and a free set of $18$ information fibres, such that the check matrices restricted to the pivot fibres retain their full rank (Methods, Theorem~\ref{thm:basis}). Each free fibre then carries a pair of seed representatives $x_i$ and $z_i$, whose translates $T^tx_i$ and $T^tz_i$ form a register of $79$ canonical logical qubits on which $T$ acts as the cyclic shift $(i,t)\mapsto(i,t+1)$ (Extended Data Fig.~\ref{edfig:basis}\textbf{a}). 
The four remaining logical qubits span an invariant sector of $T$ and form the invariant modes, for which we construct explicit canonical pairs (Methods).

Since the cost of graph surgery grows with the weight of the measured representative, we jointly optimize the basis over three kinds of freedom that preserve both canonicity and translation covariance: the fibre partition, polynomial symplectic shears that add translates of one register to another, and stabilizer dressing (Methods). 
A randomized search followed by simulated annealing of the partition reduces the total weight of the $36$ seed representatives from $11{,}580$ (first valid partition) to $4{,}060$. Greedy shears then lower it to $3{,}740$ in $17$ steps (Extended Data Fig.~\ref{edfig:basis}\textbf{b}); since they mix the registers, the final basis lies outside the standard form.
The final seeds for $X$-, $Z$- and $Y$-type logical operators have weights of $94$--$120$, $62$--$120$ and $169$--$231$, respectively (Extended Data Fig.~\ref{edfig:basis}\textbf{c}).

\begin{figure*}[t]
\centering\includegraphics[width=\textwidth]{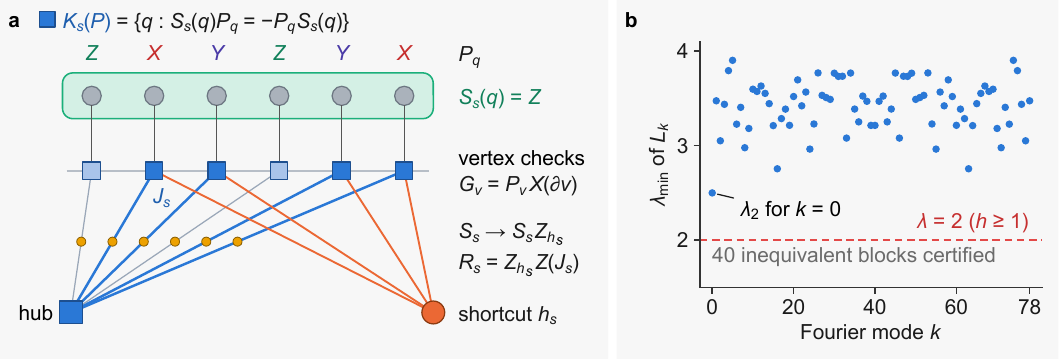}
\caption{
\textbf{Fixed extractor.}
\textbf{a},~Local rule around one old check $S_s$, shown on six of its twelve qubits. Its hub is joined to the data vertices of its support by star edges, each carrying an ancilla qubit (orange dots), and every data vertex carries the vertex check $G_v=P_vX(\partial v)$. For a target $P$, the vertices in $K_s(P)$ (dark blue), at which the local factor $P_q$ anticommutes with $S_s$, are joined by the shortcut hyperedge $h_s$; the old check becomes $S_sZ_{h_s}$, and $R_s=Z_{h_s}Z(J_s)$ is the cycle formed by the shortcut and the star edges $J_s$ from the hub to these vertices.
\textbf{b},~Smallest eigenvalue of each Fourier block $L_k$ of the extractor Laplacian, excluding the zero eigenvalue of $L_0$. Blocks $k$ and $79-k$ are complex conjugates, which leaves $40$ inequivalent blocks; exact certificates show that all these eigenvalues lie above $2$ (dashed line), so that the extractor graph has expansion $h\ge1$ and every mode preserves the code distance.
}
\label{fig:extractor}
\end{figure*}

\begin{figure*}[t]
\centering\includegraphics[width=\textwidth]{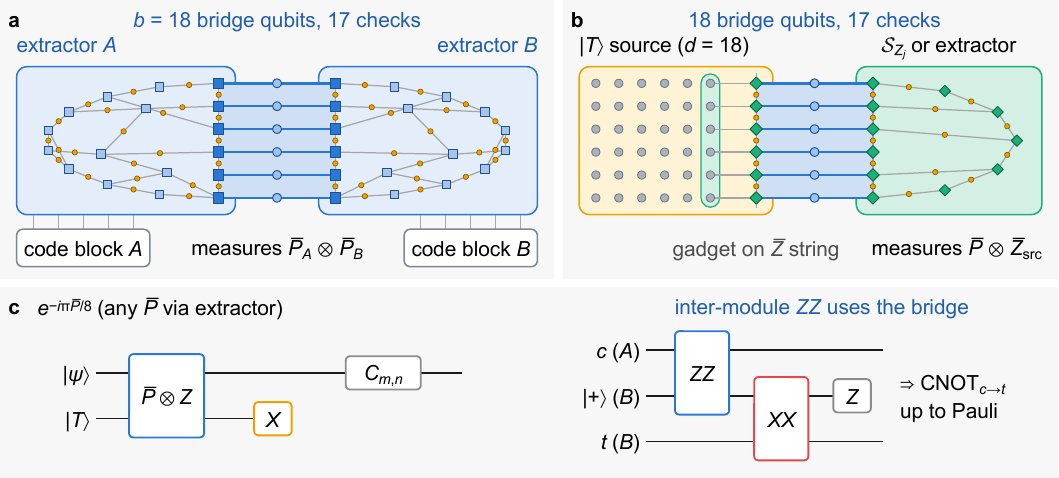}
\caption{
\textbf{Connecting modules and magic-state sources.} \textbf{a},~The extractors of two Cornucopia modules are joined by a matching bridge of $18$ ancilla qubits (light blue) between fixed ports, and $17$ square checks complete the cycle basis. For any pair of targets the merged code measures $\PP_A\otimes\PP_B$, satisfies $\wmax\le16$ and $\Dmax\le14$, and preserves the distance.
\textbf{b},~A distance-$18$ surface-code patch holding a $\ket T$ state carries an $18$-vertex gadget on its $\ZZ$ string, which is joined by the bridge to a seed gadget $\Sg_{Z_j}$ or to the extractor, so that the merged code measures $\PP\otimes\ZZ_\src$.
\textbf{c},~Protocols. Left: the rotation $e^{-i\pi\PP/8}$ for an arbitrary logical Pauli operator $\PP$, implemented by measuring $\PP\otimes\ZZ_\src$ with a $\ket T$ source, reading out the source in the $X$ basis and applying the Clifford correction. 
Right: a CNOT gate between a control $c$ in module $A$ and a target $t$ in module $B$, using a work qubit prepared in $\ket+$ in module $B$; only the $ZZ$ measurement crosses the module boundary and uses the inter-module bridge.
}
\label{fig:modules}
\end{figure*}

\begin{figure*}[t]
\centering\includegraphics[width=\textwidth]{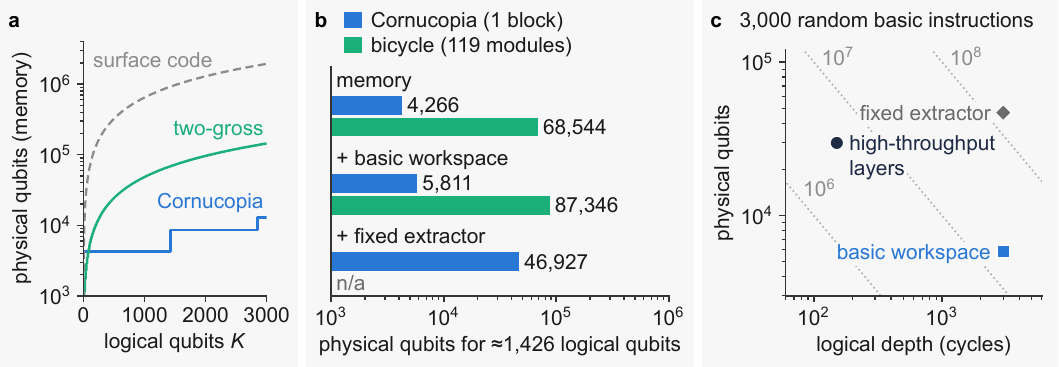}
\caption{\textbf{Physical-qubit overhead and benchmark workloads.} 
\textbf{a},~Physical qubits required to store $K$ logical qubits in Cornucopia blocks, in two-gross bicycle modules and in the rotated surface code.
\textbf{b},~Physical qubits required for about $1{,}426$ logical qubits: memory only; memory plus a workspace for the largest basic instruction (Cornucopia: on the $1{,}422$ register qubits; bicycle: one logical processing unit per module); and memory plus the fixed extractor. 
\textbf{c},~Logical depth and physical qubits required by a Cornucopia module executing random workloads of $3{,}000$ basic instructions on its register qubits with a single basic workspace, with the high-throughput set (largest layer; mean over three workloads) and with the fixed extractor. Dotted lines mark constant space--time volumes of $10^6$, $10^7$ and $10^8$ qubit-cycles, where a logical cycle is one surgery of $O(d)$ syndrome rounds (Methods).
}
\label{fig:resources}
\end{figure*}

\vspace{.2cm}
\noindent\textbf{\large{}Reusable gadgets and matching bridges} \\
\noindent 
We measure a logical Pauli operator $\PP$ by graph surgery~\cite{Cohen2022Lowoverhead,Cross2024Improved,Williamson2026Lowoverhead,Ide2025Faulttolerant,Bhardwaj2026Highrate} (Fig.~\ref{fig:toolkit}\textbf{a}). A gadget is a connected graph whose vertices carry new checks and whose edges are new ancilla qubits. Each vertex is attached to at most one qubit in the support of the physical representative of $\PP$, every old check that anticommutes locally with this representative is deformed onto ancilla qubits, and one check is added for every cycle of a complete cycle basis (Methods). 
The resulting merged code measures exactly $\PP$, and its distance is at least $d$ whenever the edge expansion of the graph is at least one (Methods, Theorem~\ref{thm:static}). 
The edge expansion is at least $\lambda_2/2$, where $\lambda_2$ is the second-smallest eigenvalue of the graph Laplacian~\cite{Mohar1989Isoperimetric}, so we certify this condition by proving $\lambda_2>2$ with an exact certificate (Methods).

Translation covariance renders these gadgets reusable. Because $T$ is an automorphism of both the code and the canonical logical basis, a certified gadget for $\XX_{(i,0)}$ translated by $T^t$ is a certified gadget for $\XX_{(i,t)}$ (Fig.~\ref{fig:toolkit}\textbf{b}), so that one seed gadget per register and Pauli type serves all $79$ logical qubits of a register. The library of $54$ seed gadgets thus covers all single-qubit Pauli measurements on the register qubits, with $179$--$752$ ancilla data qubits per gadget (Extended Data Fig.~\ref{edfig:ibdata}\textbf{a}). The degree of a gadget ancilla is dominated by the cycle checks that contain it, and fundamental cycle bases derived from spanning trees, which route many cycles through edges near the root, produce ancilla degrees of up to $256$ for our seed gadgets. 
Since any invertible transformation of the cycle checks leaves the merged stabilizer group unchanged, we instead construct each cycle basis in a load-aware manner: short cycles are added one at a time, preferably through ancilla data qubits with a low load, that is, on which few checks act so far. As a result, at most $9$ checks act on any ancilla data qubit of the gadget library (Extended Data Fig.~\ref{edfig:ibdata}\textbf{c}).

Products of two logical operators are measured by joining two gadgets with a bridge~\cite{Cross2024Improved,Swaroop2024Universal,Bhardwaj2026Highrate}. 
We call gadgets compatible if all their vertex attachments on each shared data qubit carry the same Pauli factor. 
As for the adapters of ref.~\cite{Swaroop2024Universal}, joining two expanding, compatible gadgets by a matching bridge of $b\ge d$ edges, together with $b-1$ square cycle checks, yields a merged code that measures exactly the product of the two target logical operators and has distance at least $d$ (Fig.~\ref{fig:toolkit}\textbf{c} and Methods, Proposition~\ref{thm:bridge}).
The condition $b\ge d$ cannot be relaxed, because $X$ acting on the $b$ bridge qubits is a logical operator of the merged code; our bridges of $18$ edges therefore yield merged distance exactly $18$. Two-qubit measurements are as reusable as single-qubit ones: every gadget of the library carries a fixed port of $18$ vertices, and the gadget for $\XX_j\XX_k$ consists of the two translated seeds and the $18$-edge bridge between their ports, without any pair-specific optimization. Translation covariance reduces the $2\binom{1422}{2}$ products of register qubits to $25{,}578$ inequivalent merged codes, each of which we assembled and verified exactly. Bridged products use $376$--$772$ ancilla data qubits and satisfy $\wmax\le16$ and $\Dmax\le8$ (Extended Data Fig.~\ref{edfig:ibdata}\textbf{b}).

Magic states are injected through the same mechanism. A distance-$18$ surface-code patch holding a logical $\ket T$ state, prepared in a separate factory~\cite{Bravyi2005Universal,Gidney2024Magic}, carries a certified $18$-vertex gadget on its $\ZZ$ string. Bridging this gadget to the seed gadget of register qubit $j$ measures $\ZZ_j\ZZ_\src$, and a subsequent transversal $X$-basis readout of the source teleports the magic state into $j$ up to a Pauli correction~\cite{Litinski2019GameOfSurfaceCodes}. Because the module and the source together form a code of distance $18$, and the two gadgets act on disjoint data qubits and are therefore compatible, Proposition~\ref{thm:bridge} guarantees that the merged code retains code distance $18$.

\vspace{.2cm}
\noindent\textbf{\large{}Parallel surgery} \\
\noindent 
A module stores $1{,}422$ moving logical qubits, so a processor measuring one Pauli product per logical cycle leaves most of them idle. When a program applies the same measurement to multiple logical operators, high-rate surgery measures the entire family at low space and time overhead~\cite{Zheng2025Highrate,Bhardwaj2026Highrate,Maskara2026Efficient}. 
Here we parallelize arbitrary basic instructions. Compatible certified gadgets for independent logical operators can be placed on a module simultaneously: the merged code of their direct sum measures exactly the span of the target operators and retains distance $d$ (Fig.~\ref{fig:toolkit}\textbf{d}). The gadgets interact only through the input code, where old checks deformed by several gadgets accumulate ancilla data qubits. When such a check would exceed weight 16, we route groups of its deformations through shortcut hyperedges, which leave both the measured operators and the expansion unchanged. This yields local compatibility rules that predict all check weights and qubit degrees of a merged code without assembling it (Methods, Proposition~\ref{prop:compat}). Greedy packing finds verified batches of up to $37$ simultaneous single-qubit measurements and $9$ bridged two-qubit measurements, and $20$ magic states can be injected simultaneously (Fig.~\ref{fig:parallel}\textbf{a}). 
To estimate the throughput on a generic program, we draw random workloads of $3{,}000$ basic instructions, each chosen uniformly among $\XX_j$, $\YY_j$, $\ZZ_j$, $\XX_j\XX_k$ and $\ZZ_j\ZZ_k$ on uniformly random register qubits, and schedule them into layers. Each instruction is placed into the earliest layer that follows all earlier instructions on the same logical qubits and in which it is compatible with the instructions already present (Proposition~\ref{prop:compat}); each layer is then executed as one merged code in a single logical cycle of $O(d)$ syndrome rounds. A layer contains on average $19.3$ instructions, which reduces the logical depth nearly twentyfold compared with executing one instruction per cycle (Fig.~\ref{fig:parallel}\textbf{b}).

The block-convolutional structure of Cornucopia codes provides another source of parallelism. Extending the check-tile index from three to six values defines latent rows~\cite{Kasai2026Breaking}, which are not checks of the code, and the sum of the three rows of a latent block acts on a single column of three qubits in each of the twelve data tiles (Fig.~\ref{fig:parallel}\textbf{c}). 
We show that these weight-$36$ operators form six families of mutually commuting logical operators, three of $X$ type and three of $Z$ type, and that within each family the supports of the $79$ translates partition the data qubits and span a $78$-dimensional space of logical operators (Methods, Proposition~\ref{prop:native}).
Consequently, a single merged code built from the $79$ translates of one certified gadget returns $78$ independent logical outcomes with $\wmax\le13$ and $\Dmax\le14$ (Extended Data Tab.~\ref{tab:native}), an instance of a group-invariant measurement family~\cite{Maskara2026Efficient}, realized here by translating a single certified gadget rather than by chain-map convolution. 
Each native operator acts on $680$--$742$ logical qubits of our basis, so the six families provide $468$ commuting multi-qubit observables of physical weight $36$, suited, for instance, to preparing or verifying large stabilizer states.

\vspace{.2cm}
\noindent\textbf{\large{}Fixed extractor} \\
\noindent 
An extractor is a single ancilla system that is attached to the code once and whose connections are selected in software to measure any chosen logical Pauli product~\cite{He2025Extractors,Webster2025Explicit}. 
Our extractor is built on the Tanner graph of the code. It has one vertex for each of the $2{,}844$ data qubits of the code, and one hub for each of the $1{,}422$ checks of the code (the old checks), and each hub is joined to the $12$ data vertices of its check;
a translation-covariant cycle through the data vertices adds $2{,}844$ further edges, giving $4{,}266$ vertices and $19{,}908$ edges in total. 
Every edge carries an ancilla data qubit, and $1{,}422$ further shortcut ancillas, one per old check, bring the total to $21{,}330$ ancilla data qubits. The extractor also comprises $4{,}266$ vertex checks, $15{,}643$ cycle checks of weight at most eight and $1{,}422$ relation checks, one per shortcut, $21{,}331$ checks in total (Fig.~\ref{fig:extractor}\textbf{a} and Methods).

For a target $\PP$, every data vertex carries the local factor of the physical representative of $\PP$, and every old check is deformed onto its shortcut, which acts as a hyperedge on the data vertices that anticommute with the check. 
Because the connections are fixed, counting how many of them can meet at a check or a qubit bounds the check weights and qubit degrees of the merged code of every logical Pauli target by $\wmax\le15$ and $\Dmax\le14$ (Supplementary Information Sec.~5). For six targets, including a dense random Pauli product on all $1{,}422$ register qubits, we verified the merged codes exactly (Extended Data Tab.~\ref{tab:extmodes}).

Shortcut hyperedges never decrease the boundary of a vertex set, so the distance of all $4^{1426}-1$ modes follows from a single property of the fixed graph, namely $\lambda_2>2$. Because the graph is invariant under $T$, its Laplacian decomposes into $79$ Fourier blocks of size $54\times54$, of which $40$ are inequivalent (Fig.~\ref{fig:extractor}\textbf{b}). 
An exact certificate for each block shows that all its eigenvalues other than the trivial zero exceed $2$, which proves $\lambda_2>2$ for the whole graph (numerically, $\lambda_2=2.501$; Methods), so that every mode of the extractor preserves the code distance.
The extractor also makes Pauli-based computation~\cite{Bravyi2016Trading,Litinski2019GameOfSurfaceCodes} native: to apply $e^{-i\pi\PP/8}$ for an arbitrary logical Pauli operator $\PP$, the extractor configured for $\PP$ is bridged to a $\ket T$ source, $\PP\otimes\ZZ_\src$ is measured, and the source is read out in the $X$ basis, whereupon the outcomes determine a Clifford correction (Fig.~\ref{fig:modules}\textbf{c}).

\vspace{.2cm}
\noindent\textbf{\large{}Modular architecture} \\
\noindent 
Distinct Cornucopia blocks are independent error-correction domains that can be reset, replaced and decoded separately, and a large computer is assembled from many such modules~\cite{Yoder2025Tour}. Since the direct sum of two codes of distance $d$ has distance $d$, and gadgets on different modules act on disjoint data qubits and are therefore compatible for any pair of targets, Proposition~\ref{thm:bridge} applies across modules. We can join the extractors of two modules $A$ and $B$ by a matching bridge of $18$ edges between fixed ports on their data cycles (Fig.~\ref{fig:modules}\textbf{a}). The bridge adds $18$ ancilla data qubits and $17$ ancilla checks of weight four, and for every pair of targets the merged code measures exactly $\PP_A\otimes\PP_B$, satisfies $\wmax\le16$ and $\Dmax\le14$, and retains distance $18$. We verified four inter-module modes exactly, including dense random Pauli products on both modules (Extended Data Tab.~\ref{tab:intermodule}). 

These joint measurements suffice for logic between modules. Teleporting a logical qubit requires one inter-module $\ZZ\ZZ$ measurement and one single-qubit $\XX$ measurement, and a logical CNOT gate uses a work qubit and three measurements, only one of which crosses the module boundary (Fig.~\ref{fig:modules}\textbf{c} and Methods). Magic-state factories are separate modules that attach through the same bridges (Fig.~\ref{fig:modules}\textbf{b}). 
Between identical modules, a transversal CNOT gate applies CNOT to all $1{,}426$ aligned pairs of logical qubits at once and, combined with shift automorphisms, pairs register $i$ of one module with register $i$ of the other at any relative offset.

\vspace{.2cm}
\noindent\textbf{\large{}Physical-qubit overhead} \\
\noindent 
We compare physical-qubit provisioning with that of the bicycle architecture~\cite{Yoder2025Tour} at a distance $d=18$ and a capacity of about $1{,}426$ logical qubits, excluding magic-state factories (Fig.~\ref{fig:resources} and Methods).
The comparison concerns space, not logical error rates or cycle times. A Cornucopia block stores $1{,}426$ logical qubits in $4{,}266$ physical qubits, compared with $68{,}544$ for $119$ two-gross $\cod{288}{12}{18}$ bicycle modules. A workspace for every basic instruction on the $1{,}422$ register qubits raises a Cornucopia module to $5{,}811$ physical qubits, or $4.1$ per logical qubit, whereas the bicycle modules with their logical processing units require $87{,}346$, or $61.2$ per logical qubit.
With the fixed extractor, which measures any logical Pauli product across all $1{,}426$ logical qubits, a Cornucopia module comprises $46{,}927$ physical qubits.

This comparison does not capture throughput: the logical processing units of the $119$ bicycle modules can operate simultaneously, whereas one Cornucopia workspace executes one basic instruction per logical cycle. 
Within our architecture, we quantify the trade-off between space and logical depth with random workloads of $3{,}000$ basic instructions on the register qubits (Fig.~\ref{fig:resources}\textbf{c} and Methods). A single basic workspace requires $3{,}000$ logical cycles, whereas the high-throughput set completes three workloads, drawn independently from the same distribution, in $151$ cycles on average with at most $29{,}755$ physical qubits, reducing the space--time volume by a factor of $3.9$. The extractor suits workloads of high-weight Pauli products instead, each of which it measures in a single cycle.

\vspace{0.2cm}
\noindent\textbf{\large{}Discussion} \\
In summary, we have shown that the $\cod{2844}{1426}{18}$ Cornucopia code block can be operated as a universal logical processor and that such blocks combine into a modular architecture. Building on the symmetry-adapted standard form, optimized by polynomial shears, we organized $1{,}422$ of the $1{,}426$ logical qubits into $18$ cyclic registers, so that $54$ certified gadgets, reused by translation and combined through matching bridges, address every single-qubit Pauli measurement and every two-qubit $\XX\XX$ and $\ZZ\ZZ$ measurement on these register qubits individually. Direct sums and native operator families provide parallelism, a fixed extractor measures arbitrary logical Pauli products, and bridges of $18$ ancilla qubits connect modules and magic-state sources. 
With a workspace for every basic instruction on the register qubits, a module needs on average $4.1$ physical qubits per logical qubit, compared with $3.0$ for the memory alone; apart from magic-state factories, which we do not count, the high encoding rate of the code thereby largely survives the addition of a universal instruction set.
Although we have focused on the $\cod{2844}{1426}{18}$ code, the construction relies on the column-translation symmetry shared by all Cornucopia codes, and therefore carries over straightforwardly to other codes of the family~\cite{Lu2026Quantum}.

Several directions remain for future work. The distance guarantees established here concern the merged codes themselves, whereas the circuit-level distance depends on the syndrome extraction schedule, and the logical error rate per instruction must be established by circuit-level simulation and decoding. 
The rigid-motion structure of Cornucopia syndrome extraction suggests that the translated seeds and the extractor, which inherit the translation symmetry, admit structured schedules. 
The number of syndrome rounds per surgery could be reduced with fast-surgery constructions~\cite{Baspin2025Fast}, and gadgets built on lighter representatives or directly on native operator families would lower the cost of every instruction and could bring the four invariant modes into the basic instruction set. 
Symmetric surgery with chain-map convolution~\cite{Maskara2026Efficient}, which measures all translates of a Pauli product in a few syndrome rounds and injects magic states into an entire orbit, would complement our individually addressable instructions and, together with high-rate surgery~\cite{Zheng2025Highrate,Cowtan2026Parallel}, raise the throughput. 
Mapping the Cornucopia architecture onto a concrete reconfigurable atom array, including where the ancilla atoms of each gadget are placed and how atoms are moved to realize the bridges between modules~\cite{Bluvstein2026Faulttolerant}, is a natural next step.

\vspace{2mm}
\noindent\textbf{Note added.} After completion of this work, an independent preprint by Maskara et al. appeared on arXiv~\cite{Maskara2026Efficient}, which also applies Gottesman's standard form to whole symmetry orbits to construct a symmetry-adapted logical basis of ultra-high-rate codes. Here, in this work the symmetry serves instead to translate a single certified gadget to any qubit of a register, which yields individually addressable measurements.

\bibliography{references}

\clearpage
\majorheading{Methods}

\noindent\textbf{Code construction} \\
All linear algebra is over $\F$, and checks and logical representatives are stored as row vectors. A Pauli operator on $n$ qubits is described by a pair $(x,z)\in\F^n\times\F^n$ through the Hermitian representative $P(x,z)=i^{|x\wedge z|}X(x)Z(z)$; its weight is $\wt(x\lor z)$, and two Pauli operators commute if and only if $x\cdot z'+z\cdot x'=0$. Data qubit $(b,r,c)$ with tile $b\in\{0,\dots,11\}$ (ordered as $L_0,\dots,L_5,R_0,\dots,R_5$), row $r\in\mathbb Z_3$ and column $c\in\mathbb Z_q$ has index $(3b+r)q+c$, and check $(i,r,c)$ of either type has index $(3i+r)q+c$. The Cornucopia code~\cite{Lu2026Quantum} has check matrices
\begin{align}
H_X&=\bigl[\,\mathcal A_{j-i}\;\big|\;\mathcal B_{j-i}\,\bigr],\notag\\
H_Z&=\bigl[\,\mathcal B_{i-j}^{-1}\;\big|\;\mathcal A_{i-j}^{-1}\,\bigr],
\label{eq:cornucopia}
\end{align}
with block rows $i=0,1,2$ and block columns $j=0,\dots,5$, indices taken modulo six, and left and right halves acting on the $L$ and $R$ tiles. Here $\mathcal A_j$ and $\mathcal B_j$ are the $3q\times3q$ permutation matrices of the coordinate maps $A_j,B_j:(r,c)\mapsto(\rho(r),c+s)$ on $\mathbb Z_3\times\mathbb Z_q$, with the convention of ref.~\cite{Lu2026Quantum}: $(\mathcal A_j)_{u,u'}=1$ if and only if $u'=A_j(u)$, and likewise for $\mathcal B_j$; for permutation matrices $\mathcal A_j^{-1}=\mathcal A_j^{\mathsf T}$. Thus the $X$ check at coordinate $(r,c)$ of tile $X_i$ acts on $A_{j-i}(r,c)$ in $L_j$ and on $B_{j-i}(r,c)$ in $R_j$, and the $Z$ check at $(r,c)$ of tile $Z_i$ acts on $B_{i-j}^{-1}(r,c)$ in $L_j$ and on $A_{i-j}^{-1}(r,c)$ in $R_j$. For $q=79$ the column shifts are $(s_{A_0},\dots,s_{A_5})=(6,49,55,18,40,7)$ and $(s_{B_0},\dots,s_{B_5})=(24,41,78,53,68,21)$; the row maps are $r\mapsto r+1$ for $A_0$ and $B_2$, $r\mapsto2r+2$ for $A_1$ and $B_3$, and the identity otherwise. Column shifts commute with each other and with every row map, and the only noncommuting pairs, $(A_0,B_3)$ and $(A_1,B_2)$, would enter $H_XH_Z^{\mathsf T}$ only through block-row differences $i'-i\equiv3\pmod6$, which do not occur for $i,i'\in\{0,1,2\}$; hence $H_XH_Z^{\mathsf T}=0$. Every check has weight 12, and every data qubit lies in three $X$ and three $Z$ checks. 
The checks of each check tile sum to the all-ones vector, which gives two redundancies per type, so that $\rk H_X=\rk H_Z=709$ and $k=2{,}844-2\times709=1{,}426$. The distance $d=18$ was certified in ref.~\cite{Lu2026Quantum}. 

The translation $T:(b,r,c)\mapsto(b,r,c+1)$ of the data qubits maps the support of check $(i,r,c)$ onto that of check $(i,r,c+1)$, because column shifts commute with each other and row maps do not act on columns; it is therefore an automorphism of the code. Identifying $T$ with an indeterminate $D$, the data space becomes the free module $\Rq^{36}$ over $\Rq=\F[D]/(D^q-1)$, with one coordinate per fibre. For $q=79$ the multiplicative order of 2 modulo 79 is 39, so that $\mathcal R_{79}\cong\F\times\mathbb F_{2^{39}}\times\mathbb F_{2^{39}}$, where the first factor corresponds to evaluation at $D=1$. A Hermitian logical Pauli operator $i^{|a\wedge b|}\XX^a\ZZ^b$ with $a,b\in\F^{1426}$ has the physical representative $P(x,z)$ with $x=aL_X$ and $z=bL_Z$, where the rows of $L_X$ and $L_Z$ are the canonical representatives, and the two operators differ by the sign $\sigma=(-1)^{(|a\wedge b|-|x\wedge z|)/2}$. The exponent is an integer because $x\cdot z=a\cdot b$ for a canonical basis; the sign is reported with every measurement, so that the logical outcome is recovered from the physical one.

\vspace{2mm}
\noindent\textbf{Canonical basis from the symmetry-adapted standard form} \\
Matrices $L_X,L_Z\in\F^{k\times n}$ with rows $x_j$ and $z_j$ form a canonical basis if $H_ZL_X^{\mathsf T}=0$, $H_XL_Z^{\mathsf T}=0$ and $L_XL_Z^{\mathsf T}=I_k$. The last condition implies that no nonzero combination of the $x_j$ lies in $\row H_X$, because such a combination would pair trivially with every $z_j$; the $k^2$ pairings therefore certify the independence and nontriviality of all $2k$ representatives at once. For a set $P$ of fibres, $H[:,P]$ denotes the restriction of $H$ to the columns of these fibres. We use the symmetry-adapted standard form, which extends Gottesman's binary standard form~\cite{Gottesman1997Stabilizer} to whole symmetry orbits. For the translation symmetry of Cornucopia codes, it reads as follows.

\begin{theorem}[Translation-covariant canonical basis]\label{thm:basis}
Partition the $36$ data fibres into sets $P_X$ and $P_Z$ of nine fibres each and a free set $F$ of $18$ fibres such that
\begin{equation}
\rk H_X[:,P_X]=\rk H_X,\quad \rk H_Z[:,P_Z]=\rk H_Z .
\label{eq:pivotrank}
\end{equation}
For every $f\in F$, let $e_f$ be the unit vector supported on the data qubit at column $0$ of fibre $f$, and choose $u_f$ supported on $P_Z$ and $v_f$ supported on $P_X$ with $H_Zu_f^{\mathsf T}=H_Ze_f^{\mathsf T}$ and $H_Xv_f^{\mathsf T}=H_Xe_f^{\mathsf T}$. Then the $18q$ pairs
\begin{align}
\XX_{(f,t)}&=X\bigl(T^t(e_f+u_f)\bigr),\notag\\
\ZZ_{(f,t)}&=Z\bigl(T^t(e_f+v_f)\bigr),\qquad t\in\mathbb Z_q,
\end{align}
are canonical logical operators, the supports of conjugate operators intersect in exactly one qubit, those of nonconjugate operators are disjoint, and $T$ acts as $(f,t)\mapsto(f,t+1)$. They extend to a complete canonical basis by $k-18q$ further pairs.
\end{theorem}

The disjointness of $P_X$, $P_Z$ and $F$ then yields the pairing (Supplementary Information Sec.~2.1). For $q=79$, each pivot block has $9q=711$ columns and rank $709$, so every seed equation has four affine solutions. Their differences are translation-invariant logical operators of weight $158$ rather than stabilizers, so the choice among them is a choice of basis, and we select the lightest solution. 
Conversely, stabilizer dressing cannot reduce the weight of a representative while respecting the pivot structure, because the pivot fibres detect every nonzero stabilizer.

\begin{lemma}\label{lem:nodress}
If Eq.~\eqref{eq:pivotrank} holds and $s\in\row H_X$ vanishes on $P_X$, then $s=0$.
\end{lemma}

Since $n=36q$ and $\rk H_X=\rk H_Z=9q-2$, the code has $k=18q+4$ logical qubits, and four logical pairs remain besides the moving sector, forming the fixed modes in the terminology of ref.~\cite{Lu2026Quantum}. Every logical class fixed by $T$ contains a representative that is constant on fibres, because $\Omega=\sum_tT^t$ is idempotent for odd $q$; the $T$-invariant $X$-type classes are spanned by the $18$ orbit sums $\sum_tT^tx_i$ and four further representatives, and likewise for the $Z$ type (Supplementary Information Sec.~2.1).
We construct these four pairs from fibre-constant vectors in $\ker H_Z$ and $\ker H_X$, orthogonalize them against the moving sector, select the change of basis $M\in GL_4(\F)$, acting as $X\mapsto MX$ and $Z\mapsto(M^{-1})^{\mathsf T}Z$, that minimizes their weights by exhaustive search over all $20{,}160$ elements, and finally reduce their weights by simulated-annealing stabilizer dressing. This yields $X$-type weights $(474, 692, 666, 680)$ and $Z$-type weights $(682, 674, 678, 672)$ and leaves the moving sector unchanged. 
These four pairs complete the canonical basis, so that every logical Pauli operator, including every target of the extractor, has a definite physical representative and sign; we do not equip them with gadgets. 
Moving logical qubits carry the linear index $j=79i+t$, and the invariant modes the indices $j=1{,}422,\dots,1{,}425$. 
The selected partition is $P_X=\{3,4,5,18,20,24,26,27,35\}$ and $P_Z=\{1,6,7,10,12,15,16,19,31\}$, with fibre index $3b+r$.

\vspace{2mm}
\noindent\textbf{Optimization of the basis} \\
The ancilla count of a gadget grows with the weight of its target, and all translates of a seed have the same weight, so we minimize the total seed weight $W=\sum_i(\wt x_i+\wt z_i)$. Writing $p(T)x=\sum_tp_tT^tx$ for $p\in\Rq$ and $U^\dagger$ for the transpose of $U$ combined with $D\mapsto D^{-1}$, the pairing polynomials $G_{ij}(D)=\sum_t(x_i\cdot T^tz_j)D^t$ encode all $(18q)^2$ pairings of the moving sector, which is canonical if and only if $G=I_{18}$. The substitution $x_i'=\sum_kU_{ik}(T)x_k$, $z_j'=\sum_lV_{jl}(T)z_l$ gives $G'=UGV^\dagger$, so every invertible $U\in GL_{18}(\Rq)$ with $V=(U^\dagger)^{-1}$ maps a canonical, translation-covariant basis to another one. An elementary shear
\begin{equation}
x_i\leftarrow x_i+T^tx_j,\qquad z_j\leftarrow z_j+T^{-t}z_i\qquad(i\neq j)
\label{eq:shear}
\end{equation}
changes the total weight by $\Delta=\wt(x_j)+\wt(z_i)-2\bigl(|x_i\cap T^tx_j|+|z_j\cap T^{-t}z_i|\bigr)$ and is therefore favourable when the seeds of two registers overlap substantially after translation. The search proceeds in five stages with fixed random seeds (Supplementary Information Sec.~2.2); the first two optimize the pivot assignment of the standard form, whereas the shears and the dressing step act on the representatives themselves, beyond the standard form. First, of $1{,}200$ random partitions, $229$ satisfy Eq.~\eqref{eq:pivotrank}; the first of them, $B_0$, has $W=11{,}580$, and the best has $W=4{,}840$. Second, simulated annealing over fibre swaps evaluates $3{,}839$ distinct partitions, of which $2{,}032$ are valid, and retains the $24$ lightest, with $W=4{,}038$--$4{,}070$. Third, greedy monomial shears, each strictly decreasing $W$ and creating no seed heavier than $126$, reduce these candidates to $W=3{,}740$--$3{,}888$. 
Fourth, we select the candidate that is lightest after shearing; in the third stage, $17$ shears had reduced its weight from $4{,}060$ to $3{,}740$, and its $36$ compiled $X$- and $Z$-type seed gadgets use $11{,}584$ ancilla data qubits in total, $0.3\%$ more than the smallest total in the shortlist (Supplementary Tabs.~S1 and S2). 
Fifth, a dressing step explores, for every seed, the representatives reachable by weight-nonincreasing single-check moves and keeps the one with the smallest compiled gadget; for the selected basis it changes no seed. The final basis therefore differs from the standard form only through the shears: for the register qubits, conjugate representatives still intersect in exactly one qubit, as in the standard form, whereas $3{,}160$ of the $2{,}020{,}662$ nonconjugate pairs $(x_j,z_l)$, all of which are disjoint in the standard form, now intersect in two qubits.

\vspace{2mm}
\noindent\textbf{Graph surgery for general Pauli targets} \\
Let the input code, possibly a direct sum of code blocks, have distance $d$ and stabilizer group $\mathcal S$ generated by old checks $S_s$, and let $P=\bigotimes_qP_q$ be the physical representative of the target. 
A gadget consists of four components: (i) vertices $v$, each attached to at most one data qubit $q(v)$ with a local factor $F_v$ (a gadget for $\PP$ attaches one vertex to every qubit of the support of $P$, with $F_v=P_{q(v)}$); (ii) ancilla qubits $e$, the edges of a graph with incidence matrix $\Gamma\in\F^{E\times V}$; (iii) for every old check $S_s$, an edge set $J_s$ whose boundary $\Gamma^{\mathsf T}\one_{J_s}$ equals the set $K_s$ of attached vertices whose local factors anticommute with $S_s$; and (iv) a complete basis of the cycle space $\ker\Gamma^{\mathsf T}$.
Several gadgets placed together carry the attachments of all of them and require that all attachments on every data qubit carry the same factor (compatibility, see below). The merged code is generated by
\begin{equation}
S_sZ(J_s),\qquad G_v=F_v\otimes X(\partial v),\qquad Z(\gamma),
\label{eq:merged}
\end{equation}
where $\partial v$ is the set of edges incident to $v$ and $\gamma$ ranges over the cycle basis. These generators commute, and the product of all vertex checks equals $\prod_vF_v=P$ because every edge lies in an even number of vertex checks. For a pure $X$ target this is the usual CSS merged code, for a pure $Z$ target it is a CSS code up to a Hadamard transformation of the ancilla data qubits, and for targets with $Y$ factors it is a general stabilizer code. The merged code measures the group $\mathcal M=\{\prod_{v\in t}F_v:t\in\ker\Gamma\}$; for a connected graph $\ker\Gamma=\{0,\one\}$, so exactly $\PP$ is measured and the merged code encodes $k-1$ logical qubits.

\begin{theorem}[Distance of the merged code~\cite{Cross2024Improved,Williamson2026Lowoverhead,Bhardwaj2026Highrate}]\label{thm:static}
Let $h_2(\Gamma)=\min_{t\notin\ker\Gamma}\wt(\Gamma t)/\dist(t,\ker\Gamma)$. If the cycle basis is complete and $h_2(\Gamma)\ge1$, every nontrivial logical operator of the merged code has weight at least $d$.
\end{theorem}

The proof is a trade-off between the data qubits and the ancilla data qubits. Since the cycle basis is complete, the $X$ part of a nontrivial logical operator $E$ on the ancilla data qubits is the set $\Gamma t$ of edges leaving some vertex set $t$, which we choose as small as possible. Multiplying $E$ by the vertex checks of $t$ removes this part and changes $E$ on at most $|t|$ data qubits, leaving a nontrivial logical operator of the input code, so $E$ has weight at least $d-|t|$ on the data qubits of the input code. Expansion, $h_2(\Gamma)\ge1$, gives at least $|t|$ edges in $\Gamma t$, so $E$ has weight at least $|t|$ on the ancilla data qubits and at least $d$ in total.

Supplementary Information Sec.~3.1 proves the theorem for general Pauli targets, for direct sums $\Gamma=\bigoplus_g\Gamma_g$, for which $h_2(\Gamma)=\min_gh_2(\Gamma_g)$, and in the presence of shortcut hyperedges. 
For a connected graph, $h_2$ is the Cheeger constant $h$, and the bound $h\ge\lambda_2(L)/2$ reduces the condition to $\lambda_2(L)>2$ for the graph Laplacian $L$. 
For a graph on $v$ vertices with Laplacian eigenvalues $0=\lambda_1\le\lambda_2\le\dots\le\lambda_v$, the integer matrix $B=vL-2vI+3\,\one\one^{\mathsf T}$ has eigenvalue $v$ on $\one$ and eigenvalues $v(\lambda_i-2)$ for $i\ge2$ on the orthogonal complement, so that $B$ is positive definite, written $B\succ0$, if and only if $\lambda_2>2$.

We certify $B\succ0$ in exact integer arithmetic. For each gadget, an invertible integer matrix $Q$ is computed with floating-point linear algebra and stored, and the verifier checks exactly whether $QBQ^{\mathsf T}$ is strictly diagonally dominant, that is, whether each diagonal entry is positive and larger than the sum of the absolute values of the other entries in its row. If it is, $QBQ^{\mathsf T}$ is positive definite, and hence so is $B$, because $Q$ is invertible. A gadget is accepted only if this check succeeds.

\vspace{2mm}
\noindent\textbf{Gadget compilation and load-aware cycle bases} \\
Seed gadgets follow the skeleton-and-expansion recipe~\cite{Cross2024Improved,Bhardwaj2026Highrate} (Supplementary Information Sec.~3.2). The compiler places one vertex on every qubit of the target support, pairs the vertices of each set $K_s(P)$ into skeleton edges that form $J_s$, adds a random regular expansion graph that raises the vertex degree to between seven and nine until $\lambda_2>2.025$, prunes expansion edges that contribute least to $\lambda_2$ as long as $\lambda_2>2.01$, and finally computes the integer certificate and checks the gadget exactly.
Since $T$ permutes the data qubits and the checks, translating the attachments and deformations of a certified gadget by $T^t$ while keeping its graph yields an isomorphic merged code for the translated target.

The degree of an ancilla edge counts its two vertex checks, the old checks deformed onto it and the cycle checks that contain it. Since any invertible transformation of the cycle checks leaves the merged stabilizer group invariant, we construct the cycle basis directly. The algorithm maintains the current load $\ell_e$ of every edge, initialized with its vertex and deformed checks, and passes repeatedly over the edges in random order. For an edge $e=(u,v)$, it finds the $u$-$v$ path of at most $15$ edges in $G-e$ that minimizes $\sum_{e'}(\ell_{e'}+1)^4$, a steeply increasing cost that steers new cycles away from heavily loaded edges (the exponent is a heuristic choice), and it accepts the resulting cycle if the cycle is linearly independent of those already chosen. For every gadget in this work the basis is complete after a few passes. With fundamental cycle bases, the maximum ancilla degree reaches $141$ for the $X$- and $Z$-type seeds and $256$ for the $Y$-type seeds; the load-aware construction reduces it to at most $9$, with cycle checks of weight at most $10$, while the graphs and certificates remain unchanged. 

For the bridges described below, every seed gadget also carries a fixed port, a path of $18$ vertices to which a bridge is attached. Because a bridge adds one edge to every port vertex and one square check to every port edge, the port is chosen where vertex check weights and edge degrees are still at most $15$, so that both stay within the cap of $16$ after bridging.

\vspace{2mm}
\noindent\textbf{Matching bridges} \\
A port of a gadget is a path $u_1,\dots,u_b$ of distinct vertices. Given ports $u$ of $\Sg_1$ and $w$ of $\Sg_2$, a matching bridge consists of the $b$ edges $(u_i,w_i)$ and the $b-1$ square cycles formed by $(u_i,w_i)$, $(w_i,w_{i+1})$, $(w_{i+1},u_{i+1})$ and $(u_{i+1},u_i)$. The joined graph is connected, so it measures only $\PP_1\PP_2$, and the squares, whose restrictions to the bridge edges are linearly independent, complete its cycle basis. Gadgets are compatible if all their vertex attachments on every shared data qubit carry the same local Pauli factor; otherwise vertex checks attached to that qubit anticommute. 
For single-target gadgets, compatibility means that the physical representatives of the targets agree on shared qubits. For composite gadgets, the condition concerns the attachments rather than the target: in the gadget for $\XX_j\XX_k$, a qubit in the supports of both $x_j$ and $x_k$ carries two $X$ attachments, whose factors cancel in the target but which both anticommute with a $Z$ attachment of another gadget on that qubit. Commuting targets therefore need not have compatible gadgets; for instance, $\XX_1\XX_2$ and $\ZZ_1\ZZ_2$ commute, but their bridged gadgets attach $X$ and $Z$ factors to the qubit where $x_1$ meets $z_1$. The matching bridge is a special case of the adapters of ref.~\cite{Swaroop2024Universal}, whose Lemma~3 and Theorem~3 show that joining expanding gadgets by at least $d$ edges preserves the code distance, also for compatible gadgets on the same code block. Proposition~\ref{thm:bridge} states the case of path ports used in this work, and we give a self-contained proof in Supplementary Information Sec.~3.3.

\begin{proposition}[Matching bridges]\label{thm:bridge}
Let $\Sg_1$ and $\Sg_2$ be compatible gadgets measuring distinct logical operators $\PP_1$ and $\PP_2$ of a stabilizer code, or of a direct sum of codes, with distance $d$, each satisfying the hypotheses of Theorem~\ref{thm:static}. If they are joined by a matching bridge of $b\ge d$ edges together with its square cycles, the merged code measures exactly $\PP_1\PP_2$ and has distance at least $d$.
\end{proposition}

The proof applies the argument of Theorem~\ref{thm:static} to $t=(t_1,t_2)$.
Since only $\PP_1\PP_2$ is measured, $t_1$ and $t_2$ can be complemented together but not separately, so the argument carries over if each of them contains at most half of the vertices of its gadget, or each at least half. Otherwise, let $t_i^*$ be the smaller of $t_i$ and its complement. Every bridge edge not in $\Gamma t$ has an endpoint in $t_1^*$ or $t_2^*$, so there are at most $|t_1^*|+|t_2^*|$ of them because no two bridge edges share a vertex, while expansion gives at least $|t_1^*|+|t_2^*|$ edges of $\Gamma t$ inside the gadgets; hence $\wt(\Gamma t)\ge b\ge d$. 
The condition $b\ge d$ is necessary: the vertex checks of $\Sg_1$ multiply to $P_1$ times $X$ on the bridge, so $X$ on the $b$ bridge qubits is a weight-$b$ representative of the unmeasured $\PP_1$. 
All bridges of this work join compatible gadgets: for $\XX_j\XX_k$ both gadgets carry only $X$ attachments and for $\ZZ_j\ZZ_k$ only $Z$ attachments, so they agree on every shared qubit, and bridges between modules or to magic-state sources join gadgets on disjoint data qubits.

The gadget for $\XX_j\XX_k$ in a module is the union of the translated gadgets of $j$ and $k$ and the $18$-edge bridge between their fixed ports. Every seed deforms each old check onto at most two ancillas, so that a check deformed by two seeds has weight at most $12+2+2=16$, and a data qubit carries at most two vertex attachments, so that no shortcut is needed. Because seeds and ports translate with $T$, the merged codes of the pairs $\{(a,s),(b,s+\delta)\}$ and $\{(a,0),(b,\delta)\}$ are isomorphic. For each Pauli type this leaves $\binom{18}{2}\times79+18\times39=12{,}789$ classes, $25{,}578$ in total, which cover all $2\binom{1422}{2}=2{,}020{,}662$ products of register qubits. For every class we assembled the complete merged code and verified commutation, the exact rank ($k'=1{,}425$), the vertex product, $\wmax\le16$ and $\Dmax\le8$.

\vspace{2mm}
\noindent\textbf{Direct sums, shortcuts and scheduling} \\
For compatible certified gadgets measuring operators that are independent modulo stabilizers, the incidence matrix of their union is a direct sum, so that the merged code measures exactly the span of the $m$ targets, encodes $k-m$ logical qubits and, by Theorem~\ref{thm:static} for direct sums together with Proposition~\ref{thm:bridge} for bridged components, has distance at least $d$. 
When an old check $s$ is deformed by several gadgets and $12+|J_s|>16$, its deformation edges are replaced by shortcuts. The edges are split by gadget into groups, and the groups are distributed over bins by first-fit decreasing: in order of decreasing size, each group is placed in the first bin that can still accommodate it, and opens a new bin otherwise. Each bin $J$ receives a shortcut ancilla $h$ that acts as a hyperedge on the boundary $K=\Gamma^{\mathsf T}\one_J$, the check $S_sZ(J_s)$ is replaced by $S_s\prod_hZ_h$, and the check $Z_hZ(J)$, called a relation, is added. A bin can accommodate a group as long as the relation and the shortcut stay within the cap, $|J|+1\le16$ and $|K|+2\le16$, and at most four bins are allowed, so that $S_s\prod_hZ_h$ has weight at most $16$.
The incidence row of each shortcut, $\one_K^{\mathsf T}$, is the sum of the incidence rows of the edges in $J$, so $\ker\Gamma$ is unchanged and $\wt(\Gamma t)$ does not decrease; the condition $h_2(\Gamma)\ge1$ of Theorem~\ref{thm:static} is therefore preserved.

\begin{proposition}[Compatibility]\label{prop:compat}
A set of certified gadgets forms a certified parallel batch if (i) the gadgets are compatible, that is, all vertex attachments on the same data qubit carry the same local factor, and every data qubit carries at most ten attachments; (ii) every old check $s$ either satisfies $12+|J_s|\le16$ or its deformation edges $J_s$ can be packed by gadget into at most four bins $J\subseteq J_s$, each with boundary $K=\Gamma^{\mathsf T}\one_J$ satisfying $|J|+1\le16$ and $|K|+2\le16$; and (iii) for every vertex, its check weight plus the number of shortcuts acting on it is at most 16.
\end{proposition}

These conditions involve only local counts, so they predict every check weight and qubit degree without assembling the merged code (Supplementary Information Sec.~4.1).
For each instruction type, the largest batch in Fig.~\ref{fig:parallel}\textbf{a} was found by randomized greedy packing: the candidate instructions are shuffled and added one by one whenever they remain compatible with the batch and act on new logical qubits. Workloads of $3{,}000$ instructions drawn uniformly from $\{\XX_j,\ZZ_j,\YY_j,\XX_j\XX_k,\ZZ_j\ZZ_k\}$ are list-scheduled into layers, each of which is a batch: in program order, each instruction is placed in the first compatible layer after the last layer that acts on any of its logical qubits.
The largest batches and, for one workload of each type, the largest layer were verified exactly.

\vspace{2mm}
\noindent\textbf{Native operator families} \\
Extending the check-block index of Eq.~\eqref{eq:cornucopia} to $i\in\mathbb Z_6$ defines latent rows $i=3,4,5$. Summing the three rows $(i,r,c)$, $r\in\mathbb Z_3$, of a latent block gives the weight-36 operators
\begin{align}
\mathsf P^X_{i,c}&=\prod_{j=0}^{5}\prod_{r=0}^{2}X_{L_j,r,c+s_{A_{j-i}}}X_{R_j,r,c+s_{B_{j-i}}},\notag\\
\mathsf P^Z_{i,c}&=\prod_{j=0}^{5}\prod_{r=0}^{2}Z_{L_j,r,c-s_{B_{i-j}}}Z_{R_j,r,c-s_{A_{i-j}}}.
\label{eq:native}
\end{align}

\begin{proposition}[Native families]\label{prop:native}
The operators $\mathsf P^X_{i,c}$ and $\mathsf P^Z_{i',c'}$ with $i,i'\in\{3,4,5\}$ and $c,c'\in\mathbb Z_{79}$ are logical operators of the code and commute with each other. For a fixed Pauli type and $i$, the 79 translates have disjoint supports that cover all data qubits and span a 78-dimensional space of logical operators; the six families together span 468 dimensions.
\end{proposition}

The product of the $79$ translates of a family is $X^{\otimes n}$ or $Z^{\otimes n}$, which is a stabilizer. A whole family is measured in a single \emph{full-orbit batch}, the direct sum of the $79$ translates of one certified gadget: the merged code returns $78$ independent outcomes, and the product of all $79$ outcomes, the value of a stabilizer, serves as a parity check. Each of the $711$ checks of the opposite Pauli type meets six of the translates, each in two qubits, and acquires one deformation ancilla from each of the six gadgets, which would raise its weight to $12+6=18$; one shortcut per check carries all six deformations and lowers its weight to $13$, within the cap of $16$ (Extended Data Tab.~\ref{tab:native}). 
Related latent-row operators were previously used as distance witnesses for affine-permutation codes~\cite{Kasai2026Heuristic}; here they serve as measurable instructions.

\vspace{2mm}
\noindent\textbf{Extractor construction} \\
The extractor graph $G$ has one vertex per data qubit and one hub per listed check. Its edges are the $17{,}064$ Tanner incidences between hubs and data vertices and a translation-covariant cycle of degree two on the data vertices ($2{,}844$ edges), obtained by drawing random cyclic orders of the $36$ data fibres, with random relative shifts, until one gives $\lambda_2>2.10$; the accepted arrangement has $\lambda_2=2.501$.
The cycle basis is formed by $12{,}781$ independent four-cycles of the Tanner graph, each through two hubs and two data qubits shared by an $X$ and a $Z$ check, $18$ short fundamental cycles that complete the cycle space of the Tanner graph, and, for each of the $2{,}844$ data-cycle edges, one cycle that closes it through a short path of Tanner edges; these paths are chosen under a convex load penalty, which keeps every edge in at most seven basis cycles.
For a target $P$, data vertex $q$ is attached to qubit $q$ with factor $P_q$, hubs are unattached, and each old check $S_s$ owns a shortcut $h_s$ that acts as a hyperedge on $K_s(P)$. The check becomes $S_sZ_{h_s}$, and the relation $R_s=Z_{h_s}Z(J_s)$, where $J_s$ are the star edges from hub $s$ to $K_s(P)$, closes the shortcut. 
Counting incidences gives mode-independent bounds: a data-vertex check contains at most eight ordinary edges, six shortcuts and one data factor; old checks and relations have weight at most $13$ and cycle checks at most eight; and an ordinary edge lies in at most $10$ checks, a shortcut in at most $14$ and a data qubit in seven. This yields $\wmax\le15$ and $\Dmax\le14$ for every target (Supplementary Information Sec.~5).

Adding hyperedges does not change $\ker\Gamma$ and does not decrease $\wt(\Gamma t)$, so that $h_2(\Gamma)\ge h(G)$ for every mode. The vertices of $G$ carry 54 orbit labels, namely the 36 data fibres and the 18 hub fibres, and every edge belongs to an edge class $(a,b,s)$ that joins label $a$ at column $c$ to label $b$ at column $c+s$. With $\zeta=e^{2\pi i/79}$, the Fourier transform over the columns block-diagonalizes the Laplacian $L_G$ of $G$ into the $54\times54$ blocks
\begin{equation}
L_k=D-\sum_{(a,b,s)}\bigl(\zeta^{ks}E_{ab}+\bar\zeta^{ks}E_{ba}\bigr),\qquad k\in\mathbb Z_{79},
\end{equation}
where $D$ is the diagonal degree matrix and $E_{ab}$ is a matrix unit. Since $L_{79-k}=\overline{L_k}$, the $40$ blocks $k=0,\dots,39$ determine the spectrum. The bound $\lambda_2(L_G)>2$ is equivalent to $L_k-2I\succ0$ for $k\ne0$ together with $L_0-2I+\tfrac{3}{54}\one\one^{\mathsf T}\succ0$, where the rank-one term moves the eigenvalue of the constant vector from $-2$ to $1$ and leaves the others unchanged. 
Since eigenvalues computed in floating point carry rounding errors, we prove these conditions with integer arithmetic alone. Each block is rounded to integers with a bounded error, and for each of the $40$ blocks we store a certificate, checkable with integer operations only, showing that the block remains positive definite even with the largest possible rounding error (Supplementary Information Sec.~5).

\vspace{2mm}
\noindent\textbf{Magic-state injection and Pauli-based computation} \\
A magic-state source (Supplementary Information Sec.~6.1) is a distance-$18$ unrotated surface code with $613$ data qubits and $612$ checks, whose $\ZZ$ string of $18$ qubits carries a certified gadget with $41$ ancilla data qubits and an $18$-vertex port. For injection, logical qubit $j$ is prepared in $\ket+$, $\ZZ_j\ZZ_\src$ is measured with outcome $m$, and the source is measured transversally in the $X$ basis with outcome $n$; this leaves $j$ in the state $\ket T$ up to a Pauli correction determined by $(m,n)$~\cite{Litinski2019GameOfSurfaceCodes}. The interface adds $238$--$432$ ancilla data qubits; the largest parallel batch injects $20$ magic states in parallel with $\wmax\le15$ and $\Dmax\le16$. For Pauli-based computation, measuring $\PP\otimes\ZZ_\src$ on $\ket\psi\otimes\ket T$ and reading out the source in the $X$ basis yields $C\,e^{-i\pi\PP/8}\ket\psi$ with the byproduct $C=\PP^{\,n}e^{im\pi\PP/4}$ determined by the outcomes $m$ and $n$; the Pauli part is absorbed into the Pauli frame, and the Clifford rotation $e^{i\pi\PP/4}$ is undone by a further PPM or commuted through the subsequent Pauli-based operations, which changes only the Pauli operators measured later. We verified this protocol on all outcome branches, and the merged codes of module, extractor, bridge and source were verified exactly ($\wmax\le16$, $\Dmax\le14$).

\vspace{2mm}
\noindent\textbf{Universality and inter-module operations} \\
Each instruction set is universal on the register qubits (Supplementary Information Sec.~6.2). Stabilizer states are prepared by PPMs followed by Pauli corrections. The rotations $e^{-i\pi\ZZ_j/4}$ and $e^{-i\pi\XX_j/4}$ are implemented by measuring $\ZZ_j\ZZ_a$ or $\XX_j\XX_a$ with a work qubit $a$ prepared in an eigenstate of $\YY_a$, followed by a single-qubit measurement of $a$, and together with the CNOT gate below they generate the Clifford group; injected $\ket T$ states complete a universal gate set by gate teleportation~\cite{Bravyi2005Universal,Litinski2019GameOfSurfaceCodes}. The set $\IH$ contains $\IB$, and $\IF$ supports Pauli-based computation directly~\cite{Bravyi2016Trading}. A logical qubit $j$ is teleported to a qubit $k$ prepared in $\ket+$, possibly in another module, by measuring $\ZZ_j\ZZ_k$ (outcome $m$) and $\XX_j$ (outcome $n$) and applying the correction $\XX_k^m\ZZ_k^n$. A CNOT gate with control $c$ and target $t$ uses a work qubit $a$ prepared in $\ket+$: the measurements $\ZZ_c\ZZ_a$, $\XX_a\XX_t$ and $\ZZ_a$, with outcomes $m_1$, $m_2$ and $m_3$, implement $\mathrm{CNOT}_{c\to t}$ up to the correction $\ZZ_c^{m_2}\XX_t^{m_1+m_3}$~\cite{Litinski2019GameOfSurfaceCodes}, and only the first measurement crosses the module boundary if $a$ resides with $t$. 
Each inter-module port is a path of $18$ consecutive data vertices along the data cycle of an extractor, so every bridge square uses two data-cycle edges. The bridge raises the weight of the port vertex checks by one, to at most $16$, and the degree of the data-cycle edges by one, which keeps them within the extractor bound of $14$; hence $\wmax\le16$ and $\Dmax\le14$ for every pair of targets, and the merged code of two modules encodes $2\times1{,}426-1=2{,}851$ logical qubits. Two seed gadgets in different modules can be bridged in the same way; for instance, the inter-module measurement $\XX_{(3,17)}\otimes\XX_{(11,40)}$ uses $657$ ancilla data qubits.

\vspace{2mm}
\noindent\textbf{Resource accounting} \\
A gadget that adds $n_a$ ancilla data qubits and $m_a$ ancilla checks occupies $n_a+m_a$ physical qubits, with one measurement qubit per check; the ancilla qubits of a gadget are the edges of its graph, $n_a=|E|$, and a connected gadget with a complete cycle basis has $|V|$ vertex checks and $|E|-|V|+1$ cycle checks, so that $m_a=n_a+1$. The bare code module occupies $4{,}266$ physical qubits, or $3.0$ per logical qubit. 
The basic workspace is that of the largest instruction on the $1{,}422$ register qubits, a bridged product with $772$ ancilla data qubits, which gives $4{,}266+2\times772+1=5{,}811$ physical qubits, or $4.1$ per logical qubit.
For the bicycle architecture we use two-gross code modules of $576$ physical qubits ($288$ data qubits and $288$ checks), each with a logical processing unit of $158$ physical qubits~\cite{Yoder2025Tour}. The logical processing unit measures any logical Pauli operator on two fixed logical qubits of its module and reaches the other logical qubits through shift automorphisms, so we count it as the basic workspace of a bicycle module.
The fixed extractor for the Cornucopia processor adds $21{,}330$ ancilla data qubits and $21{,}331$ ancilla checks, which gives $46{,}927$ physical qubits per module; the bicycle architecture has no fixed-extractor configuration in this comparison. The comparison concerns space only and does not account for the different logical error rates, cycle times and decoders of the architectures (Supplementary Information Sec.~7 and Supplementary Tab.~S6).

\vspace{2mm}
\noindent\textbf{Workload benchmark} \\
A benchmark workload is a sequence of $3{,}000$ instructions, each drawn uniformly from $\XX_j$, $\YY_j$, $\ZZ_j$, $\XX_j\XX_k$ and $\ZZ_j\ZZ_k$ on uniformly random register qubits, the domain of the basic workspace. 
In all configurations, a logical cycle is one surgery of $O(d)$ syndrome rounds, and the space--time volume is the number of provisioned physical qubits times the number of logical cycles.
A single basic workspace executes one instruction per logical cycle, and so does the fixed extractor, so both require $3{,}000$ logical cycles.
For the high-throughput set, a list scheduler places every instruction in the earliest layer that follows all earlier instructions on its logical qubits and can accommodate it under the compatibility rule (Proposition~\ref{prop:compat}); each layer is executed as one merged code in one logical cycle. A layer occupies the $4{,}266$ qubits of the bare module and the ancilla data qubits and ancilla checks that its merged code adds. Since the layers are executed one after another, the high-throughput configuration provides as many physical qubits as the largest layer occupies. The three workloads required $149$--$153$ cycles, with $19.9$ instructions per layer on average and at most $29{,}755$ physical qubits (Supplementary Tab.~S5).
For the basic workspace and the high-throughput set, these counts assume that one pool of ancilla qubits serves every instruction or layer in turn, and they do not model the connectivity that this requires.

\vspace{2mm}
\noindent\textbf{Exact verification} \\
Every merged code reported in this work was verified by exact computation: commutation of all generators; the rank over $\F$, which fixes the number of encoded logical qubits and thereby excludes unintended measurements; the product of the vertex checks; all row and column weights; and the integer or rational spectral certificates of every gadget. 
A single verification program checks the canonical basis (all pairings and exact translation covariance), the gadget library, the bridged classes (all $25{,}578$ classes), the extractor certificates and modes, the inter-module modes, magic-state injection and the parallel batches.

\vspace{.4cm}
\noindent\textbf{\large{}Data availability}\\
The data presented in the figures and that support the other findings of this study will be made publicly available for download on Zenodo upon publication.

\vspace{.4cm}
\noindent\textbf{\large{}Code availability}\\
The data analysis and numerical simulation codes will be made publicly available for download on Zenodo upon publication.

\vspace{.4cm}
\noindent\textbf{\large{}Acknowledgments}\\
This work is supported by the National Natural Science Foundation of China (T2225008), the Quantum Science and Technology-National Science and Technology Major Project (2021ZD0302203), the Fundamental and Interdisciplinary Disciplines Breakthrough Plan of the Ministry of Education of China (JYB2025XDXM112), the Shanghai Qi Zhi Institute Innovation Program (SQZ202318), and the Tsinghua University Dushi Program. B.W., W.L., and Z.-Z.S. acknowledge in addition support from the China Postdoctoral Science Foundation (2025T180926). 
The authors acknowledge the use of the AI tool Claude, which assisted with the calculations, verification and editing of the manuscript. The authors reviewed all AI-assisted output and take full responsibility for the content of this manuscript.

\vspace{.4cm}
\noindent\textbf{\large{}Author contributions} \\
Z.L. and W.L. designed the surgery gadgets for the Cornucopia processor under the supervision of D.-L.D..
Z.L., W.L., Y.-H.M., H.Y., B.W., Y.-X.M. and Z.-Z.S. developed the code that constructs and  verifies all gadgets and merged codes.
All authors contributed to writing the manuscript.

\vspace{.4cm}
\noindent\textbf{\large{}Competing interests} \\
The authors declare no competing interests.

\clearpage
\onecolumngrid
\renewcommand{\figurename}{Extended Data Fig.}
\renewcommand{\tablename}{Extended Data Tab.}
\setcounter{figure}{0}
\setcounter{table}{0}
\renewcommand{\theHfigure}{ED.\arabic{figure}}
\renewcommand{\theHtable}{ED.\arabic{table}}
\makeatletter
\setlength{\@fptop}{0pt}\setlength{\@dblfptop}{0pt}
\makeatother

\begin{datatable}
\caption{\textbf{Instruction set of a Cornucopia module.} For each instruction: logical action, gadget, number $n_a$ of ancilla data qubits added to the module, number of added checks, and the maximum check weight $\wmax$ and qubit degree $\Dmax$ of the merged code ($12$ and $6$ for the bare code). Ranges are over all instances (all registers, pairs and targets). $^\dagger$A distance-$18$ unrotated surface-code source adds $613$ data qubits and $612$ checks and carries an $18$-vertex gadget. $s$: number of shortcut ancilla data qubits of the batch (Methods).}
\label{tab:isa}
\begin{ruledtabular}
\begin{tabular}{llllrcc}
Instruction & Action & Gadget & $n_a$ & checks & $\wmax$ & $\Dmax$\\\hline
Shift $\overline U^{s}$ & $(i,t)\mapsto(i,t{+}s)$ & data-tile translation $T^s$ & 0 & 0 & 12 & 6\\
$\XX_j$, $\ZZ_j$ & PPM & seed $T^t\Sg_{X_i},T^t\Sg_{Z_i}$ & 179--377 & $n_a{+}1$ & 14 & 8\\
$\YY_j$ & PPM & seed $T^t\Sg_{Y_i}$ & 547--752 & $n_a{+}1$ & 14 & 9\\
$\XX_j\XX_k$ & PPM & two seeds + bridge & 586--772 & $n_a{+}1$ & 16 & 8\\
$\ZZ_j\ZZ_k$ & PPM & two seeds + bridge & 376--764 & $n_a{+}1$ & 16 & 8\\
$\ket T\to j$ & $\ZZ_j\ZZ_{\rm src}$, $X_{\rm src}$ & seed + bridge + source$^\dagger$ & 238--432 & $n_a{+}1$ & 14 & 8\\
Parallel batch & $m$ PPMs & direct sum & $\sum_g n_{a,g}+s$ & $\sum_g (n_{a,g}{+}1)+s$ & 16 & 16\\
Native family & 78 PPMs & 79 translates & 8{,}374--8{,}690 & $n_a{+}79$ & 13 & 14\\
Any $\PP$ & PPM & fixed extractor & 21{,}330 & 21{,}331 & 15 & 14\\
$e^{-i\pi\PP/8}$ & $\PP\ZZ_{\rm src}$, $X_{\rm src}$ & extractor + bridge + source$^\dagger$ & $+${}59 & $+${}59 & 16 & 14\\
$\PP_A\otimes\PP_B$ & PPM & two extractors + bridge & $+18$ & $+17$ & 16 & 14\\
\end{tabular}
\end{ruledtabular}
\end{datatable}

\begin{datatable}
\caption{\textbf{Native operator families and their full-orbit batches.} For each of the six families $\mathsf P^X_i$ and $\mathsf P^Z_i$ ($i=3,4,5$), the full-orbit batch is the direct sum of the $79$ translates of one certified gadget and returns $78$ independent logical outcomes. Every check of the opposite Pauli type is deformed by six translates, and one shortcut per such check, $711$ in total, keeps every check weight at most $16$. Gadget $n_a$: ancilla qubits of one gadget; batch $n_a$: ancilla qubits of the batch, $79$ times the gadget value plus the $711$ shortcuts.}
\label{tab:native}
\begin{ruledtabular}
\begin{tabular}{lrrrrr}
Family & gadget $n_a$ & batch $n_a$ & shortcuts & $\wmax$ & $\Dmax$\\\hline
$\mathsf P^X_{3}$ & 97 & 8{,}374 & 711 & 13 & 14\\
$\mathsf P^X_{4}$ & 101 & 8{,}690 & 711 & 13 & 14\\
$\mathsf P^X_{5}$ & 97 & 8{,}374 & 711 & 13 & 14\\
$\mathsf P^Z_{3}$ & 97 & 8{,}374 & 711 & 13 & 14\\
$\mathsf P^Z_{4}$ & 101 & 8{,}690 & 711 & 13 & 14\\
$\mathsf P^Z_{5}$ & 97 & 8{,}374 & 711 & 13 & 14\\
\end{tabular}
\end{ruledtabular}
\end{datatable}

\begin{datatable}
\caption{
\textbf{Verified extractor modes.} For each target on a single module, the merged code was verified exactly: the product of the vertex checks equals the target, the code encodes $k'=1{,}425$ logical qubits, and the check weights and qubit degrees are as listed. Every merged code has $24{,}174$ data qubits ($2{,}844$ data and $21{,}330$ ancilla data qubits) and $22{,}753$ checks ($1{,}422$ old and $21{,}331$ new checks). $|P|$: weight of the physical representative $P$ of the target; sign: the sign $\sigma=\pm1$ by which the logical operator differs from $P$ (Methods).
Dense random: a logical Pauli product with uniformly random $X$ and $Z$ on all $1{,}422$ register qubits.
}
\label{tab:extmodes}
\begin{ruledtabular}
\begin{tabular}{lrrrr}
Target & $|P|$ & sign & $\wmax$ & $\Dmax$\\\hline
$\XX_{(0,0)}$ & 102 & $+$ & 13 & 9\\
$\ZZ_{(9,40)}$ & 94 & $+$ & 13 & 9\\
$\YY_{(0,0)}$ & 217 & $+$ & 15 & 9\\
$\XX_{(0,0)}\XX_{(12,53)}$ & 142 & $+$ & 13 & 9\\
$\ZZ_{(0,13)}\ZZ_{(17,78)}$ & 148 & $+$ & 13 & 9\\
dense random & 1{,}798 & $+$ & 15 & 12\\
\end{tabular}
\end{ruledtabular}
\end{datatable}

\begin{datatable}
\caption{
\textbf{Verified inter-module modes.} 
First four rows: the extractors of two modules joined by an $18$-edge matching bridge; every merged code encodes $k'=2{,}851=2\times1{,}426-1$ logical qubits. Last row: a basic instruction across modules, realized by two seed gadgets joined by an $18$-edge bridge ($657$ ancilla qubits, $k'=2{,}851$). $|P_A|,|P_B|$: weights of the physical representatives of the targets on modules $A$ and $B$.
Dense $\otimes$ dense: independent dense random products on the two modules.
}
\label{tab:intermodule}
\begin{ruledtabular}
\begin{tabular}{lrrr}
Target & $|P_A|,|P_B|$ & $\wmax$ & $\Dmax$\\\hline
$\XX_{(0,0)}\otimes\XX_{(0,5)}$ & 102, 102 & 13 & 9\\
$\YY_{(0,4)}\otimes\YY_{(9,40)}$ & 217, 197 & 15 & 9\\
$\ZZ_{(0,13)}\otimes\XX_{(17,78)}$ & 116, 114 & 13 & 9\\
dense $\otimes$ dense & 1{,}778, 1{,}761 & 16 & 12\\
$\XX_{(3,17)}\otimes\XX_{(11,40)}$ (seeds) & 112, 94 & 14 & 8\\
\end{tabular}
\end{ruledtabular}
\end{datatable}

\begin{figure*}[p]
\centering\includegraphics[width=\textwidth]{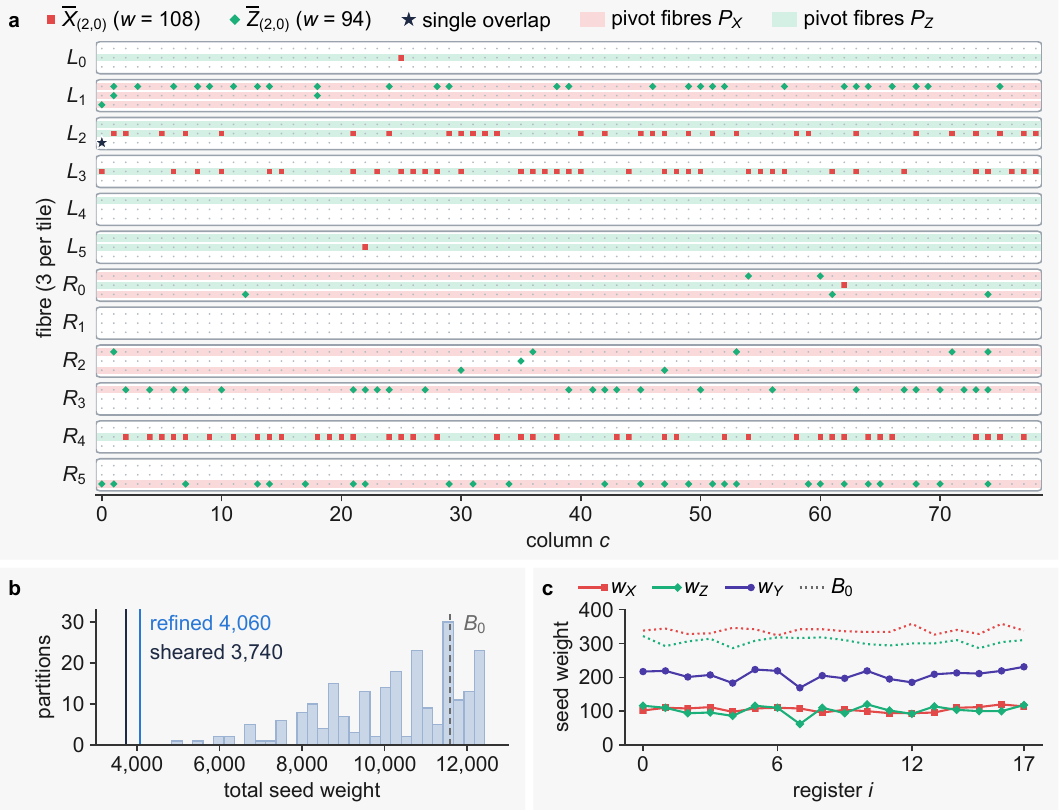}
\caption{\textbf{Canonical logical basis.} 
\textbf{a},~Supports of the conjugate seed pair $\XX_{(2,0)}$ and $\ZZ_{(2,0)}$ of the optimized basis on the $36\times79$ grid of data qubits (rows: the $36$ fibres, three per tile; columns: $c\in\mathbb Z_{79}$). Red and green shading marks the $X$- and $Z$-pivot fibres $P_X$ and $P_Z$; unshaded rows are the free fibres. The two supports intersect in a single qubit (star), so that $x\cdot z=1$.
\textbf{b},~Total weight $W$ of the $36$ $X$- and $Z$-type seed representatives: histogram over the $229$ valid random fibre partitions, the first valid partition $B_0$ (dashed line), and the selected partition after refinement by simulated annealing and after $17$ shears (solid lines).
\textbf{c},~Final seed weights $w_X$, $w_Z$ and $w_Y=\wt(x\lor z)$ of the $18$ registers; dotted lines show $w_X$ and $w_Z$ of $B_0$.
}
\label{edfig:basis}
\end{figure*}

\begin{figure*}[p]
\centering\includegraphics[width=\textwidth]{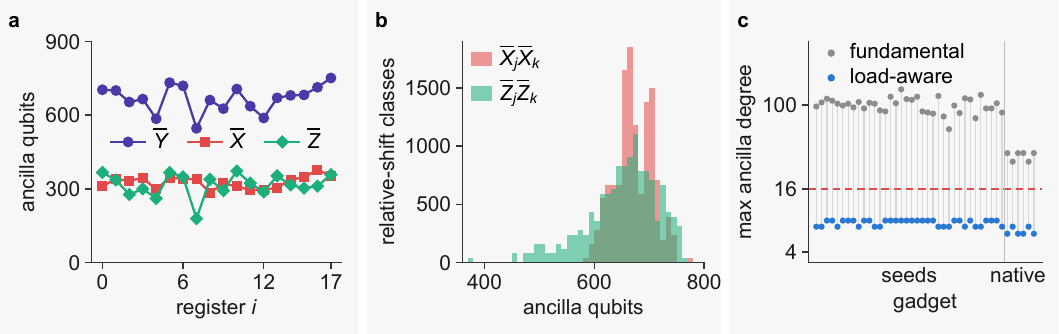}
\caption{
\textbf{Costs of the basic instruction set.}
\textbf{a},~Ancilla data qubits of the $54$ seed gadgets, one $\XX$, $\ZZ$ and $\YY$ seed for each of the $18$ registers. \textbf{b},~Distribution of the ancilla counts of the bridged gadgets $\XX_j\XX_k$ and $\ZZ_j\ZZ_k$ over the $12{,}789$ translation-inequivalent pairs of each type. 
\textbf{c},~Maximum ancilla degree, that is, the largest number of checks acting on a gadget ancilla data qubit, of the $36$ $X$- and $Z$-type seed gadgets and the six native gadgets (logarithmic scale), with fundamental (spanning-tree) cycle bases and with the load-aware cycle bases used in this work; the dashed line marks the degree cap of $16$.}
\label{edfig:ibdata}
\end{figure*}

\end{document}


\renewcommand{\tablename}{Tab.}
\renewcommand{\figurename}{Fig.}

\title{Supplementary Information for ``A modular quantum computer based on Cornucopia codes''}
\maketitle

\tableofcontents

\section{Conventions and code parameters}\label{sec:conventions}
This section fixes the notation, recalls the definition of Cornucopia codes, and introduces the translation automorphism and the fibres on which the whole construction rests.

\subsection{Binary Pauli notation}
Up to a phase, a Pauli operator on $n$ qubits is described by a pair of binary vectors $(x,z)\in\F^n\times\F^n$. Throughout this work we use the Hermitian representative
\begin{equation}
P(x,z)=i^{|x\wedge z|}X(x)Z(z),\qquad X(x)=\bigotimes_jX_j^{x_j},\quad Z(z)=\bigotimes_jZ_j^{z_j},
\label{eq:pauli}
\end{equation}
where $|x\wedge z|$ is the number of positions at which both $x$ and $z$ equal one. Since $iXZ=Y$, the operator $P(x,z)$ is a tensor product of single-qubit operators $I,X,Y,Z$, and its weight is $\wt(x\lor z)$; a $Y$ factor counts as a single qubit. Two Pauli operators commute if and only if their symplectic product $x\cdot z'+z\cdot x'$ vanishes modulo two, and products obey $X(x)Z(z)X(x')Z(z')=(-1)^{z\cdot x'}X(x+x')Z(z+z')$.

A CSS code is specified by check matrices $H_X\in\F^{m_X\times n}$ and $H_Z\in\F^{m_Z\times n}$ with $H_XH_Z^{\mathsf T}=0$; its stabilizer group is generated by $X(h)$ for the rows $h$ of $H_X$ and by $Z(h')$ for the rows $h'$ of $H_Z$. The $X$-type logical operators are the vectors $x\in\ker H_Z$ modulo $\row H_X$, and similarly for the $Z$ type,
\begin{equation}
\mathcal L_X=\ker H_Z/\row H_X,\qquad \mathcal L_Z=\ker H_X/\row H_Z,\qquad k=n-\rk H_X-\rk H_Z .
\end{equation}
The distance $d$ is the minimum weight of a nontrivial logical operator, that is, of a Pauli operator that commutes with all checks but does not belong to the stabilizer group.

\subsection{The Cornucopia code}
The data qubits are arranged in twelve tiles $L_0,\dots,L_5,R_0,\dots,R_5$ and the checks in six tiles $X_0,X_1,X_2,Z_0,Z_1,Z_2$; every tile is a $3\times q$ array with coordinates $(r,c)\in\mathbb Z_3\times\mathbb Z_q$. Data qubit $(b,r,c)$, with $b=0,\dots,11$ enumerating $L_0,\dots,L_5,R_0,\dots,R_5$, has index $(3b+r)q+c$, and check $(i,r,c)$ of either type has index $(3i+r)q+c$. The blocks $\mathcal A_j$ and $\mathcal B_j$ ($j\in\mathbb Z_6$) of Eq.~(1) in the main text are the $3q\times3q$ permutation matrices of the coordinate maps $A_j,B_j:(r,c)\mapsto(\rho(r),c+s)$, with rows and columns indexed by the coordinates of a tile and with the convention of ref.~\cite{Lu2026Quantum}: $(\mathcal A_j)_{u,u'}=1$ if and only if $u'=A_j(u)$, and likewise for $\mathcal B_j$. Row $u$ of $\mathcal A_j$ therefore has its single nonzero entry in column $A_j(u)$, and $\mathcal A_j^{-1}=\mathcal A_j^{\mathsf T}$ is the matrix of the coordinate map $A_j^{-1}$. For the Cornucopia code $\cod{2844}{1426}{18}$, the column shifts and row maps are
\begin{equation}
\begin{gathered}
(s_{A_0},\dots,s_{A_5})=(6,49,55,18,40,7),\qquad (s_{B_0},\dots,s_{B_5})=(24,41,78,53,68,21),\\
\rho_{A_0}=\rho_{B_2}:r\mapsto r+1,\qquad \rho_{A_1}=\rho_{B_3}:r\mapsto2r+2,\qquad \rho=\mathrm{id}\ \text{otherwise}.
\end{gathered}
\label{eq:shifts}
\end{equation}
All three row maps are affine permutations of $\mathbb Z_3$: $r\mapsto r+1$ is a cyclic shift, and $r\mapsto2r+2$ exchanges rows $0$ and $2$ and fixes row $1$. The check matrices of Eq.~(1) in the main text are $3\times12$ block matrices whose block rows are check tiles and whose block columns are data tiles. With the above convention, the $X$ check at coordinate $(r,c)$ of tile $X_i$ acts on the qubit at coordinate $A_{j-i}(r,c)$ of $L_j$ and on the qubit at $B_{j-i}(r,c)$ of $R_j$, and the $Z$ check at coordinate $(r,c)$ of tile $Z_i$ acts on the qubits at $B_{i-j}^{-1}(r,c)$ of $L_j$ and $A_{i-j}^{-1}(r,c)$ of $R_j$.

Every row of $H_X$ has exactly one nonzero entry in each of the twelve data tiles, because every block is a permutation matrix, so every check has weight $12$. Every column has one nonzero entry in each of the three $X$ check tiles, so every data qubit lies in exactly three $X$ checks and three $Z$ checks, and the qubit degree of the bare code is six. 
We obtain that $H_XH_Z^{\mathsf T}=0$~\cite{Lu2026Quantum}.

\subsection{Parameters and redundancies}
For $q=79$ and the parameters of Eq.~\eqref{eq:shifts}, the code has $n=36q=2{,}844$ data qubits and $3\times3q=711$ checks of each type. Both check matrices have rank $709$, so that
\begin{equation}
k=2{,}844-709-709=1{,}426 .
\end{equation}
The two redundancies per type have a simple origin. The $3q$ checks of one check tile form a permutation on every data tile, so their sum is the all-ones vector $\one_n$. Since the checks of each of the three $X$ check tiles sum to $\one_n$, the relations $\sum X_0+\sum X_1=0$ and $\sum X_0+\sum X_2=0$ are independent, and similarly for the $Z$ type. In particular, $X^{\otimes n}$ and $Z^{\otimes n}$ are stabilizers. The distance $d=18$ was certified in ref.~\cite{Lu2026Quantum}.

\subsection{Translation automorphism}
The translation $T:(b,r,c)\mapsto(b,r,c+1)$ is a permutation of the data qubits that shifts every data tile by one column. Because column shifts commute with each other and row maps do not act on columns, $T$ maps the support of check $(i,r,c)$ onto the support of check $(i,r,c+1)$. Thus $T$ maps the stabilizer group onto itself and is an automorphism of the code, which maps logical operators to logical operators.

The $q$ data qubits with a fixed tile and row form a \emph{fibre}, labelled $f=3b+r\in\{0,\dots,35\}$; similarly there are nine $X$-check fibres and nine $Z$-check fibres. The translation $T$ acts on every data fibre as a cyclic shift of $\mathbb Z_q$, and its induced relabelling acts in the same way on every check fibre. Identifying $T$ with an indeterminate $D$, the data space becomes the free module $\Rq^{36}$ over the ring
\begin{equation}
\Rq=\F[D]/(D^q-1),
\end{equation}
and $H_X$ becomes a $9\times36$ matrix over $\Rq$ whose entries are monomials $D^s$ or zero. For $q=79$ the multiplicative order of $2$ modulo $79$ is $39$, so that $D^{79}-1=(D-1)\Phi_{79}(D)$, where the cyclotomic polynomial $\Phi_{79}$ factors over $\F$ into two irreducible polynomials of degree $39$. Hence
\begin{equation}
\mathcal R_{79}\cong\F\times\mathbb F_{2^{39}}\times\mathbb F_{2^{39}} ,
\label{eq:ringsplit}
\end{equation}
where the first factor corresponds to evaluation at $D=1$, that is, to the $T$-invariant part.
In neutral-atom implementations, the shift $T^s$ only requires moving the data tiles by $s$ columns relative to the check tiles, a collective column translation of the same type as those that drive syndrome extraction~\cite{Lu2026Quantum}. The check qubits are reset in every round and carry no logical information, so they need not move: after the shift, the check qubit at $(i,r,c)$ measures the check labelled $(i,r,c-s)$, and the measured stabilizer group is unchanged.

\subsection{Logical Pauli operators and signs}
Logical qubits are labelled by the linear index $j=79i+t$ for the moving logical qubit $(i,t)$ in register $i\in\{0,\dots,17\}$ at position $t\in\mathbb Z_{79}$, and by $j=1{,}422,\dots,1{,}425$ for the four invariant modes (Sec.~\ref{sec:basis}).
A Hermitian logical Pauli operator is written as $\PP=i^{|a\wedge b|}\XX^a\ZZ^b$ with $a,b\in\F^{1426}$. Its physical representative is $X(x)Z(z)$ with $x=aL_X$ and $z=bL_Z$, because $\XX^a=\prod_jX(x_j)^{a_j}=X(aL_X)$ and similarly for $\ZZ^b$. Hence
\begin{equation}
\PP=i^{|a\wedge b|}X(x)Z(z)=i^{|a\wedge b|-|x\wedge z|}P(x,z)=\sigma P(x,z),\qquad \sigma=(-1)^{(|a\wedge b|-|x\wedge z|)/2}.
\label{eq:sign}
\end{equation}
The exponent $|a\wedge b|-|x\wedge z|$ is even. Indeed, both $\PP$ and $P(x,z)$ are Hermitian, so they can differ only by a sign; explicitly, $|x\wedge z|\equiv x\cdot z=\sum_{j,l}a_jb_l\,x_j\cdot z_l=a\cdot b\equiv|a\wedge b|\pmod2$, where the canonical condition $x_j\cdot z_l=\delta_{jl}$ was used. Multiplying a measured outcome of $P(x,z)$ by $\sigma$ yields the logical outcome. For products of only $X$ or only $Z$ factors $\sigma=1$, whereas products with $Y$ factors require tracking $\sigma$.

\section{Translation-covariant canonical logical basis}\label{sec:canonical}
A logical processor requires a fixed logical coordinate system: every logical qubit $j$ has definite physical representatives of $\XX_j$ and $\ZZ_j$, every logical Pauli product can be converted into a physical Pauli operator together with its sign, and every measured outcome can be converted back. We obtain such a basis from the symmetry-adapted standard form (Sec.~\ref{sec:basis}) and then reduce the weights of its representatives (Sec.~\ref{sec:optimization}).

\subsection{Canonical basis}\label{sec:basis}
Theorem~1 of the main text states the symmetry-adapted standard form, which applies Gottesman's standard form~\cite{Gottesman1997Stabilizer} to whole symmetry orbits, for the translation symmetry of Cornucopia codes; we include its proof for completeness. We also prove Lemma~1 of the main text, analyse the affine freedom of the construction and construct the invariant sector.

\begin{definition}\label{def:canonical}
Matrices $L_X,L_Z\in\F^{k\times n}$ with rows $x_j$ and $z_j$ form a \emph{canonical basis} if
\begin{equation}
H_ZL_X^{\mathsf T}=0,\qquad H_XL_Z^{\mathsf T}=0,\qquad L_XL_Z^{\mathsf T}=I_k .
\label{eq:canonical}
\end{equation}
\end{definition}
The first two conditions only state that every $X(x_j)$ and $Z(z_j)$ commutes with all checks, so a single row could still be a stabilizer. The third condition rules this out, as the following lemma shows.

\begin{lemma}\label{lem:pairing}
If Eq.~\eqref{eq:canonical} holds, no nonzero combination of the rows $x_j$ lies in $\row H_X$, and no nonzero combination of the rows $z_j$ lies in $\row H_Z$. Consequently $\{x_j\}$ and $\{z_j\}$ are bases of $\mathcal L_X$ and $\mathcal L_Z$, the operators $\XX_j=X(x_j)$ and $\ZZ_j=Z(z_j)$ anticommute, and operators with different indices commute.
\end{lemma}
\begin{proof}
Suppose that $aL_X\in\row H_X$. Every vector of $\row H_X$ is orthogonal to every vector of $\ker H_X$, and $z_j\in\ker H_X$, so $(aL_X)z_j^{\mathsf T}=a_j=0$ for all $j$, that is, $a=0$. The $k$ vectors $x_j$ are therefore linearly independent in the $k$-dimensional space $\mathcal L_X$ and form a basis; the same argument applies to the $z_j$. The remaining statements are an entrywise reading of $L_XL_Z^{\mathsf T}=I_k$.
\end{proof}
Checking the $k^2$ pairings thus certifies the independence and nontriviality of all $2k$ representatives; for $k=1{,}426$ these are $2{,}033{,}476$ pairings.

\begin{restated}{Theorem~1}{Translation-covariant canonical basis}
Partition the $36$ data fibres into sets $P_X$ and $P_Z$ of nine fibres each and a free set $F$ of $18$ fibres such that $\rk H_X[:,P_X]=\rk H_X$ and $\rk H_Z[:,P_Z]=\rk H_Z$. For every $f\in F$, let $e_f$ be the unit vector supported on the data qubit at column $0$ of fibre $f$, and choose $u_f$ supported on $P_Z$ and $v_f$ supported on $P_X$ with $H_Zu_f^{\mathsf T}=H_Ze_f^{\mathsf T}$ and $H_Xv_f^{\mathsf T}=H_Xe_f^{\mathsf T}$. Then the $18q$ pairs $\XX_{(f,t)}=X(T^t(e_f+u_f))$ and $\ZZ_{(f,t)}=Z(T^t(e_f+v_f))$, $t\in\mathbb Z_q$, are canonical logical operators, the supports of conjugate operators intersect in exactly one qubit, those of nonconjugate operators are disjoint, and $T$ acts as $(f,t)\mapsto(f,t+1)$. They extend to a complete canonical basis by $k-18q$ further pairs.
\end{restated}

\begin{proof}
\emph{Solvability.} The column space of $H_Z[:,P_Z]$ is contained in that of $H_Z$, and the two have equal rank, so they coincide. The syndrome $H_Ze_f^{\mathsf T}$ lies in the column space of $H_Z$ and is therefore realized by some $u_f$ supported on $P_Z$. The same argument applies to the $X$ side.

\emph{Commutation with the checks.} In characteristic two, $H_Z(e_f+u_f)^{\mathsf T}=H_Ze_f^{\mathsf T}+H_Zu_f^{\mathsf T}=0$, so $x_f=e_f+u_f\in\ker H_Z$, and similarly $z_f=e_f+v_f\in\ker H_X$. Since $T$ is an automorphism of the code, all translates commute with the checks as well.

\emph{Pairing.} The support of $T^tu_f$ lies in $P_Z$, that of $T^{t'}v_{f'}$ in $P_X$, and $T^te_f$ is the single qubit $(f,t)$ of the free set $F$. Because the three sets of fibres are disjoint,
\begin{equation}
\supp(T^tx_f)\cap\supp(T^{t'}z_{f'})=\{(f,t)\}\cap\{(f',t')\},
\end{equation}
so that $(T^tx_f)\cdot(T^{t'}z_{f'})=\delta_{ff'}\delta_{tt'}$: conjugate representatives intersect in exactly one qubit, and all others are disjoint. By Lemma~\ref{lem:pairing}, the $18q$ pairs are independent, nontrivial logical classes.

\emph{Action of $T$.} By construction, $T$ maps the pair $(f,t)$ to $(f,t+1)$.

\emph{Completion.} The $18q$ pairs span a symplectic subspace $W$ of the $2k$-dimensional symplectic space of logical classes. Its symplectic complement $W^\perp$ has dimension $2(k-18q)$ and is nondegenerate, so it admits a canonical basis, which together with the pairs of $W$ forms a complete canonical basis.
\end{proof}

The dimensions are consistent with the code parameters: with $n=36q$ and $\rk H_X=\rk H_Z=9q-2$ one has $k=36q-2(9q-2)=18q+4$. The $18$ free fibres contribute the $18q$ moving logical qubits, and the two missing ranks of each pivot block, whose $9q$ columns have rank $9q-2$, leave the four logical qubits of the invariant sector, called fixed modes in ref.~\cite{Lu2026Quantum}.

\runin{Affine solutions}
Each pivot block $H_X[:,P_X]$ has $9q=711$ columns and rank $709$, so its null space $N_X$, consisting of vectors $a$ supported on $P_X$ with $H_Xa^{\mathsf T}=0$, is two-dimensional, and the solutions of each seed equation form a coset $v_f+N_X$ with four elements.

\begin{lemma}\label{lem:affine}
Every nonzero element of $N_X$ is a $T$-invariant $Z$-type logical operator or a $T$-invariant stabilizer. For the selected partition, the three nonzero elements are nontrivial $T$-invariant logical operators of weight $158$, that is, of two complete fibres. The same holds for $N_Z$.
\end{lemma}
\begin{proof}
An element $a\in N_X$ satisfies $H_Xa^{\mathsf T}=0$, so $Z(a)$ commutes with all $X$ checks. The translation $T$ maps $P_X$ to itself and preserves $H_X$ up to a permutation of rows, so it acts linearly on the two-dimensional space $N_X$. The order of this action divides $q=79$,
whereas every element of $GL_2(\F)$, a group of six elements, has order $1$, $2$ or $3$; hence $T$ acts trivially on $N_X$, and every element of $N_X$ is $T$-invariant. Whether an element is a stabilizer is decided by exact computation: for the selected partition, none of the three nonzero elements lies in $\row H_Z$, and all three have weight $158$.
\end{proof}

Choosing one of the four affine solutions therefore amounts to choosing a basis, which differs by an invariant logical operator, rather than reducing the weight within a logical class. We select the lightest element of each coset.

Adding a stabilizer to a representative, $x\mapsto x+s$ with $s\in\row H_X$ (and likewise $z\mapsto z+s$ with $s\in\row H_Z$), does not change its logical class but can change its weight; we call this \emph{stabilizer dressing}. The $X$-type seeds of Theorem~1 vanish on $P_X$, and the following lemma shows that any nonzero dressing would put weight on $P_X$.

\begin{restated}{Lemma~1}{No dressing within a pivot structure}
If $\rk H_X[:,P_X]=\rk H_X$ and $s\in\row H_X$ vanishes on $P_X$, then $s=0$.
\end{restated}
\begin{proof}
Write $s=aH_X$. Then $aH_X[:,P_X]=0$. Two matrices with the same column space have the same left kernel, since the left kernel is the annihilator of the column space. Hence $aH_X=0$, that is, $s=0$.
\end{proof}

Effective stabilizer dressing must therefore move a representative outside the support structure of the partition. This is why dressing is applied only after the shears, which mix the registers (Sec.~\ref{sec:optimization}), and why it has little effect: for the selected basis, it leaves every seed unchanged.

The translation $T$ acts linearly on the logical classes. On the moving sector it is the cyclic shift of the $18$ registers, which in the language of $\Rq$ is multiplication by $D$ on $\Rq^{18}$. The only elements $p\in\Rq$ with $Dp=p$ are $0$ and $\Omega=1+D+\dots+D^{q-1}$, so the $T$-invariant $X$-type classes of the moving sector are spanned by the $18$ \emph{orbit sums}
\begin{equation}
\sigma_i=\sum_{t\in\mathbb Z_q}T^tx_i,\qquad i=0,\dots,17 .
\end{equation}
Together with the four invariant pairs, the $T$-invariant $X$-type logical classes span $22$ dimensions. In the decomposition of Eq.~\eqref{eq:ringsplit}, $\Omega$ corresponds to the factor $\F$ obtained by evaluation at $D=1$.

\begin{lemma}\label{lem:fibreconstant}
Every $T$-invariant logical class contains a representative that is constant on fibres, that is, a vector of the form $v_u=\sum_{f\in u}\one_f$ with $u\subseteq\{0,\dots,35\}$, where $\one_f$ is the indicator vector of fibre $f$. In particular, its weight is a multiple of $q$.
\end{lemma}
\begin{proof}
For odd $q$, $\Omega^2=q\Omega=\Omega$ is idempotent. If the class of $v$ is $T$-invariant, then $Tv=v+s$ with a stabilizer $s$, and by induction $T^tv=v+s_t$ with stabilizers $s_t$, because the stabilizer group is $T$-invariant. Hence $\Omega v=\sum_tT^tv=qv+\sum_ts_t$ differs from $v$ by a stabilizer, since $q$ is odd. The vector $\Omega v$ satisfies $T\Omega v=\Omega v$ bitwise, and since $T$ acts transitively on every fibre, it is constant on every fibre.
\end{proof}

\emph{Construction.} The four invariant pairs are constructed in four steps.
\begin{enumerate}
\item \emph{Fibre-constant kernel vectors.} The condition $v_u\in\ker H_Z$ can be expressed at the level of fibres. Every check in $Z$-check fibre $(i,r)$ intersects data fibre $f$ in the same number $\mu_{(i,r),f}\in\{0,1\}$ of qubits, because the $q$ checks of a check fibre are related by translations. Hence $H_Zv_u^{\mathsf T}=0$ if and only if $\mu u=0$ for the binary $9\times36$ matrix $\mu$, which is obtained from the first check of every check fibre.
\item \emph{Orthogonalization.} A vector $v_u$ need not commute with the moving $Z$-type representatives. Since $v_u$ is $T$-invariant, its product with $T^tz_i$ does not depend on $t$, and if it anticommutes with $z_i$ we add the orbit sum $\sigma_i$. Because $\sigma_i\cdot T^tz_{i'}=\sum_sT^sx_i\cdot T^tz_{i'}=\delta_{ii'}$, this step flips only the anticommutation with register $i$. The orbit sum $\sigma_i$ is bitwise $T$-invariant and hence constant on fibres, so the orthogonalized vectors remain constant on fibres.
\item \emph{Selection.} The orthogonalized vectors are tested in turn for independence modulo $\row H_X$ and the $1{,}422$ moving $X$-type representatives, and the first four independent vectors are kept. The $Z$ side is treated symmetrically.
\item \emph{Canonicalization.} Let $\xi_1,\dots,\xi_4$ and $\zeta_1,\dots,\zeta_4$ be the selected $X$- and $Z$-type vectors with Gram matrix $G_{ab}=\xi_a\cdot\zeta_b$. Both sets are orthogonal to the moving sector and together with it span all logical classes, so $G$ is invertible, and replacing $\zeta$ by $\zeta'=(G^{-1})^{\mathsf T}\zeta$ yields $\xi_a\cdot\zeta'_b=\delta_{ab}$.
\end{enumerate}
The resulting four pairs are strictly $T$-invariant, with weights $(1422, 1422, 1106, 790)$ for the $X$ type and $(1422, 1264, 1106, 948)$ for the $Z$ type, all multiples of $79$.

\emph{Change of basis.} Within the four-dimensional invariant sector, any symplectic change of basis $X'=MX$, $Z'=(M^{-1})^{\mathsf T}Z$ with $M\in GL_4(\F)$ preserves canonicity, since $X'Z'^{\mathsf T}=MXZ^{\mathsf T}M^{-1}=I$, and leaves the moving sector unchanged. 
We enumerate all $20{,}160$ elements of $GL_4(\F)$ and compare the eight new weights by their maximum and then by their sum; ties are broken by the smallest first row of $M$, then the smallest second row, and so on. The optimal $M$ replaces $(x_0,x_1,x_2,x_3)$ by $(x_0+x_1,\,x_1+x_2,\,x_0+x_1+x_2,\,x_0+x_2+x_3)$ and, correspondingly, $(z_0,z_1,z_2,z_3)$ by $(z_1+z_2+z_3,\,z_0+z_1+z_3,\,z_0+z_1+z_2,\,z_3)$, which yields weights $(474, 948, 1106, 948)$ and $(948, 1264, 790, 948)$.

\emph{Dressing.} Because the invariant representatives consist of entire fibres, they overlap strongly with many checks and admit substantial weight reduction within their classes. For each representative $r$ we run simulated annealing in $r+\row H$: each step flips a randomly chosen check $s$ of the same type, which changes the weight by $\delta=12-2|s\cap r|$, and is accepted according to the Metropolis rule at a temperature that decreases geometrically from $4$ to $0.02$ over $1.5\times10^6$ steps; the lightest visited representative is kept. The program maintains the overlap of every check with the current representative and, when a bit is flipped, updates only the three checks containing it. The random seeds are $11+j$ for the $X$ side and $21+j$ for the $Z$ side. The final weights are $(474, 692, 666, 680)$ ($X$) and $(682, 674, 678, 672)$ ($Z$). The difference between each new representative and the original one is verified to lie in $\row H_X$ or $\row H_Z$, so the logical classes and the canonical pairing are unchanged; the representatives are no longer bitwise invariant, but their classes are, which the verification checks through $x+Tx\in\row H_X$.

The invariant representatives are considerably heavier than the moving seeds, because their logical information is distributed over the entire block: a fibre-constant class contains each fibre either completely or not at all, and dressing breaks this structure only partially. We thus do not equip the invariant modes with gadgets: the basic and high-throughput instruction sets act on the $1{,}422$ register qubits. The invariant pairs complete the canonical basis, so that every logical Pauli operator has a definite physical representative and sign, and the invariant modes remain accessible to the fixed extractor, whose distance guarantee covers every logical Pauli product (Sec.~\ref{sec:extractor}).

\subsection{Optimization of the basis}\label{sec:optimization}
Theorem~1 yields a translation-covariant canonical basis for every partition that satisfies the rank condition, but the seed weights differ substantially between different partitions. 
We describe the objective, the structure-preserving freedoms, every stage of the search, and the results. 

\runin{Objective}
A gadget measuring $\XX_{(i,t)}$ places one vertex on every qubit of the target support and one or more skeleton edges on every old check that anticommutes locally with the target (Sec.~\ref{sec:library}), so its ancilla and check counts grow in proportion to the weight of the target; the seed gadgets use roughly three ancilla qubits per qubit of the target (Tab.~\ref{tab:registers}). The $79$ representatives of a register are translates of each other and have equal weights, so the weights of all $1{,}422$ moving representatives are determined by the $36$ seeds. We take
\begin{equation}
W=\sum_{i=0}^{17}\bigl(\wt(x_i)+\wt(z_i)\bigr)
\label{eq:objective}
\end{equation}
as the primary objective and $\max_i\max(\wt x_i,\wt z_i)$ as the secondary objective; the $Y$-type weights $\wt(x_i\lor z_i)$ decrease accordingly. The actual ancilla counts of compiled gadgets serve as a cross-check and for a final refinement.

\runin{Structure-preserving freedoms}
Three kinds of freedom preserve both canonicity and translation covariance.

\emph{Fibre partitions and affine solutions.} There are $\binom{36}{9}\binom{27}{9}\approx4.4\times10^{14}$ partitions $(P_X,P_Z,F)$, of which roughly one fifth satisfy the rank condition. For every valid partition, the seeds are fixed by Theorem~1 and the selection rule of Lemma~\ref{lem:affine}.

\emph{Paired shears.} We regard the $18$ $X$-type and the $18$ $Z$-type seeds as vectors in $\Rq^{18}$. For $p(D)=\sum_tp_tD^t\in\Rq$ we write $p(T)x=\sum_tp_tT^tx$ and $\bar p(D)=p(D^{-1})$, and for a matrix $U$ over $\Rq$ we write $U^\dagger=\bar U^{\mathsf T}$. The pairing polynomials
\begin{equation}
G_{ij}(D)=\sum_{t\in\mathbb Z_q}\bigl(x_i\cdot T^tz_j\bigr)D^t\in\Rq
\end{equation}
encode all $(18q)^2$ pairings of the moving sector, because $T^sx_i\cdot T^{s'}z_j=x_i\cdot T^{s'-s}z_j$; the moving sector is canonical if and only if $G=I_{18}$.
\begin{proposition}\label{prop:shear}
Let $U,V\in\Rq^{18\times18}$, $x_i'=\sum_kU_{ik}(T)x_k$ and $z_j'=\sum_lV_{jl}(T)z_l$. Then $G'=UGV^\dagger$. In particular, if $G=I$ and $U$ is invertible, the choice $V=(U^\dagger)^{-1}$ yields a canonical and translation-covariant moving sector.
\end{proposition}
\begin{proof}
Writing $a\in U_{ik}$ for the exponents of the monomials of $U_{ik}$, we have $x_i'\cdot T^tz_j'=\sum_{k,l}\sum_{a\in U_{ik}}\sum_{b\in V_{jl}}T^ax_k\cdot T^{t+b}z_l=\sum_{k,l,a,b}x_k\cdot T^{t+b-a}z_l$. Multiplying by $D^t$, summing over $t$ and substituting $t'=t+b-a$ gives $G'_{ij}=\sum_{k,l}\sum_{a,b}D^{a-b}G_{kl}=\sum_{k,l}U_{ik}G_{kl}\bar V_{jl}$, that is, $G'=UGV^\dagger$. For $G=I$ one obtains $G'=UV^\dagger$, which equals $I$ if and only if $V=(U^\dagger)^{-1}$. The new representatives are again translates of seeds, so translation covariance is preserved.
\end{proof}
The simplest invertible transformations are elementary shears $U=I+pE_{ij}$ with $i\neq j$. Since $E_{ji}^2=0$, one has $(U^\dagger)^{-1}=(I+\bar pE_{ji})^{-1}=I+\bar pE_{ji}$, and the elementary shear acts as
\begin{equation}
x_i\leftarrow x_i+\sum_{t\in p}T^tx_j,\qquad z_j\leftarrow z_j+\sum_{t\in p}T^{-t}z_i ,
\label{eq:elementary}
\end{equation}
leaving all other seeds unchanged; for a monomial $p=D^t$ this is Eq.~(4) in the main text. The shear preserves all $1{,}422^2$ pairings at once: adding translates of $x_j$ to $x_i$ creates anticommutation with the corresponding translates of $z_j$, which is cancelled exactly by adding the inverse translates of $z_i$ to $z_j$. Since $\wt(x_i+T^tx_j)=\wt(x_i)+\wt(x_j)-2|x_i\cap T^tx_j|$, and similarly for the $Z$ side, a monomial shear changes the objective by
\begin{equation}
\Delta=\wt(x_j)+\wt(z_i)-2\bigl(|x_i\cap T^tx_j|+|z_j\cap T^{-t}z_i|\bigr).
\label{eq:sheardelta}
\end{equation}
A shear is favourable when the $X$-type seed of register $i$ overlaps strongly with a translate of the seed of register $j$, and the $Z$-type seeds overlap correspondingly. This situation is common for bases constructed from a fibre partition: the correction parts of all $X$-type seeds are supported on the same nine fibres of $P_Z$, where the seeds of different registers often share long identical segments, which the shears cancel against each other. A sequence of shears composes to $U_X=E_K\cdots E_1$ and $U_Z=(U_X^\dagger)^{-1}$.

\emph{Stabilizer dressing.} Replacing a seed by $x_i+aH_X$ or $z_i+bH_Z$ leaves its logical class and therefore all pairings unchanged; since the stabilizer group is $T$-invariant, dressing commutes with translation and preserves covariance. Flipping a single check $s$ changes the weight of $x$ by $12-2|s\cap x|$, so only checks that overlap $x$ in at least six qubits can avoid increasing the weight. 

\runin{Search stages}
The search consists of five stages, each driven by a fixed random seed. The first two optimize the pivot assignment of the standard form, whereas the shears and the dressing step act on the representatives themselves, beyond the standard form.
\begin{enumerate}
\item \emph{Random partitions.} A pseudorandom generator produces $1{,}200$ random permutations of the $36$ fibres; the first nine fibres form $P_X$, the next nine $P_Z$ and the remaining $18$ the free set $F$. The $12$ lightest valid partitions are retained. The first valid partition is the baseline $B_0$.
\item \emph{Simulated annealing of the partition.} A local search over partitions performs four restarts, the $r$-th starting from the $r$-th partition of the current pool, with $1{,}000$ steps each. A step selects two of the three sets at random and exchanges one randomly chosen fibre between them. Partitions that occurred before, identified by $(P_X,P_Z)$, are skipped, and invalid partitions are rejected. A valid proposal $C'$ is accepted with the Metropolis probability $\min(1,e^{-\delta/\tau})$ for $\delta=W(C')-W(C)$ at temperature $\tau=55(1-s/1000)+1$ in step $s$, and every new partition lighter than the heaviest member of the pool enters the pool, which retains the $24$ lightest distinct partitions.
\item \emph{Greedy shears.} For every partition of the pool, the monomial shear that minimizes $\wt(x_i')+\wt(z_j')-\wt(x_i)-\wt(z_j)$ is applied repeatedly, as long as it strictly decreases $W$, for at most $60$ steps. New representatives may not exceed weight $126$, which prevents the creation of individual heavy seeds in exchange for a lower total weight. Candidates are compared by the key $\bigl(\Delta,\max(\wt a,\wt b),(\wt a)^2+(\wt b)^2\bigr)$ for $a=x_i+T^tx_j$ and $b=z_j+T^{-t}z_i$. Each step compares $18\times17\times79=24{,}174$ candidates.
\item \emph{Selection.} The candidate with the smallest $(W,\max)$ after the shears is selected. To confirm that seed weight is a good proxy for gadget cost, the $36$ $X$- and $Z$-type seed gadgets of every candidate are compiled with the seed compiler (Sec.~\ref{sec:library}), $864$ gadgets in total, and their total and maximum ancilla counts are recorded (Tab.~\ref{tab:shortlist}).
\item \emph{Gadget-aware dressing.} Starting from every seed, a breadth-first search enumerates all representatives that are reachable by single-check moves that never increase the weight, up to $128$ states. If this neutral basin contains more than one representative, a gadget is compiled for each of them, and the representative with the fewest ancilla qubits, subject to merged check weights of at most $14$, is kept.
\end{enumerate}

\runin{Results}
Of the $1{,}200$ random permutations, $229$ are valid. The first valid permutation, number $16$ of the sequence, defines $B_0$, with $W=11{,}580$ ($X$ side $6{,}088$, individual weights $324$--$358$; $Z$ side $5{,}492$, individual weights $286$--$322$); 
the lightest random partition has $W=4{,}840$. Simulated annealing proposes $3{,}839$ distinct partitions, of which $2{,}032$ are valid, and lowers the lightest partition to $W=4{,}038$; the final pool of $24$ partitions spans $W=4{,}038$--$4{,}070$ (Tab.~\ref{tab:shortlist}). After the greedy shears, these candidates have $W=3{,}740$--$3{,}888$. The lightest is candidate $10$, whose weight decreases from $4{,}060$ to $3{,}740$ through $17$ shears (Tab.~\ref{tab:shears}). Its $36$ seed gadgets use $11{,}584$ ancilla qubits in total, the third smallest total among the $24$ candidates and $0.3\%$ more than the smallest, $11{,}547$ for candidate $1$ (Tab.~\ref{tab:shortlist}). We keep the selection by seed weight, which is exact and independent of the random choices of the compiler. Notably, the lightest partition before the shears ($W=4{,}038$) is not the best after them: whether shears reduce the weight depends on whether the seeds of different registers cancel large parts of each other after translation. Applying the shears to a pool of candidates rather than to a single optimal partition is thus essential.

The selected partition is
\begin{gather}
P_X=\{3,4,5,18,20,24,26,27,35\},\qquad P_Z=\{1,6,7,10,12,15,16,19,31\},\notag\\
F=\{0,2,8,9,11,13,14,17,21,22,23,25,28,29,30,32,33,34\},
\end{gather}
and register $i$ is carried by the free fibre $F[i]$, with $F$ in ascending order. Before the shears, the two sides have total weights $2{,}062$ ($X$, individual weights $106$--$122$) and $1{,}998$ ($Z$, $96$--$124$); after the $17$ shears they have $1{,}896$ and $1{,}844$. The final weights $w_X$, $w_Z$ and $w_Y$ lie in the ranges $94$--$120$, $62$--$120$ and $169$--$231$, respectively (Tab.~\ref{tab:registers}). The dressing stage leaves all $36$ seeds unchanged.

The final basis therefore differs from the standard form only through the shears. The supports of conjugate representatives $\XX_{(i,t)}$ and $\ZZ_{(i,t)}$ still intersect in exactly one qubit, as in the standard form, whereas $3{,}160$ of the $2{,}020{,}662$ nonconjugate pairs of moving representatives, all of which are disjoint in the standard form, now intersect in two qubits.

\begin{datatable}
\caption{\textbf{Fibre partitions.} The $24$ partitions retained after refinement, ordered by their total seed weight after the greedy shears: index, total seed weight of the partition, number of accepted shears, total and maximum seed weight after the shears, and total and largest ancilla data qubit count of the $36$ $X$ and $Z$ seed gadgets compiled for the sheared seeds. The selected candidate is shown in bold.}
\label{tab:shortlist}
\begin{ruledtabular}
\begin{tabular}{rrrrrrr}
candidate & partition & shears & sheared & max & gadget sum & gadget max\\\hline
\textbf{10} & 4{,}060 & 17 & 3{,}740 & 120 & 11{,}584 & 381\\
3 & 4{,}052 & 15 & 3{,}766 & 122 & 11{,}663 & 372\\
0 & 4{,}038 & 17 & 3{,}778 & 122 & 11{,}557 & 376\\
1 & 4{,}040 & 18 & 3{,}790 & 118 & 11{,}547 & 367\\
4 & 4{,}054 & 16 & 3{,}792 & 116 & 11{,}613 & 362\\
5 & 4{,}056 & 17 & 3{,}800 & 120 & 11{,}631 & 378\\
20 & 4{,}068 & 18 & 3{,}806 & 122 & 11{,}697 & 382\\
9 & 4{,}060 & 18 & 3{,}812 & 124 & 11{,}699 & 381\\
7 & 4{,}058 & 18 & 3{,}816 & 126 & 11{,}677 & 395\\
17 & 4{,}066 & 24 & 3{,}822 & 122 & 11{,}747 & 378\\
23 & 4{,}070 & 17 & 3{,}822 & 122 & 11{,}692 & 380\\
8 & 4{,}060 & 19 & 3{,}822 & 126 & 11{,}699 & 389\\
2 & 4{,}048 & 13 & 3{,}824 & 124 & 11{,}750 & 386\\
6 & 4{,}056 & 18 & 3{,}826 & 116 & 11{,}745 & 358\\
15 & 4{,}066 & 14 & 3{,}826 & 124 & 11{,}750 & 388\\
18 & 4{,}066 & 16 & 3{,}830 & 122 & 11{,}750 & 395\\
11 & 4{,}064 & 16 & 3{,}838 & 120 & 11{,}736 & 369\\
12 & 4{,}064 & 18 & 3{,}846 & 116 & 11{,}788 & 363\\
21 & 4{,}068 & 15 & 3{,}846 & 118 & 11{,}845 & 375\\
16 & 4{,}066 & 14 & 3{,}858 & 120 & 11{,}898 & 378\\
14 & 4{,}066 & 14 & 3{,}874 & 124 & 11{,}989 & 386\\
22 & 4{,}068 & 17 & 3{,}874 & 124 & 12{,}018 & 388\\
13 & 4{,}064 & 16 & 3{,}876 & 122 & 11{,}956 & 375\\
19 & 4{,}068 & 18 & 3{,}888 & 124 & 11{,}992 & 386\\
\end{tabular}
\end{ruledtabular}
\end{datatable}

\begin{datatable}
\caption{\textbf{Accepted shears.} The $17$ accepted shears in chronological order: step, registers $(i,j)$, shift $t$ of the move $x_i\leftarrow x_i+T^tx_j$, $z_j\leftarrow z_j+T^{-t}z_i$, and the change $\Delta$ of $\wt(x_i)+\wt(z_j)$.}
\label{tab:shears}
\begin{ruledtabular}
\begin{tabular}{rrrrr@{\hspace{2.2em}}rrrrr}
step & $i$ & $j$ & $t$ & $\Delta$ & step & $i$ & $j$ & $t$ & $\Delta$\\\hline
1 & 7 & 3 & 10 & -26 & 10 & 11 & 6 & 48 & -24\\
2 & 4 & 1 & 46 & -26 & 11 & 1 & 9 & 3 & -12\\
3 & 1 & 12 & 5 & -38 & 12 & 12 & 10 & 43 & -10\\
4 & 10 & 17 & 47 & -24 & 13 & 14 & 15 & 16 & -8\\
5 & 7 & 4 & 18 & -22 & 14 & 14 & 0 & 62 & -12\\
6 & 16 & 7 & 0 & -36 & 15 & 0 & 5 & 20 & -10\\
7 & 8 & 15 & 8 & -22 & 16 & 13 & 0 & 47 & -8\\
8 & 9 & 5 & 56 & -20 & 17 & 7 & 1 & 7 & -4\\
9 & 11 & 2 & 44 & -18 &  &  &  &  & \\
\end{tabular}
\end{ruledtabular}
\end{datatable}
\begin{datatable}
\caption{\textbf{Registers of the canonical basis.} For each register $i$: the free fibre (tile, row) carrying its seeds, the seed weights of the selected partition before the shears, the final weights $w_X$, $w_Z$ and $w_Y=\wt(x\lor z)$, and the ancilla counts of the three seed gadgets.}
\label{tab:registers}
\begin{ruledtabular}
\begin{tabular}{rlrrrrrrrr}
$i$ & fibre & $w_X^{\rm part}$ & $w_Z^{\rm part}$ & $w_X$ & $w_Z$ & $w_Y$ & $n_a(\Sg_X)$ & $n_a(\Sg_Z)$ & $n_a(\Sg_Y)$\\\hline
0 & 0 (L0,0) & 120 & 112 & 102 & 116 & 217 & 311 & 367 & 704\\
1 & 2 (L0,2) & 120 & 116 & 110 & 110 & 219 & 339 & 336 & 701\\
2 & 8 (L2,2) & 108 & 110 & 108 & 94 & 201 & 334 & 277 & 654\\
3 & 9 (L3,0) & 112 & 118 & 112 & 96 & 207 & 344 & 300 & 666\\
4 & 11 (L3,2) & 118 & 110 & 98 & 86 & 183 & 298 & 261 & 586\\
5 & 13 (L4,1) & 108 & 114 & 108 & 116 & 223 & 346 & 367 & 733\\
6 & 14 (L4,2) & 110 & 114 & 110 & 110 & 219 & 341 & 349 & 720\\
7 & 17 (L5,2) & 114 & 96 & 108 & 62 & 169 & 339 & 179 & 547\\
8 & 21 (R1,0) & 114 & 110 & 96 & 110 & 205 & 284 & 339 & 662\\
9 & 22 (R1,1) & 118 & 116 & 104 & 94 & 197 & 322 & 293 & 627\\
10 & 23 (R1,2) & 118 & 112 & 100 & 120 & 219 & 310 & 373 & 707\\
11 & 25 (R2,1) & 116 & 102 & 94 & 102 & 195 & 295 & 323 & 637\\
12 & 28 (R3,1) & 112 & 110 & 94 & 92 & 185 & 296 & 288 & 589\\
13 & 29 (R3,2) & 106 & 114 & 96 & 114 & 209 & 305 & 353 & 671\\
14 & 30 (R4,0) & 120 & 104 & 110 & 104 & 213 & 337 & 316 & 681\\
15 & 32 (R4,2) & 112 & 116 & 112 & 100 & 211 & 348 & 303 & 683\\
16 & 33 (R5,0) & 122 & 100 & 120 & 100 & 219 & 377 & 311 & 714\\
17 & 34 (R5,1) & 114 & 124 & 114 & 118 & 231 & 354 & 358 & 752\\
\hline sum & & 2{,}062 & 1{,}998 & 1{,}896 & 1{,}844 & 3{,}722 & 5{,}880 & 5{,}693 & 12{,}034\\
\end{tabular}
\end{ruledtabular}
\end{datatable}
\section{Reusable gadgets and matching bridges}\label{sec:gadgets}
This section covers the constructions behind the basic instruction set: the general form of the merged codes used throughout this work and their distance (Sec.~\ref{sec:surgery}), the reusable gadget library (Sec.~\ref{sec:library}), and the matching bridges that join two library gadgets to measure a two-logical-qubit product (Sec.~\ref{sec:bridge}).

\subsection{Graph surgery and the distance of merged codes}\label{sec:surgery}
We give the general form of the merged codes of graph surgery~\cite{Cohen2022Lowoverhead,Cross2024Improved,Williamson2026Lowoverhead,Ide2025Faulttolerant,Bhardwaj2026Highrate} used throughout this work, characterize the measured operators and the number of encoded logical qubits. Theorem~2 of the main text is the distance bound of refs.~\cite{Cross2024Improved,Williamson2026Lowoverhead,Bhardwaj2026Highrate}; for completeness, we prove it for general Pauli targets, direct sums and shortcut hyperedges.

\runin{Merged codes for general Pauli targets}
Let the input code, possibly a direct sum of code blocks, act on $n$ data qubits, and let its stabilizer group $\mathcal S$ be generated by the old checks $S_s$. The target is a Hermitian logical Pauli operator $\PP$ with physical representative $P=\bigotimes_qP_q$ given by Eq.~\eqref{eq:sign}.
\begin{definition}\label{def:system}
A \emph{surgery system} consists of
\begin{enumerate}
\item vertices $v\in V$, each attached to at most one data qubit $q(v)$ and carrying a local factor $F_v\in\{X,Y,Z\}$ on this qubit, with $F_v=I$ for unattached vertices, such that all vertex attachments on the same data qubit carry the same factor;
\item ancilla qubits $e\in E$, the edges of a graph or hypergraph on $V$; ordinary edges join two vertices and hyperedges join an even number of vertices, and $\Gamma\in\F^{E\times V}$ denotes the incidence matrix;
\item for every old check $S_s$, the set $K_s$ of attached vertices whose local factor anticommutes with $S_s$, and an edge set $J_s$ whose boundary is $\Gamma^{\mathsf T}\one_{J_s}=\one_{K_s}$, which exists because $K_s$ meets every connected component in an even number of vertices whenever $S_s$ commutes with the product of the factors of each component; the check $S_s$ is deformed to $S_sZ(J_s)$;
\item a complete basis of the cycle space $\ker\Gamma^{\mathsf T}$, the edge sets with vanishing boundary, with one check $Z(\gamma)$ per basis element $\gamma$.
\end{enumerate}
\end{definition}
A gadget for a single target $\PP$ attaches one vertex to every qubit of the support of its physical representative $P$, with $F_v=P_{q(v)}$, so that $\prod_vF_v=P$. A system of several gadgets, such as a bridged pair or a parallel batch, carries the attachments of all of them, and a data qubit may then receive several attachments. The merged code is generated by
\begin{equation}
S_sZ(J_s)\ \text{(all old checks)},\qquad G_v=F_v\otimes X(\partial v)\ (v\in V),\qquad Z(\gamma)\ \text{($\gamma$ in the cycle basis)},
\label{eq:mergedsi}
\end{equation}
where $\partial v$ denotes the set of edges incident to $v$.

\begin{lemma}\label{lem:mergedcommute}
The generators of Eq.~\eqref{eq:mergedsi} commute pairwise, and the product of all vertex checks equals $\prod_vF_v$, which is $P$ for a single target, acting trivially on the ancilla qubits.
\end{lemma}
\begin{proof}

Two vertex checks carry equal local factors on a shared data qubit, by item~1 of Definition~\ref{def:system}, or factors on different qubits, and $X$-type operators on the ancillas, so they commute. A vertex check and a cycle check overlap on the ancillas in $|\partial v\cap\gamma|$ qubits, whose parity is the boundary of $\gamma$ at $v$ and hence zero. A vertex check and a deformed check anticommute on the data qubits exactly when $v\in K_s$; on the ancillas, $X(\partial v)$ and $Z(J_s)$ overlap with the parity of the boundary of $J_s$ at $v$, which by construction is one exactly when $v\in K_s$, so the two signs cancel. Deformed checks and cycle checks act as $Z$-type operators on the ancillas and as commuting old checks on the data qubits. Finally, every ordinary edge lies in exactly two vertex checks and every hyperedge in an even number of them, so the ancilla parts cancel in the product of all vertex checks, and the data part is $\prod_vF_v$.
\end{proof}
For a pure $X$ target, every generator is purely $X$- or purely $Z$-type, and Eq.~\eqref{eq:mergedsi} is the usual CSS merged code. For a pure $Z$ target, the vertex checks $Z_q\otimes X(\partial v)$ and the deformed $X$-type checks $S_sZ(J_s)$ contain both $X$ and $Z$ factors, so the generators as written are not of CSS form; a Hadamard transformation of every ancilla data qubit maps them to $Z_q\otimes Z(\partial v)$ and $S_sX(J_s)$, and the cycle checks to $X(\gamma)$, which is the usual CSS merged code for a $Z$-type target.  Physically, the Hadamard transformation only exchanges the preparation and measurement bases of the ancilla qubits. For targets with $Y$ factors the merged code is a general stabilizer code. 
Our programs build every merged code in this general form, storing each generator as a pair $(x,z)$ as in Sec.~\ref{sec:conventions}, so that $X$-, $Z$- and $Y$-type targets need no separate treatment.

For $t\in\F^V$ let $A(t)=\prod_{v\in t}F_v$, so that $\wt A(t)\le\wt(t)$. The product of the vertex checks in $t$ is $\prod_{v\in t}G_v=A(t)\otimes X(\Gamma t)$, which belongs to the merged stabilizer group and acts trivially on the ancillas if and only if $\Gamma t=0$. The operators of the input code that the merged code measures therefore form the group
\begin{equation}
\mathcal M=\{A(t):t\in\ker\Gamma\}.
\end{equation}
For a connected graph $\ker\Gamma=\{0,\one\}$ and $\mathcal M=\{1,\prod_vF_v\}$: for a single target, exactly $\PP$ is measured, and for a bridged pair, exactly the product of the two targets.

\begin{lemma}\label{lem:count}
If the graph is connected, the cycle basis is complete and $P$ is not a stabilizer, the merged code encodes $k-1$ logical qubits.
\end{lemma}
\begin{proof}
The merged code has $n+|E|$ qubits, and we compute the rank of its generators class by class. First consider the deformed old checks and the cycle checks. A relation $\prod_{s\in R}S_s=I$ among old checks becomes $\prod_{s\in R}S_sZ(J_s)=Z(\sum_{s\in R}J_s)$ after deformation. The boundary of $\sum_{s\in R}J_s$ is $\sum_{s\in R}\one_{K_s}$, whose value at every vertex is the commutation parity of $\prod_{s\in R}S_s=I$ with the local factor and hence zero; thus $\sum_{s\in R}J_s$ is a cycle in the span of the cycle checks. Modulo the cycle checks, the deformed old checks therefore satisfy exactly the relations of the old checks, and the two classes together have rank $(n-k)+(|E|-|V|+1)$. Next, if a product $\prod_{v\in t}G_v$ lies in the span of the first two classes, comparing the $X$ parts on the ancillas gives $\Gamma t=0$, that is, $t\in\{0,\one\}$; and $t=\one$ would express $P$ as a stabilizer times a $Z$-type operator on the ancillas, contradicting the nontriviality of $P$. The $|V|$ vertex checks therefore contribute rank $|V|$, the total rank is $n-k+|E|+1$, and the number of logical qubits is $n+|E|-(n-k+|E|+1)=k-1$.
\end{proof}

\runin{Distance}
For a connected graph, $\dist(t,\ker\Gamma)=\min(|t|,|V|-|t|)$, and the relative Cheeger constant $h_2(\Gamma)$ reduces to the edge expansion $h(G)=\min_{|S|\le|V|/2}|\partial S|/|S|$.
\begin{restated}{Theorem~2}{Distance of the merged code~\cite{Cross2024Improved,Williamson2026Lowoverhead,Bhardwaj2026Highrate}}
Let $h_2(\Gamma)=\min_{t\notin\ker\Gamma}\wt(\Gamma t)/\dist(t,\ker\Gamma)$. If the cycle basis is complete and $h_2(\Gamma)\ge1$, every nontrivial logical operator of the merged code has weight at least $d$.
\end{restated}
\begin{proof}
Let $E$ be a Pauli operator that commutes with all merged generators but does not belong to the merged stabilizer group, with data part $D$ and with $X$ component $u\in\F^E$ on the ancillas.

\emph{Step 1: moving the ancilla $X$ part to the data qubits.} Since $E$ commutes with every cycle check $Z(\gamma)$, $u$ is orthogonal to every $\gamma\in\ker\Gamma^{\mathsf T}$, so $u\in(\ker\Gamma^{\mathsf T})^\perp=\im\Gamma$. We write $u=\Gamma t$ and choose $t$ of minimum weight in $t+\ker\Gamma$, so that $\wt(t)=\dist(t,\ker\Gamma)$. The operator $E'=E\prod_{v\in t}G_v$ differs from $E$ by a merged stabilizer, acts on the ancillas only through a $Z$-type operator, and has data part $D'=DA(t)$.

\emph{Step 2: $D'$ commutes with the input stabilizers.} Let $Z(w)$ be the ancilla part of $E'$. Commutation of $E'$ with a deformed check $S_sZ(J_s)$, whose ancilla part is also $Z$-type, requires $D'$ to commute with $S_s$. Hence $D'$ lies in the centralizer of the input code.

\emph{Step 3: the trivial case.} Suppose that $D'\in\mathcal S\mathcal M$, that is, $D'=S\,A(c)$ with $S\in\mathcal S$ and $c\in\ker\Gamma$. Multiplying $E'$ by the corresponding deformed old checks and by $\prod_{v\in c}G_v$, which acts trivially on the ancillas, leaves an operator $Z(w')$ supported on the ancillas that commutes with every vertex check, so that $\Gamma^{\mathsf T}w'=0$. Thus $w'$ is a cycle and $Z(w')$ is a product of cycle checks, which would make $E$ a merged stabilizer, a contradiction.

\emph{Step 4: weight bound.} Therefore $D'\notin\mathcal S\mathcal M$; in particular $D'\notin\mathcal S$, so $D'$ is a nontrivial logical operator of the input code and $\wt(D')\ge d$. Since $A(t)$ acts on at most $\wt(t)$ qubits, $\wt(D)\ge\wt(D')-\wt(t)$, and
\begin{equation}
\wt(E)\ge\wt(D)+\wt(u)\ge d-\wt(t)+\wt(\Gamma t)\ge d+(h_2-1)\dist(t,\ker\Gamma)\ge d .\qedhere
\end{equation}
\end{proof}

\runin{Direct sums}
When several gadgets act simultaneously, $\Gamma=\bigoplus_g\Gamma_g$ and $\ker\Gamma=\bigoplus_g\ker\Gamma_g$. For $t=(t_g)_g$, both $\wt(\Gamma t)=\sum_g\wt(\Gamma_gt_g)$ and $\dist(t,\ker\Gamma)=\sum_g\dist(t_g,\ker\Gamma_g)$ are sums over the components, so that $h_2(\Gamma)=\min_gh_2(\Gamma_g)$. The proof above applies: if every component satisfies $h_2\ge1$, the merged code of the direct sum has distance at least $d$, and the measured group is $\{\prod_gP_g^{c_g}\}$, where $P_g=\prod_{v\in V_g}F_v$ is the target of component $g$. This is the origin of the distance guarantee for parallel batches (Sec.~\ref{sec:parallel}).

\runin{Shortcut hyperedges}
When several gadgets deform the same old check, as in parallel batches and full-orbit batches, its weight $12+|J_s|$ may exceed $16$. We then split the deformation edges $J_s$ of such a check into at most $16-12=4$ groups, called bins, keeping the edges of each gadget together, and replace each bin $J$ by a single new ancilla $h$, the shortcut, which acts as a hyperedge on the boundary $K=\Gamma^{\mathsf T}\one_J$ of the bin. The check $S_sZ(J_s)$ becomes $S_s\prod_{\text{bins}}Z_h$, and for each bin we add the check $Z_hZ(J)$, called its relation, which ties the shortcut to the edges it replaces: the product of $S_s\prod_hZ_h$ and all relations is $S_sZ(J_s)$. Each bin must satisfy $|J|+1\le16$ and $|K|+2\le16$.

\begin{lemma}\label{lem:shortcut}
Theorem~2 remains valid after the addition of shortcuts: $\ker\Gamma$ is unchanged, $\wt(\Gamma t)$ does not decrease for any $t$, and the enlarged cycle basis is complete.
\end{lemma}
\begin{proof}
The incidence row of the hyperedge $h$ is $\one_K^{\mathsf T}=(\Gamma^{\mathsf T}\one_J)^{\mathsf T}$, a sum of existing rows; adding such a row leaves $\ker\Gamma$ unchanged and can only increase the weight of $\Gamma t$. The boundary of $Z_h$ is $K$, equal to that of $Z(J)$, so the new deformation satisfies the same boundary condition. The relation $Z_hZ(J)$ has boundary $K+K=0$ and is therefore a cycle. The cycle space grows by one dimension with every shortcut, since one edge is added and $\ker\Gamma$ is unchanged, and the new relation is the only cycle check that contains $h$, so it is independent of the previous cycle basis, which remains complete.
\end{proof}
The cost of a shortcut is the weight $1+|J|$ of its relation and the degree $|K|+2$ of the hyperedge ancilla data qubit, consisting of $|K|$ vertex checks, the deformed check and the relation; the two conditions on each bin, $|J|+1\le16$ and $|K|+2\le16$, keep both within the cap of $16$.

\runin{Spectral certificates for small graphs}
For a connected graph, the Cheeger inequality $h(G)\ge\lambda_2(L)/2$~\cite{Mohar1989Isoperimetric}, where $L$ is the graph Laplacian with parallel edges counted with multiplicity, reduces the condition $h\ge1$ to $\lambda_2(L)>2$, which in fact gives $h>1$. We prove this inequality in integer arithmetic. For a graph on $v$ vertices let
\begin{equation}
B=vL-2vI+3\,\one\one^{\mathsf T},
\end{equation}
where $\one\one^{\mathsf T}$ is the all-ones matrix. The vector $\one$ is a null vector of $L$, so $B\one=(0-2v+3v)\one=v\one$; on the orthogonal complement of $\one$, $\one\one^{\mathsf T}$ vanishes and $B$ has eigenvalues $v(\lambda_i-2)$. Hence $B\succ0$ if and only if $\lambda_2>2$. A certificate is an integer lower-triangular matrix $Q$ with nonzero diagonal. The verifier computes $C=QBQ^{\mathsf T}$ exactly in $64$-bit integer arithmetic, after checking that $v\max|B|\,(\max_i\sum_j|Q_{ij}|)^2<2^{62}$: every entry of $C$, and every partial sum formed while computing it, is bounded by $\max|B|\,(\max_i\sum_j|Q_{ij}|)^2$, and every row sum of $|C|$ by $v$ times this bound, so no step overflows. It then checks that $C$ is symmetric with positive diagonal and strictly diagonally dominant, $C_{ii}>\sum_{j\neq i}|C_{ij}|$. By Gershgorin's theorem $C\succ0$, and since $Q$ is invertible, Sylvester's law of inertia gives $B\succ0$. 
The certificate $Q$ is found with floating-point arithmetic. Since $B\succ0$, it has a Cholesky factorization $B=RR^{\mathsf T}$ with $R$ lower triangular, and $R^{-1}B(R^{-1})^{\mathsf T}=I$. We therefore round $2^{16}R^{-1}$ to integers, or $2^{14}R^{-1},\dots,2^{8}R^{-1}$ if the overflow bound above requires smaller entries, so that $QBQ^{\mathsf T}\approx2^{32}I$ is strongly diagonally dominant. Whether $Q$ is accepted is decided by the integer check alone, so rounding errors cannot produce a false certificate.

\runin{Verification of merged codes}
For every surgery system, the verification assembles the sparse generator matrices of the merged code and checks
\begin{enumerate}
\item commutation of all generators, through the sparse product $S_XS_Z^{\mathsf T}+S_ZS_X^{\mathsf T}=0$ modulo two;
\item the row and column weights, that is, check weights and qubit degrees, against the cap of $16$;
\item the exact rank of the generators over $\F$, separately for the $X$ and $Z$ blocks of CSS generator sets and by symplectic elimination otherwise, which yields $k'=n_{\rm tot}-\rk$ and is required to equal $k-m$;
\item the measured operator: the product of all vertex checks acts trivially on the ancillas and equals the target on the data qubits.
\end{enumerate}
The distance follows from Theorem~2, or Proposition~1 for bridged systems, together with the integer certificate of every gadget.

\subsection{Gadget library and load-aware cycle bases}\label{sec:library}

The single-qubit part of the basic instruction set is realized by a reusable gadget library: one seed gadget per register and Pauli type, whose translates serve the remaining $78$ logical qubits of the register. The four invariant modes are not equipped with gadgets. We describe the gadget compiler, the reuse by translation, the construction of cycle bases, the native gadgets, the bridge ports and the statistics of the library.

\runin{Seed compiler}
A certified gadget for a physical Pauli operator $P(x,z)$ is constructed by the skeleton-and-expansion method~\cite{Cross2024Improved,Bhardwaj2026Highrate} in five steps.
\begin{enumerate}
\item \emph{Vertices.} One vertex is placed on every qubit of the target support $\supp(x\lor z)$, with the local factor $X$, $Y$ or $Z$ of the target on this qubit, so that the number of vertices equals the target weight $w$.
\item \emph{Skeleton edges.} For every old check $S_s$ that anticommutes locally with the target, the vertices of $K_s$, an even set, are shuffled and paired, and every pair is joined by an edge; these edges form $J_s$, whose boundary is $K_s$. For the seeds of the library, $|K_s|$ equals two in almost all cases and four in a few, so every old check is deformed onto one or two ancillas ($10{,}876$ deformations with $|J_s|=1$ and $172$ with $|J_s|=2$ over the $54$ seeds). For $X$- and $Z$-type seeds, every target qubit lies in three checks of the opposite type, so every vertex carries at most three skeleton edges.
\item \emph{Random expansion.} For a target degree $\delta\in\{7,8,9\}$, skipping odd values of $(\delta-3)w$, a random $(\delta-3)$-regular graph is added and the second-smallest Laplacian eigenvalue $\lambda_2$ of the resulting graph is computed; the graph is accepted if $\lambda_2>2.025$ and otherwise resampled, with at most $64$ attempts per $\delta$.
\item \emph{Pruning.} Repeatedly, the Fiedler vector $f$ is computed, and among the remaining expansion edges the $16$ edges with the smallest $(f_u-f_v)^2$ are removed one at a time; an edge is deleted if $\lambda_2$ stays above $2.01$, after which the next round starts, for at most $w$ rounds. Skeleton edges are never removed.
\item \emph{Certificate and cycle basis.} For the final graph, the integer certificate (Sec.~\ref{sec:surgery}) and the load-aware cycle basis (below) are computed, and the gadget is checked exactly: the graph is connected ($\rk\Gamma=|V|-1$), the boundary of every $J_s$ equals $K_s$, every cycle has vanishing boundary and the cycle basis consists of $|E|-|V|+1$ independent elements, and the product of the local factors equals the target.
\end{enumerate}
Random numbers only propose candidates; acceptance is decided by exact checks, and the thresholds $2.025$ and $2.01$ leave a margin for the integer certificate. The weight of a vertex check is its degree plus one for the attached data qubit, so expansion degrees of at most nine keep vertex checks at weight around ten, and pruning further reduces the number of ancilla qubits.

\runin{Reuse by translation}
Translating a gadget by $T^t$ replaces every attached data qubit by its image under $T^t$ and every deformed check by its image, while the graph, the cycle basis and the certificate are unchanged.
\begin{proposition}\label{prop:translate}
If $\Sg$ is a certified gadget measuring $P$, then $T^t\Sg$ is a certified gadget measuring $T^tP$, and the two merged codes are isomorphic.
\end{proposition}
\begin{proof}
$T^t$ permutes the data qubits and maps checks to checks (Sec.~\ref{sec:conventions}). Extending it to the merged code by leaving the ancillas fixed maps the deformed check $S_sZ(J_s)$ to $S_{T^ts}Z(J_s)$, the vertex check of $v$ to the vertex check attached to $T^tq(v)$, and every cycle check to itself. This is a weight-preserving isomorphism between the two merged codes, so commutation, rank, weights, degrees and distance coincide, and the vertex product of $T^t\Sg$ is $T^tP$.
\end{proof}
Hence one seed per register and Pauli type suffices: the seed $\Sg_{X_i}$ translated by $t$ measures $\XX_{(i,t)}$.

\runin{Load-aware cycle bases}
The degree of an ancilla edge $e$ is the number of checks acting on it: its two vertex checks, the old checks deformed onto it, and every cycle check that contains it. The first two contributions are fixed by the graph and the deformations, whereas the third depends on the choice of cycle basis. The common fundamental basis of a spanning tree takes, for every non-tree edge, the cycle closed by the tree path, so the edges near the root are traversed by many cycles. For the graphs of our library, fundamental bases of breadth-first trees produce ancilla degrees of up to $141$ for the $X$- and $Z$-type seeds and the native gadgets and up to $256$ for the $Y$-type seeds, far above the cap of $16$.

Since any invertible linear transformation of the cycle checks leaves the merged stabilizer group unchanged, the cycle basis can be chosen freely, and we construct it directly from short, load-balanced cycles.
\begin{enumerate}
\item Initialize the load $\ell_e$ of every edge with the contributions of its vertex checks and deformed checks, and let $\mathcal B=\emptyset$.
\item For at most six passes, visit the edges $e=(u,v)$ in random order. Skip $e$ if $\ell_e+1>16$. Otherwise, find by Dijkstra's algorithm the $u$--$v$ path $\pi$ of at most $15$ edges in $G-e$ that minimizes $\sum_{e'\in\pi}(\ell_{e'}+1)^4$ and avoids edges with $\ell_{e'}+1>16$. If $\pi$ exists and the cycle $\pi+e$ is linearly independent of $\mathcal B$, as decided by exact elimination, add it to $\mathcal B$ and increase the load of its edges by one. Stop as soon as $|\mathcal B|=|E|-|V|+1$.
\item If necessary, complete $\mathcal B$ by fundamental cycles.
\end{enumerate}
The maximum ancilla data qubit degree of all $60$ gadgets, namely the $54$ seeds of the library and the six native gadgets, decreases to at most $9$, with $8$ for the $X$- and $Z$-type seeds, $9$ for the $Y$-type seeds and $7$ for the native gadgets, and the longest cycle check has weight $10$ (Extended Data Fig.~2c in the main text). The graphs, certificates and merged stabilizer groups are unchanged; only the generating set of the cycle space differs. 

\runin{Native gadgets}
The gadgets of the six native operator families (Sec.~\ref{sec:native}) are obtained from two compiled templates, for $\mathsf P^X_{5,0}$  and for $\mathsf P^Z_{4,0}$. The skeleton of a native operator is highly regular: every anticommuting check of the opposite type meets its support in exactly two qubits, which yields a skeleton graph with $36$ vertices and $54$ edges, all of degree three. These skeleton graphs fall into two isomorphism classes, one formed by the four families with $i=3$ and $i=5$ and one by the two families with $i=4$, so that two templates suffice. 
For every family, the VF2 algorithm~\cite{Cordella2004Subgraph} finds an isomorphism from the skeleton of a template to that of the family; the template's edges, cycle basis and certificate are kept, and only the attachments and local factors of the vertices and the indices of the deformed checks are relabelled. The graph itself is unchanged, and the isomorphism of the skeletons ensures that every skeleton edge of the template corresponds to an anticommuting check of the new family. The four families with $i=3$ and $i=5$ use the first template, and the two with $i=4$ use the second. The resulting gadgets have $97$ or $101$ ancilla data qubits, respectively.

\runin{Bridge ports}
Every seed gadget carries a fixed \emph{port}, a simple path $u_1,\dots,u_{18}$ of $18$ vertices used by the bridges of Sec.~\ref{sec:bridge}. Ports are found by depth-first search, which requires the current degree of every path edge to be at most $15$ and the check weight of every path vertex to be at most $15$, so that both remain within $16$ after a bridge adds one edge to every port vertex and one square cycle to every path edge; the search prefers edges with low load and vertices with low weight.

Tab.~\ref{tab:library} summarizes the $54$ seeds of the library together with the six native gadgets, and Tab.~\ref{tab:registers} lists the seeds of every register. All merged codes of single-qubit instructions satisfy $\wmax\le14$ and $\Dmax\le9$.

\begin{table}[tbp]
\caption{\textbf{Gadget library and native gadgets.} The merged code of every gadget was assembled and verified exactly. $n_a$: ancilla data qubits; $\wmax$ and $\Dmax$: maximum check weight and qubit degree of the merged code.}
\label{tab:library}
\begin{ruledtabular}
\begin{tabular}{lrrrrr}
Gadgets & number & target weight & $n_a$ & $\wmax$ & $\Dmax$\\\hline
$X$-type seeds & 18 & 94--120 & 284--377 & 14 & 8\\
$Z$-type seeds & 18 & 62--120 & 179--373 & 14 & 8\\
$Y$-type seeds & 18 & 169--231 & 547--752 & 14 & 9\\
native families ($\mathsf P^X_i$, $\mathsf P^Z_i$) & 6 & 36 & 97--101 & 13 & 7\\
\end{tabular}
\end{ruledtabular}
\end{table}

\subsection{Matching bridges and two-logical-qubit measurements}\label{sec:bridge}
The two-logical-qubit measurements $\XX_j\XX_k$ and $\ZZ_j\ZZ_k$ form the largest part of the basic instruction set. Joining two certified gadgets by a matching bridge of at least $d$ edges measures the product of their targets without reducing the distance, so that two-logical-qubit measurements are composed from library gadgets and a bridge of fixed shape. This is a special case of the adapter theorem of ref.~\cite{Swaroop2024Universal} (Lemma~3 and Theorem~3 there); for completeness, we give a self-contained proof (Proposition~1 of the main text).

\runin{Ports and matching bridges}
A \emph{port} of a gadget $\Sg$ is a simple path $u_1,\dots,u_b$ of $b$ distinct vertices, where consecutive vertices are joined by a graph edge $c_i=(u_i,u_{i+1})$. Given a port $u$ of $\Sg_1$ and a port $w$ of $\Sg_2$, the \emph{matching bridge} consists of the $b$ new edges $\beta_i=(u_i,w_i)$ together with the $b-1$ square cycles
\begin{equation}
\square_i=\beta_i+c^{(2)}_i+\beta_{i+1}+c^{(1)}_i,\qquad i=1,\dots,b-1,
\end{equation}
where $c^{(1)}_i$ and $c^{(2)}_i$ are the port edges of the two gadgets. The joined graph is connected, so it measures only $\PP_1\PP_2$. The $b$ new edges enlarge the cycle space by $b-1$ dimensions, and the restrictions of the squares to the bridge edges, $\beta_i+\beta_{i+1}$, are linearly independent, so the squares complete the cycle basis. Every square check has weight four.

\runin{Compatibility}
Two gadgets can be joined only if their vertex checks commute. A vertex check $G_v=F_v\otimes X(\partial v)$ carries the local factor of its attachment, so two vertex checks attached to a common data qubit with different factors anticommute, and no bridge can remove this conflict. For example, $\XX_1\XX_2$ and $\ZZ_1\ZZ_2$ commute, but their bridged gadgets attach $X$ and $Z$ factors to the qubit where $x_1$ meets $z_1$. We therefore call a set of gadgets \emph{compatible} if all their vertex attachments on every data qubit carry the same local Pauli factor. For single-target gadgets, which attach one vertex to every qubit of the target support, this means that the physical representatives of the targets agree on every shared qubit. For composite gadgets, the condition must be imposed on the attachments rather than on the targets: in the gadget for $\XX_j\XX_k$ (below), a data qubit in the supports of both $x_j$ and $x_k$ carries two $X$ attachments, whose factors cancel in the target $x_j+x_k$, but each of which anticommutes with a $Z$ attachment of another gadget on that qubit. 

\begin{restated}{Proposition~1}{Matching bridges}
Let $\Sg_1$ and $\Sg_2$ be compatible gadgets measuring distinct logical operators $\PP_1$ and $\PP_2$ of a stabilizer code, or of a direct sum of codes, with distance $d$, each satisfying the hypotheses of Theorem~2. If they are joined by a matching bridge of $b\ge d$ edges together with its square cycles, the merged code measures exactly $\PP_1\PP_2$ and has distance at least $d$.
\end{restated}
\begin{proof}
Let $V=V_1\sqcup V_2$, and let the bridge join $u_i\in V_1$ to $w_i\in V_2$ for $i=1,\dots,b$, with pairwise distinct endpoints. By compatibility, the joined system is a surgery system, so its generators commute (Lemma~\ref{lem:mergedcommute}), and $A(t)=\prod_{v\in t}F_v$ acts on at most $|t|$ qubits even when vertices of both gadgets are attached to the same data qubit. The joined graph is connected, so $\ker\Gamma=\{0,\one\}$ and $\mathcal M=\{1,\PP_1\PP_2\}$.

We use the notation of the proof of Theorem~2: a nontrivial logical operator $E$ of the merged code has ancilla $X$ component $u=\Gamma t$, and $D'=DA(t)$ lies in the centralizer of the input code but not in $\mathcal S\mathcal M$. In contrast to that proof, we now allow any representative $t=(t_1,t_2)$ of the coset $t+\ker\Gamma=\{t,t+\one\}$ and examine the two halves separately. Let
\begin{equation}
t_i^*=\begin{cases}t_i,&|t_i|\le|V_i|/2,\\ t_i+\one_{V_i},&\text{otherwise},\end{cases}\qquad c_i=[t_i^*\neq t_i],
\end{equation}
so that $t_i^*$ is the smaller of $t_i$ and its complement in $V_i$. By the expansion of the graphs $G_i$ of the two gadgets, the internal boundaries satisfy
\begin{equation}
\wt(\Gamma_it_i)=\wt(\Gamma_it_i^*)\ge h(G_i)\,|t_i^*|\ge|t_i^*| ,
\label{eq:internal}
\end{equation}
where $\Gamma_i$ is the incidence matrix of $G_i$ itself, whose boundaries are invariant under complementation.

\emph{Case $c_1=c_2$.} Let $D''=D'P_1^{c_1}P_2^{c_2}$. For $c_1=c_2=0$, $D''=D'$; for $c_1=c_2=1$, $D''=D'\PP_1\PP_2$ differs from $D'$ by an element of $\mathcal M$. In both cases $D''$ is a nontrivial logical operator of the input code, so $\wt(D'')\ge d$. On the other hand, $D''=DA(t^*)$ up to a phase, and $A(t^*)$ acts on at most $|t_1^*|+|t_2^*|$ qubits, so that, using Eq.~\eqref{eq:internal},
\begin{equation}
\wt(E)\ge\wt(D)+\wt(u)\ge d-|t_1^*|-|t_2^*|+\wt(\Gamma_1t_1)+\wt(\Gamma_2t_2)\ge d .
\end{equation}

\emph{Case $c_1\neq c_2$}, say $c_1=1$ and $c_2=0$. Then $t_1$ contains the larger part of $V_1$, and $t_2=t_2^*$ the smaller part of $V_2$. A bridge edge $\beta_i$ belongs to $u=\Gamma t$ if and only if exactly one of its endpoints lies in $t$. If $u_i\notin t_1^*$, that is, $u_i\in t_1$, and $w_i\notin t_2$, then exactly one endpoint of $\beta_i$, namely $u_i$, lies in $t$, and $\beta_i$ is counted in $u$. Since the endpoints are pairwise distinct, at most $|t_1^*|+|t_2^*|$ indices $i$ violate this condition, so $u$ contains at least $b-|t_1^*|-|t_2^*|$ bridge edges. Together with the internal boundaries,
\begin{equation}
\wt(E)\ge\wt(u)\ge|t_1^*|+|t_2^*|+\bigl(b-|t_1^*|-|t_2^*|\bigr)=b\ge d .\qedhere
\end{equation}
\end{proof}
Shortcut hyperedges, where they are added to the joined system (Sec.~\ref{sec:surgery}), leave $\ker\Gamma$ unchanged and do not decrease any boundary, so the proof applies unchanged.

The condition $b\ge d$ cannot be relaxed. The product $\prod_{v\in V_1}G_v=P_1\otimes X(\text{bridge})$ is a merged stabilizer, so $X^{\otimes b}$ on the bridge qubits is equivalent, modulo merged stabilizers, to $\PP_1$. It commutes with all merged generators, since every square cycle contains two bridge edges and no other $Z$-type generator acts on the bridge. It is not a merged stabilizer, because the merged stabilizers supported on data qubits belong to $\mathcal S\mathcal M$, whereas $\PP_1\notin\mathcal S\mathcal M$ for distinct nontrivial $\PP_1$ and $\PP_2$. It is therefore a nontrivial logical operator of weight $b$, and the merged distance is at most $b$. For $b=d$, which is the case for every bridge in this work, the merged distance is therefore exactly $d$. In general the merged distance need not be bounded by $d$, since minimum-weight logical operators of the input code may anticommute locally with the attached factors and then acquire ancilla support in the merged code. 
The proof never uses the Pauli type of either target, so the two gadgets may measure operators of different types, for example $\XX$ on one code block and $\ZZ$ on another, as long as they are compatible.

\runin{Two-logical-qubit measurements within a module}
Every seed gadget carries a fixed $18$-vertex port (Sec.~\ref{sec:library}). The gadget measuring $\XX_j\XX_k$ for register qubits $j\neq k$ consists of the translated seed gadgets of $j$ and $k$ and the $18$-edge bridge with $17$ squares between their fixed ports; it involves no optimization specific to the pair. Both gadgets attach only $X$ factors (only $Z$ factors for $\ZZ_j\ZZ_k$), so they are compatible.

Every seed deforms each old check onto at most two ancillas, so a check deformed by two seeds has weight at most $12+2+2=16$, and no shortcut is needed. A data qubit carries at most two vertex attachments, one from each gadget, and therefore has degree at most $6+2=8$. Ports are chosen on edges and vertices with slack, so that the bridge, which raises the weight of every port vertex and the degree of every port edge by one, keeps them within $16$.

Since $T$ is an automorphism of the code and of the basis, and seeds and ports translate with $T$, the merged code for the pair $\{(a,s),(b,s+\delta)\}$ is isomorphic to that for $\{(a,0),(b,\delta)\}$. It therefore suffices to verify one representative of every translation class. For registers $a<b$, the representatives are $\{(a,0),(b,\delta)\}$ with $\delta\in\mathbb Z_{79}$, that is, $\binom{18}{2}\times79=12{,}087$ classes; for $a=b$, the pairs $\{(a,0),(a,\delta)\}$ and $\{(a,0),(a,-\delta)\}$ are equivalent (by a translation by $-\delta$), so $\delta=1,\dots,39$, that is, $18\times39=702$ classes. This gives $12{,}789$ classes per Pauli type and $25{,}578$ classes in total, which cover all $2\binom{1422}{2}=2{,}020{,}662$ products of register qubits.

For every class, we verified that the merged codes satisfy $\wmax\le16$ and $\Dmax\le8$ and use $586$--$772$ ancilla data qubits for $\XX\XX$ and $376$--$764$ for $\ZZ\ZZ$ (Extended Data Fig.~2b in the main text). The distance of every class follows from Proposition~1 and the certificates of the two gadgets.

\section{Parallel surgery and native operator families}\label{sec:throughput}
A module stores $1{,}422$ moving logical qubits, so a processor that measures one Pauli product per logical cycle leaves most of them idle. This section describes two ways of measuring many logical operators in a single merged code: direct sums of compatible gadgets, which execute basic instructions in parallel (Sec.~\ref{sec:parallel}), and the native operator families, whose $79$ translates are measured together (Sec.~\ref{sec:native}).

\subsection{Parallel surgery}\label{sec:parallel}
We prove Proposition~2 of the main text, describe how compatible batches are assembled, report the maximal batches and parallel magic-state injection.

\runin{Direct sums}
Let $\Sg_1,\dots,\Sg_m$ be compatible certified gadgets, that is, all their vertex attachments on every data qubit carry the same local Pauli factor (Sec.~\ref{sec:bridge}), measuring logical operators $\PP_1,\dots,\PP_m$ that are linearly independent modulo stabilizers. For a bridged pair, the condition includes both attachments of a data qubit in the supports of its two seeds, although their factors cancel in its target. Placed on a module simultaneously, their incidence matrices form a direct sum, and $\ker\Gamma$ is spanned by the indicator vectors of the components, so the merged code measures exactly $\langle\PP_1,\dots,\PP_m\rangle$ and encodes $k-m$ logical qubits. By the direct-sum form of Theorem~2 (Sec.~\ref{sec:surgery}), applied together with the argument of Proposition~1 for bridged components, its distance is at least $d$. Different gadgets interact only through the input code: an old check deformed by several gadgets accumulates ancillas, and a data qubit attached to several gadgets accumulates vertex checks.

\begin{restated}{Proposition~2}{Compatibility}
A set of certified gadgets forms a certified parallel batch if (i) the gadgets are compatible, that is, all vertex attachments on the same data qubit carry the same local factor, and every data qubit carries at most ten attachments; (ii) every old check $s$ either satisfies $12+|J_s|\le16$ or its deformation edges $J_s$ can be packed by gadget into at most four bins $J\subseteq J_s$, each with boundary $K=\Gamma^{\mathsf T}\one_J$ satisfying $|J|+1\le16$ and $|K|+2\le16$; and (iii) for every vertex, its check weight plus the number of shortcuts acting on it is at most $16$.
\end{restated}

\begin{proof}
By the direct-sum form of Theorem~2 (Sec.~\ref{sec:surgery}), together with Proposition~1 for bridged components and Lemma~\ref{lem:shortcut} for the shortcuts, the merged code measures exactly the span of the targets and has distance at least $d$, as discussed above. Its generators commute by Lemma~\ref{lem:mergedcommute}: condition (i) is the compatibility required there, also for the attachments of a composite gadget whose factors cancel in its target, and the shortcuts keep the boundary condition, because the boundaries of the bins of a check add up to $K_s$. It remains to bound the weights and degrees, class by class. An old check has weight $12+|J_s|\le16$ without shortcuts and $12+(\text{number of bins})\le16$ with shortcuts, by (ii); a relation $Z_hZ(J)$ has weight $|J|+1\le16$. A vertex check has its weight in the merged code of its own gadget plus the number of shortcuts acting on it, which is at most $16$ by (iii). Cycle checks are internal to the gadgets and already satisfy the cap. A data qubit lies in at most six old checks and in one vertex check per attachment, so its degree is at most $6+10=16$ by (i). An ordinary ancilla data qubit keeps the degree it has in the merged code of its own gadget, because a shortcut only moves it from the deformed check into the relation; a shortcut ancilla data qubit has degree $|K|+2\le16$ by (ii).
\end{proof}

\runin{Maximal batches and scheduling}
For every instruction type, we ran a randomized greedy procedure $12$ times. Each run goes through the candidate instructions in a new random order and adds an instruction to the batch whenever it acts only on logical qubits that the batch does not yet use and remains compatible with the batch; we kept the largest of the $12$ resulting batches.
The resulting batches contain $17$ measurements $\XX_j$, $20$ measurements $\ZZ_j$, $17$ measurements $\YY_j$, $37$ mixed $\XX_j$ and $\ZZ_j$ measurements on distinct qubits, and $8$ bridged products $\XX_j\XX_k$ and $9$ bridged products $\ZZ_j\ZZ_k$, selected from $3{,}000$ random pairs (Fig.~3a in the main text). 

To estimate the throughput on a generic workload, sequences of $3{,}000$ instructions, drawn uniformly from the five types $\XX_j$, $\ZZ_j$, $\YY_j$, $\XX_j\XX_k$ and $\ZZ_j\ZZ_k$ with uniformly random register qubits, are list-scheduled into parallel layers: each instruction is placed into the first layer, after the last layer that acts on one of its logical qubits, that can accommodate it compatibly. A layer is one merged code, executed in one logical cycle of $O(d)$ syndrome rounds. Averaged over three independent workloads, a layer contains $19.3$ instructions for this mixed workload, $31.8$ for workloads of single-qubit $\XX_j$ and $\ZZ_j$ measurements, $23.9$ for single-qubit measurements including $\YY_j$, and $15.0$ for bridged products (Fig.~3b in the main text). 
The three benchmark workloads of Sec.~\ref{sec:resources}, drawn independently from the same distribution, yield $19.9$ instructions per layer on average. The parallelism is limited by the accumulation of deformations on old checks, which shortcuts relieve up to four bins per check.

\runin{Parallel magic-state injection}
The injection interfaces (Sec.~\ref{sec:injection}) are $\ZZ$-type seeds on the processor bridged to independent source patches, so they can be packed in the same way. The largest batch found injects $20$ magic states into distinct register qubits in parallel.

\subsection{Native operator families}\label{sec:native}
We derive the support of the native operators, prove Proposition~3 of the main text, locate the native families within the logical space, and describe their full-orbit batches.

\runin{Latent rows and the support of native operators}
Extending the check-tile index of Eq.~(1) in the main text from $i\in\{0,1,2\}$ to $i\in\mathbb Z_6$ defines parent matrices $\widehat H_X$ and $\widehat H_Z$. Their rows with $i=3,4,5$ are called \emph{latent rows}~\cite{Kasai2026Breaking}; they are not checks, and individual latent rows do not, in general, commute with all checks, because the noncommuting pairs $(A_0,B_3)$ and $(A_1,B_2)$ contribute to $\widehat H_XH_Z^{\mathsf T}$ when $i'-i\equiv3\pmod 6$ (see the proof of Proposition~3 below). For a latent block $i$ and a column $c$, summing the three rows $(i,r,c)$, $r=0,1,2$, yields the operators $\mathsf P^X_{i,c}$ and $\mathsf P^Z_{i,c}$ of Eq.~(6) in the main text. To derive their supports, note that the $(i,L_j)$ block of $\widehat H_X$ is $\mathcal A_{j-i}$, so the row $(r,c)$ has its nonzero entry in $L_j$ at $A_{j-i}(r,c)=(\rho(r),c+s_{A_{j-i}})$. Summing over $r$, the row $\rho(r)$ runs through all three rows, which gives the entire column $c+s_{A_{j-i}}$ of $L_j$. The same reasoning applies to the $R$ tiles and, with the inverse blocks and opposite shifts, to the $Z$ type. Native operators thus occupy one entire column of three qubits in each of the twelve data tiles and have weight $36$.

\begin{restated}{Proposition~3}{Native families}
The operators $\mathsf P^X_{i,c}$ and $\mathsf P^Z_{i',c'}$ with $i,i'\in\{3,4,5\}$ and $c,c'\in\mathbb Z_{79}$ are logical operators of the code and commute with each other. For a fixed Pauli type and $i$, the $79$ translates have disjoint supports that cover all data qubits and span a $78$-dimensional space of logical operators; the six families together span $468$ dimensions.
\end{restated}
\begin{proof}
Let $\mathbf u_c$ denote the indicator row vector of column $c$ (all three rows) of a $3\times q$ array. For the permutation matrix $\mathcal M$ of a map $f=(\rho,s)$, with $\mathcal M_{u,u'}=1$ if and only if $u'=f(u)$, the vector $\mathbf u_c\mathcal M$ is the indicator of the coordinates $f(u)$ with $u$ in column $c$, that is, $\mathbf u_{c+s}$, independently of the row map $\rho$. The commutation parities of $\mathsf P^X_{i,c}$ with the $Z$ checks are given by $\mathbf u_c(\widehat H_XH_Z^{\mathsf T})_{i,i'}$. By Eq.~(1) in the main text and $(\mathcal M^{-1})^{\mathsf T}=\mathcal M$ for permutation matrices, the block $(i,i')$ of $\widehat H_XH_Z^{\mathsf T}$ is $\sum_{j\in\mathbb Z_6}\bigl(\mathcal A_{j-i}\mathcal B_{i'-j}+\mathcal B_{j-i}\mathcal A_{i'-j}\bigr)=\sum_{k\in\mathbb Z_6}\bigl(\mathcal A_k\mathcal B_{m-k}+\mathcal B_{m-k}\mathcal A_k\bigr)$ with $m=i'-i$, substituting $k=j-i$ in the first term and $k=i'-j$ in the second, so the parities are
\begin{equation}
\sum_{k\in\mathbb Z_6}\mathbf u_c\bigl(\mathcal A_k\mathcal B_{m-k}+\mathcal B_{m-k}\mathcal A_k\bigr)=\sum_{k\in\mathbb Z_6}\bigl(\mathbf u_{c+s_{A_k}+s_{B_{m-k}}}+\mathbf u_{c+s_{B_{m-k}}+s_{A_k}}\bigr)=0 .
\end{equation}
Row maps have no effect on column sums, so even the noncommuting terms cancel in pairs. Hence $\mathsf P^X_{i,c}$ commutes with all $Z$ checks, and by the same argument $\mathsf P^Z_{i',c'}$ commutes with all $X$ checks. The commutation parity between $\mathsf P^X_{i,c}$ and $\mathsf P^Z_{i',c'}$ is $\mathbf u_c(\widehat H_X\widehat H_Z^{\mathsf T})_{i,i'}\mathbf u_{c'}^{\mathsf T}=0$ by the same cancellation. Translating $c$ moves the occupied column of every tile through all $79$ columns, so the $79$ translates of a family partition the data qubits, and their product is $X^{\otimes n}$ (or $Z^{\otimes n}$), the product of all checks of one check tile and hence a stabilizer; consequently the rank of a family is at most $78$. 

Whether these operators are stabilizers is decided by exact computation over $\F$: we compute the logical content of every operator, its vector of pairings with all $\ZZ_j$ (or $\XX_j$), and eliminate. Each native operator anticommutes with $680$--$742$ of the $\ZZ_j$ (or $\XX_j$), so none of them is a stabilizer, and the ranks are $78$ per family, $234$ per Pauli type and $468$ in total.
\end{proof}

In the canonical basis of this work, the native operators have vanishing coefficients on the invariant modes, and their coefficient polynomials on every register vanish at $D=1$. They therefore lie in the subspace associated with the two factors $\mathbb F_{2^{39}}$ of the decomposition in Eq.~\eqref{eq:ringsplit} and are independent of the orbit sums and the invariant modes. Each native operator acts on $680$--$742$ logical qubits of the basis.

\runin{Full-orbit batches}
Every family has a certified gadget with $97$ or $101$ ancilla data qubits (Sec.~\ref{sec:library}). The direct sum of its $79$ translates measures the entire family: one merged code returns $78$ independent outcomes, and the product of all $79$ outcomes, the value of a stabilizer, serves as a parity check.
Every anticommuting check meets each of six translates in two qubits and is therefore deformed onto one ancilla by each of six gadgets, which would raise its weight to $12+6=18$. A single shortcut bin per check, with $|J|=6$ and $|K|=12$, reduces the weight to $13$. 
All $711$ anticommuting checks receive one shortcut, and the merged codes use $8{,}374$--$8{,}690$ ancilla qubits, satisfy $\wmax\le13$ and $\Dmax\le14$; all were verified exactly (Extended Data Tab.~2 in the main text). The native families are best regarded as a fixed set of $468$ commuting multi-qubit observables that can be measured at a high rate, for instance to prepare or verify large stabilizer states of the stored logical qubits.

\section{The fixed extractor}\label{sec:extractor}
An extractor is a single ancilla system that is attached to the code once and whose connections are selected in software to measure any chosen logical Pauli product~\cite{He2025Extractors,Webster2025Explicit}. It trades workspace for expressiveness and fixed wiring, which suits hardware whose connectivity cannot be reprogrammed quickly. This section describes the construction of our extractor, its modes, the mode-independent bounds on weights and degrees, and the exact Fourier certificate of its expansion.

\subsection{Construction of the hardware}
The fixed graph $G$ is constructed in four steps.
\begin{enumerate}
\item \emph{Tanner graph.} There is one vertex per data qubit and one \emph{hub} per listed check, $|V|=2{,}844+1{,}422=4{,}266$. Every Tanner incidence between hub $s$ and a data vertex in $\supp S_s$ is an edge, $1{,}422\times12=17{,}064$ edges in total; the twelve edges of hub $s$ form its \emph{star}.
\item \emph{Cycle basis of the Tanner graph.} For every pair of an $X$ check $a$ and a $Z$ check $b$ whose supports intersect, necessarily in an even number of qubits $p_0,p_1,\dots$, the four-cycles $\mathrm{hub}_a$--$p_0$--$\mathrm{hub}_b$--$p_\ell$--$\mathrm{hub}_a$ with $\ell\ge1$ are generated, and exact elimination keeps $12{,}781$ independent ones. The cycle rank of the Tanner graph is $17{,}064-4{,}266+1=12{,}799$, and the remaining $18$ dimensions are completed by short fundamental cycles of length six or eight.
\item \emph{Translation-covariant data cycle.} The Tanner edges alone do not provide sufficient expansion, so a translation-covariant cycle is added on the data vertices. The $36$ data fibres are arranged in a random cyclic order $f_0,\dots,f_{35}$, a random shift $s_i$ is chosen for every pair of consecutive fibres, and all edges $\{(f_i,c),(f_{i+1},c+s_i)\}_{c\in\mathbb Z_q}$ of the edge class $(f_i,f_{i+1},s_i)$ are added. Every data vertex thereby receives exactly two new edges, $2{,}844$ in total. For every candidate, $\lambda_2$ of the whole graph is computed through its Fourier decomposition (below), and the first candidate with $\lambda_2>2.10$ is accepted; it has $\lambda_2=2.501$.
\item \emph{Cycles for the new edges.} Every new edge $(u,v)$ requires a cycle through it. In random order, a $u$--$v$ path through Tanner edges is found for every new edge by a layered search of depth at most $15$ with edge cost $(\ell+1)^6-\ell^6$, where $\ell$ is the number of cycles already using the Tanner edge, and with loads capped at $13$. The path closed by the new edge is its cycle check. The resulting $2{,}844$ cycles have length five ($1{,}343$ cycles) or seven ($1{,}501$).
\end{enumerate}
In total, $|E|=17{,}064+2{,}844=19{,}908$, and the cycle basis consists of $12{,}781+18+2{,}844=15{,}643$ cycles of length at most eight, with every edge in at most $7$ cycles. In addition, every old check $S_s$ owns a \emph{shortcut} ancilla $h_s$, so the fixed hardware comprises $19{,}908+1{,}422=21{,}330$ ancilla data qubits.

\subsection{Modes}
For a target with physical representative $P$ (Eq.~\eqref{eq:sign}), the extractor gadget is configured as follows.
\begin{itemize}
\item Data vertex $q$ is attached to data qubit $q$ with local factor $P_q$, or left unattached if $P_q=I$; hubs are unattached.
\item All ordinary edges and cycle checks are kept.
\item For every old check $S_s$, let $K_s(P)$ be the set of data qubits on which $S_s$ anticommutes locally with $P$, an even set. The shortcut $h_s$ becomes a hyperedge on $K_s(P)$, the old check becomes $S_sZ_{h_s}$, and the relation $R_s=Z_{h_s}Z(J_s)$ is added, where $J_s$ consists of the star edges from hub $s$ to $K_s(P)$.
\end{itemize}
The boundary of $J_s$ is $K_s(P)$ together with $|K_s(P)|$ incidences at hub $s$; the latter number is even, so the boundary is exactly $K_s(P)$, equal to that of $h_s$, and $R_s$ is a cycle. The incidence row of a shortcut is a sum of rows of the ordinary incidence matrix $B$, so $\ker\Gamma=\ker B=\{0,\one\}$ and exactly $\PP$ is measured. Every relation $R_s$ is the only cycle check containing $h_s$, so the relations together with the fixed cycle basis form a complete cycle basis (Lemma~\ref{lem:shortcut}). 

The weights and degrees of the merged code can be bounded simultaneously for all targets:
\begin{itemize}
\item a data-vertex check contains at most eight ordinary edges (six Tanner edges and two data-cycle edges), at most six shortcuts (one for each check containing the qubit) and one data factor, so its weight is at most $15$; a hub check has weight $12$;
\item old checks and relations have weight at most $12+1=13$, and cycle checks have weight at most eight;
\item a data qubit lies in six old checks and one vertex check, so its degree is seven; an ordinary edge lies in two vertex checks, at most $7$ cycle checks and, as a star edge, at most one relation, so its degree is at most $10$; a shortcut lies in at most $|K_s|\le12$ vertex checks, its deformed check and its relation, so its degree is at most $14$.
\end{itemize}
Consequently, every logical Pauli target yields a merged code with
\begin{equation}
\wmax\le15,\qquad \Dmax\le14 .
\end{equation}
The merged code has $24{,}174$ data qubits and $22{,}753$ check qubits and encodes $1{,}425$ logical qubits. This was verified exactly for six targets, including a dense random Pauli product on all $1{,}422$ register qubits (Extended Data Tab.~3 in the main text).

\subsection{Exact Fourier certificate}
Adding hyperedges leaves $\ker\Gamma$ unchanged and does not decrease $\wt(\Gamma t)$, so $h_2(\Gamma)\ge h(G)$ for every mode. By Theorem~2, the distance of all $4^{1426}-1$ modes thus reduces to a single property of the fixed graph, $\lambda_2(L_G)>2$.

The graph $G$ is invariant under $T$. Its vertices carry $54$ orbit labels, namely $36$ data fibres and $18$ hub fibres with $q$ vertices each, and every edge belongs to an edge class $(a,b,s)$ consisting of all edges $\{(a,c),(b,c+s)\}$; there are $252$ edge classes, $216$ from the Tanner graph and $36$ from the data cycle. With $\zeta=e^{2\pi i/q}$, the Laplacian $L_G$ decomposes according to the characters of $\mathbb Z_q$ into $79$ Hermitian blocks of size $54\times54$,
\begin{equation}
L_k=D-\sum_{(a,b,s)}\bigl(\zeta^{ks}E_{ab}+\bar\zeta^{ks}E_{ba}\bigr),\qquad k\in\mathbb Z_{79},
\end{equation}
where $D$ is the diagonal degree matrix. Since $L_{q-k}=\overline{L_k}$, it suffices to consider the $40$ blocks $k=0,\dots,39$. The bound $\lambda_2(L_G)>2$ is equivalent to $L_k-2I\succ0$ for $k\neq0$ together with $L_0-2I+\tfrac{3}{54}\one\one^{\mathsf T}\succ0$, where the rank-one term moves the eigenvalue of the constant vector from $-2$ to $1$ and leaves all other eigenvalues unchanged.

\emph{Certificate.} The blocks $L_k$ contain the irrational numbers $\zeta^{ks}$ and therefore cannot be represented exactly by integers. We round them to integers with a rigorously bounded error and form, for every block, an integer matrix $B_k$ that approximates a scaled copy of $L_k-2I$, with the extra term $\tfrac{3}{54}\one\one^{\mathsf T}$ for $k=0$. As for the small graphs (Sec.~\ref{sec:surgery}), the certificate is an integer lower-triangular matrix $Q_k$, now with complex entries, and the verifier checks with exact integer arithmetic that $Q_kB_kQ_k^\dagger$ is diagonally dominant by a margin larger than any change the rounding could cause, so that the block is positive definite also without rounding. All $40$ blocks pass this check by a wide margin, which proves $\lambda_2(L_G)>2$; numerically, $\lambda_2=2.501$ (Fig.~4b in the main text). Every mode of the extractor therefore preserves the code distance.

\section{Magic-state injection, inter-module operations and universality}\label{sec:modular}
This section describes how magic states are injected into a module (Sec.~\ref{sec:injection}) and how modules are joined (Sec.~\ref{sec:modules}), and shows that each instruction set is universal.

\subsection{Magic-state injection and Pauli-based computation}\label{sec:injection}
We describe the magic-state source and its gadget, the injection protocol, and the Pauli-based computation protocol of the fixed extractor.

\runin{Source patches}
Magic states are produced by separate factories, by distillation or cultivation~\cite{Bravyi2005Universal,Gidney2024Magic}, and stored in a distance-$18$ unrotated surface code with $613$ data qubits and $612$ checks, called a \emph{source}. The $\ZZ$ string of the source, a row of $18$ qubits, carries a certified gadget compiled with target degrees $\{6,7,8\}$. The resulting gadget has $18$ vertices, $41$ ancilla qubits and $24$ cycle checks.

\runin{Injection}
The source gadget is bridged by $18$ edges to the gadget that measures $\ZZ_j$ on the processor, so that the merged code measures $\ZZ_j\ZZ_\src$. The processor and the source form a direct sum of codes with distance $18$, and the two gadgets act on disjoint data qubits and are therefore compatible, so by Proposition~1 the merged code has distance at least $18$. The injection protocol prepares $j$ in $\ket+$, measures $\ZZ_j\ZZ_\src$ with outcome $m$, and then measures the source transversally in the $X$ basis with outcome $n$; logical qubit $j$ is then in the state $\ket T$ up to a Pauli correction determined by $(m,n)$~\cite{Litinski2019GameOfSurfaceCodes}. 
The interface, consisting of the $Z$-type seed, the bridge and the source gadget, adds $238$--$432$ ancilla qubits.

\runin{Pauli-based computation}
To apply $e^{-i\pi\PP/8}$ for an arbitrary logical Pauli operator $\PP$, the extractor configured for $\PP$ is bridged to the gadget of a $\ket T$ source, $\PP\otimes\ZZ_\src$ is measured with outcome $m$, and the source is read out transversally in the $X$ basis with outcome $n$.
\begin{proposition}\label{prop:pbc}
Let the input state be $\ket\psi\otimes\ket T$ with $\ket T=(\ket0+e^{i\pi/4}\ket1)/\sqrt2$. After measuring $P\otimes Z$ with outcome $m$ and $X$ on the second qubit with outcome $n$, the state of the first system is $C_{m,n}\,e^{-i\pi P/8}\ket\psi$ up to a global phase, with the byproduct $C_{m,n}=P^ne^{im\pi P/4}$.
\end{proposition}
\begin{proof}
Write $\ket\psi=\ket{\psi_+}+\ket{\psi_-}$ with $P\ket{\psi_\pm}=\pm\ket{\psi_\pm}$, so that $\ket\psi\otimes\ket T\propto\ket{\psi_+}\ket0+e^{i\pi/4}\ket{\psi_+}\ket1+\ket{\psi_-}\ket0+e^{i\pi/4}\ket{\psi_-}\ket1$. For $m=0$, the projection onto $P\otimes Z=+1$ keeps $\ket{\psi_+}\ket0+e^{i\pi/4}\ket{\psi_-}\ket1$, and the subsequent measurement of $X$ with outcome $n$ leaves $\ket{\psi_+}+(-1)^ne^{i\pi/4}\ket{\psi_-}\propto P^ne^{-i\pi P/8}\ket\psi$. For $m=1$, the projection keeps $e^{i\pi/4}\ket{\psi_+}\ket1+\ket{\psi_-}\ket0$, and the measurement of $X$ leaves $(-1)^ne^{i\pi/4}\ket{\psi_+}+\ket{\psi_-}\propto P^ne^{+i\pi P/8}\ket\psi=P^ne^{i\pi P/4}e^{-i\pi P/8}\ket\psi$. In both cases the state is $P^ne^{im\pi P/4}e^{-i\pi P/8}\ket\psi$ up to a global phase.
\end{proof}
The Pauli byproduct $P^n$ is absorbed into the Pauli frame, and the Clifford byproduct $e^{i\pi P/4}$ for $m=1$ is undone by a further PPM or commuted through the subsequent operations of a Pauli-based computation, which changes only the Pauli operators measured later.

\subsection{Inter-module operations and universality}\label{sec:modules}
Distinct Cornucopia code blocks are independent error-correction domains, and a large computer consists of many such modules. We describe the bridge between modules, the logical operations across modules, and the universality of the instruction sets.

\runin{Bridges between extractors}
The direct sum of two codes of distance $d$ has distance $d$, and gadgets on different modules act on disjoint data qubits and are therefore compatible for any pair of targets, so Proposition~1 applies directly to two modules: the extractors of modules $A$ and $B$ are joined by a matching bridge of $18$ edges (Fig.~5a in the main text). Each port consists of $18$ consecutive data vertices along the data cycle of an extractor, so that every bridge square uses two data-cycle edges. The bridge adds $18$ ancilla qubits and $17$ checks of weight four. It raises the check weight of the port vertices by one, to at most $16$, and the degree of each port edge, a data-cycle edge that lies in two vertex checks and in its own cycle check, from three to four, far below the extractor bound of $14$. Hence, for every pair of targets,
\begin{equation}
\wmax\le16,\qquad \Dmax\le14 .
\end{equation}
The merged code of two modules measures exactly $\PP_A\otimes\PP_B$, encodes $2\times1{,}426-1=2{,}851$ logical qubits and has distance at least $18$. Seed gadgets of different modules can also be bridged directly. For example, joining the seed of $\XX_{(3,17)}$ in module $A$ to that of $\XX_{(11,40)}$ in module $B$ by $18$ edges gives an inter-module $\IB$ instruction with $657$ ancilla data qubits.

\runin{Logical operations across modules}
Joint measurements across modules suffice for logic between them.

\emph{Teleportation.} Let logical qubit $k$ of module $B$ be prepared in $\ket+$. Measuring $\ZZ^A_j\ZZ^B_k$ across the modules with outcome $m$ and then $\XX^A_j$ with outcome $n$ transfers the state of $j$ to $k$, with the Pauli correction $\XX_k^m\ZZ_k^n$. This is the general form of the injection protocol of Sec.~\ref{sec:injection}.

\emph{CNOT gate.} Let the control $c$ reside in module $A$ and the target $t$ in module $B$, and let a work qubit $a$, in either module, be prepared in $\ket+$. The measurements $\ZZ_c\ZZ_a$, $\XX_a\XX_t$ and $\ZZ_a$, with outcomes $m_1$, $m_2$ and $m_3$, implement $\mathrm{CNOT}_{c\to t}$ up to the Pauli correction $\ZZ_c^{m_2}\XX_t^{m_1+m_3}$~\cite{Litinski2019GameOfSurfaceCodes}. This follows by tracking the logical operators through the measurements. After the first measurement, the stabilizer $(-1)^{m_1}\ZZ_c\ZZ_a$ replaces $\XX_a$, and $\XX_c$ is represented by $\XX_c\XX_a$. After the second, the stabilizer $(-1)^{m_2}\XX_a\XX_t$ replaces $\ZZ_c\ZZ_a$, so that $\XX_c\XX_a$ is equivalent to $(-1)^{m_2}\XX_c\XX_t$, and $\ZZ_t$ is represented by $(-1)^{m_1}\ZZ_c\ZZ_a\ZZ_t$. After the third, the stabilizer $(-1)^{m_3}\ZZ_a$ replaces $\XX_a\XX_t$, so that $\ZZ_t$ is represented by $(-1)^{m_1+m_3}\ZZ_c\ZZ_t$, whereas $\ZZ_c$ and $\XX_t$ are unchanged. These are the conjugation rules of $\mathrm{CNOT}_{c\to t}$ up to the stated signs, which the Pauli correction removes. Only the first measurement crosses the module boundary if $a$ resides with $t$ (Fig.~5c in the main text).

\emph{Transversal operations.} Between two identical modules, transversal CNOT gates on the $2{,}844$ pairs of corresponding data qubits apply CNOT to all $1{,}426$ pairs of aligned logical qubits simultaneously, since a transversal CNOT between identical CSS codes preserves the canonical basis. Combined with shift automorphisms $\overline U^s$, this pairs register $i$ of $A$ with register $i$ of $B$ at any relative offset $s$. Such block operations complement the addressable instructions for data-parallel subroutines.

\emph{Networks.} A network of modules executes bridges along a matching of its module graph in parallel. Each module has one extractor, and a bridge occupies $18$ of its data vertices, so that different ports on the same extractor can serve different partners at different times. Magic-state factories are separate modules that attach through the same bridges (Sec.~\ref{sec:injection}).

\runin{Universality of the instruction sets}
Each of the three instruction sets is universal on the $1{,}422$ register qubits. We give the argument for $\IB$.
\begin{enumerate}
\item Stabilizer states are prepared by PPMs; for instance, measuring $\XX_j$ and applying a Pauli correction prepares $\ket+$, and measuring $\YY_j$ prepares an eigenstate of $\YY_j$.
\item Let a work qubit $a$ be prepared in the eigenstate $\ket{+i}$ of $\YY_a$. Measuring $\ZZ_j\ZZ_a$ and then $\XX_a$ applies $S_j=\mathrm{diag}(1,i)$ or $S_j^\dagger=\ZZ_jS_j$ to qubit $j$, depending on the outcomes; up to a Pauli correction, this implements the rotation $e^{-i\pi\ZZ_j/4}$. Exchanging the roles of $X$ and $Z$, measuring $\XX_j\XX_a$ and then $\ZZ_a$ with $a$ in an eigenstate of $\YY_a$ implements $e^{-i\pi\XX_j/4}$ up to a Pauli correction. These two rotations generate the single-qubit Clifford group, and together with the CNOT gate above they generate the full Clifford group.
\item An injected $\ket T$ state implements the $T$ gate by gate teleportation, which together with the Clifford group yields a universal gate set~\cite{Bravyi2005Universal,Litinski2019GameOfSurfaceCodes}.
\end{enumerate}

The set $\IH$ contains $\IB$, since a batch may consist of a single instruction, and $\IF$ supports Pauli-based computation directly, since every rotation $e^{-i\pi\PP/8}$ costs one measurement of $\PP\otimes\ZZ_\src$ and a single-qubit readout of the source (Sec.~\ref{sec:injection})~\cite{Bravyi2016Trading,Litinski2019GameOfSurfaceCodes}.

\section{Resource accounting}\label{sec:resources}
This section details the provisioning numbers compared in Fig.~6 in the main text. All configurations are compared at distance $d=18$ and a capacity of about $1{,}426$ logical qubits. Every check is counted with one measurement qubit, classical control is not counted, and magic-state factories are not counted. We compare with the bicycle architecture of two-gross modules~\cite{Yoder2025Tour}, whose gadgets are counted as reported there. The comparison concerns space only and does not account for the different logical error rates, cycle times and decoders of the architectures.

\emph{Gadget convention.} A gadget that adds $n_a$ ancilla data qubits and $m_a$ ancilla checks occupies $n_a+m_a$ physical qubits in addition to the module. The ancilla qubits of a gadget are the edges of its graph, so that $n_a=|E|$, whereas its vertices carry checks. A connected single-target gadget with a complete cycle basis has $|V|$ vertex checks and $|E|-|V|+1$ cycle checks, the dimension of the cycle space of a connected graph, so that $m_a=|V|+(|E|-|V|+1)=n_a+1$ irrespective of $|V|$, and the gadget occupies $2n_a+1$ physical qubits.

\emph{Memory.} A Cornucopia module stores $1{,}426$ logical qubits in $2{,}844+1{,}422=4{,}266$ physical qubits, $3.0$ per logical qubit. A fleet of $119$ two-gross $\cod{288}{12}{18}$ bicycle modules stores $1{,}428$ logical qubits in $119\times576=68{,}544$ physical qubits, $48$ per logical qubit, and a rotated distance-$18$ surface code requires $2\times18^2-1=647$ physical qubits per logical qubit (Fig.~6a in the main text). The advantage of a single large block is discrete: for up to $84$ logical qubits, seven or fewer two-gross modules require fewer physical qubits.

\emph{Basic processor.} The largest basic instruction on the $1{,}422$ register qubits is a bridged product with $772$ ancilla data qubits, so providing its workspace requires $2\times772+1=1{,}545$ physical qubits, for a total of $5{,}811$ physical qubits or $4.1$ per logical qubit. 
The $119$ bicycle modules, each with a logical processing unit of $158$ physical qubits~\cite{Yoder2025Tour}, require $119\times(576+158)=87{,}346$ physical qubits ($61.2$ per logical qubit; Fig.~6b in the main text). 
The logical processing unit measures any logical Pauli operator on two fixed logical qubits of its module and reaches the other logical qubits through shift automorphisms, so we count it as the basic workspace of a bicycle module. 
These numbers describe provisioning, not throughput: the bicycle fleet can execute one instruction in each of its $119$ modules simultaneously, whereas a single Cornucopia workspace executes one instruction at a time.

\emph{Benchmark workloads.} We quantify the trade-off between space and logical depth within our architecture by executing three random workloads of $3{,}000$ basic instructions on the register qubits with one basic workspace, with the high-throughput instruction set and with the fixed extractor (Methods and Tab.~\ref{tab:workload}). The high-throughput set reduces the logical depth by a factor of $20$ with $5.1$ times the physical qubits of the basic configuration, and thereby reduces the space--time volume by a factor of $3.9$ (Fig.~6c in the main text). 
The fixed extractor needs as many logical cycles as the basic workspace on these workloads, because it measures one product per cycle; its advantage lies in high-weight products, each of which it measures in a single cycle, whereas the basic instruction set must compile such a product into several measurements. 
The benchmark compares only configurations of the Cornucopia architecture, which execute the same instructions and count time in the same logical cycles. We make no such comparison with the bicycle architecture: the depth of a workload on $119$ bicycle modules depends on how the logical qubits are distributed over the modules and on how the operations between modules are carried out.

\begin{datatable}
\caption{\textbf{Benchmark workloads.} Three random workloads of $3{,}000$ basic instructions ($\XX_j$, $\YY_j$, $\ZZ_j$, $\XX_j\XX_k$ and $\ZZ_j\ZZ_k$ on uniformly random register qubits) executed with one basic workspace, with the high-throughput instruction set and with the fixed extractor. Cycles: logical depth in surgery cycles of $O(d)$ syndrome rounds; per cycle: mean and largest number of instructions per cycle; qubits: physical qubits to be provisioned, which for the high-throughput set is the largest layer of the workload; volume: provisioned qubits times cycles.}
\label{tab:workload}
\begin{ruledtabular}
\begin{tabular}{lrrrr}
Configuration & cycles & per cycle & qubits & volume ($10^6$ qubit-cycles)\\\hline
basic workspace & 3{,}000 & 1 & 5{,}811 & 17.4\\
high-throughput, workload 1 & 149 & 20.1 / 25 & 29{,}211 & 4.4\\
high-throughput, workload 2 & 151 & 19.9 / 25 & 29{,}755 & 4.5\\
high-throughput, workload 3 & 153 & 19.6 / 24 & 29{,}263 & 4.5\\
fixed extractor & 3{,}000 & 1 & 46{,}927 & 140.8\\
\end{tabular}
\end{ruledtabular}
\end{datatable}

\emph{Fixed extractor.} With its extractor, a Cornucopia module comprises $46{,}927$ physical qubits, namely $24{,}174$ data qubits and $22{,}753$ check qubits, and measures products across all $1{,}426$ logical qubits of the module. The bicycle architecture has no fixed-extractor configuration in this comparison. Connecting two modules requires $18+17=35$ physical qubits per bridge in the Cornucopia architecture.

Tab.~\ref{tab:resources} summarizes the configurations of the two architectures. With a basic workspace, a Cornucopia module requires $4.1$ physical qubits per logical qubit on its register qubits, compared with $61.2$ for bicycle modules, so the high encoding rate largely survives the addition of a basic instruction set. 
The gadget cost per logical operation is higher: the canonical $X$- and $Z$-type representatives have weights $62$--$120$, so the workspace of the largest basic instruction occupies $1{,}545$ physical qubits, compared with $158$ for a logical processing unit. 
This is compensated by a single workspace serving all $1{,}422$ register qubits of the module, whereas a logical processing unit serves the logical qubits of its own module. 
The balance depends on the workload: throughput-bound computations favour more parallel workspace or more modules, whereas memory-bound computations favour few large blocks.

\begin{table}[tbp]
\caption{\textbf{Physical-qubit provisioning at distance $18$.} Physical qubits for about $1{,}426$ logical qubits in the Cornucopia and bicycle architectures, with one measurement qubit per check and without magic-state factories. The bicycle architecture has no fixed-extractor configuration in this comparison. $^{a}$For a Cornucopia module, the workspace serves the $1{,}422$ register qubits; the logical processing unit of a bicycle module serves the logical qubits of that module~\cite{Yoder2025Tour}.}
\label{tab:resources}
\begin{ruledtabular}
\begin{tabular}{lrr}
Configuration & Cornucopia (1 module) & bicycle (119 modules)\\\hline
logical qubits & 1,426 & 1,428\\
memory & 4,266 & 68,544\\
memory and basic workspace$^{a}$ & 5,811 & 87,346\\
memory and fixed extractor & 46,927 & --\\
\end{tabular}
\end{ruledtabular}
\end{table}

\bibliography{references}